\documentclass[12pt]{article}
\usepackage{amsfonts}
\usepackage{amsmath, amssymb}
\usepackage{youngtab}
\usepackage{hyperref}
\usepackage[dvips]{color}
\usepackage{psfrag}
\usepackage{feynmp}
\usepackage[pdftex]{graphicx}
\usepackage{braket}
\usepackage{bm}
\usepackage{subfigure}

\usepackage{ulem}
\usepackage{comment}
\includecomment{pdffig}

\allowdisplaybreaks[1]

\Yboxdim{5pt}

\makeatletter
    
    \@addtoreset{equation}{section}
  \makeatother

\usepackage{color}

\def\rem#1{}

\renewcommand{\title}[1]{\vbox{\center\LARGE{#1}}\vspace{5mm}}
\renewcommand{\author}[1]{\vbox{\center\large#1}\vspace{5mm}}

\begin{document}
\bibliographystyle{utphys}

\begin{titlepage}
\begin{center}
\vspace{5mm}
\hfill {\tt 
IPMU19-0176
}\\
\vspace{20mm}

\title{\LARGE  
3d $\mathcal{N}=2$ Chern--Simons--matter theory, Bethe ansatz, and quantum $K$-theory of Grassmannians  
}
\vspace{7mm}

Kazushi Ueda$^a$ and  Yutaka Yoshida$^b$

\vspace{6mm}

$^a$Graduate School of Mathematical Sciences,
The University of Tokyo, \\
  3-8-1 Komaba Meguro-ku, Tokyo 153-8914
Japan \\
\href{kazushi@ms.u-tokyo.ac.jp}{\tt kazushi@ms.u-tokyo.ac.jp}

\vspace{3mm}
$^b$Kavli IPMU (WPI), UTIAS, The University of Tokyo \\
 Kashiwa, Chiba 277-8583, Japan \\
\href{mailto:yutaka.yoshida@ipmu.jp}{\tt yutaka.yoshida@ipmu.jp}

\end{center}

\vspace{7mm}
\abstract{
We study a correspondence between
3d $\mathcal{N}=2$ topologically twisted Chern--Simons--matter theories on $S^1 \times \Sigma_g$
and quantum $K$-theory of Grassmannians.  
Our starting point is a Frobenius algebra depending on a parameter $\beta$
associated with an algebraic Bethe ansatz introduced by Gorbounov--Korff. 
They showed that the Frobenius algebra with $\beta=-1$ is isomorphic
to the (equivariant) small  quantum $K$-ring of the Grassmannian,
and the Frobenius algebra with  $\beta=0$ is isomorphic
to the equivariant small quantum cohomology of the Grassmannian.
We apply supersymmetric localization formulas
to the correlation functions of supersymmetric Wilson loops
in the Chern--Simons--matter theory
and show that the algebra of Wilson loops is isomorphic
to the Frobenius algebra with $\beta=-1$.
This allows us to identify the algebra of Wilson loops
with the quantum $K$-ring of the Grassmannian.
We also show that
correlation functions of Wilson loops on  $S^1 \times \Sigma_g$
satisfy the axiom of 2d TQFT.  
For $\beta=0$,
we show the correspondence between an A-twisted GLSM,
the Frobenius algebra for $\beta=0$,
and the quantum cohomology of the Grassmannian.
We also discuss deformations of Verlinde algebras,
omega-deformations,
and the $K$-theoretic $I$-functions of Grassmannians with level structures.
\begin{center}
\end{center}
}
\vfill

\end{titlepage}

\tableofcontents

\section{Introduction and summary}
$\mathcal{N}=(2,2)$ supersymmetric gauged linear sigma models (GLSM) in two dimensions \cite{Witten:1993yc}
play important roles in connection with quantum cohomology, Gromov--Witten invariants, and mirror symmetry.
Correlation functions of the vector multiplet scalars of abelian A-twisted GLSM on the two-sphere $S^2$ was studied
in \cite{Morrison:1994fr}. 
When the Higgs branch of $\mathcal{N}=(2,2)$ GLSM is a toric Fano manifold,
the three-point correlation functions give the three-point genus-zero Gromov--Witten invariants
of that manifold.
When the Higgs branch is a Calabi--Yau 3-fold,
the identification of the three-point functions
with the B-model Yukawa couplings of the mirror
is known as toric residue mirror symmetry
\cite{MR1988969, MR2019144, MR2104791, MR2099774, MR2147350}.
Three-point functions of non-abelian A-twisted GLSMs on $S^2$,
possibly with the $\Omega$-background,
are computed in  \cite{Closset:2015rna} using supersymmetric  localization.
It was suggested in \cite{Ueda:2016wfa} that the resulting formulas
give genus-zero quasimap invariants of the space of Higgs branch vacua of the GLSM,
and proved mathematically in \cite{2016arXiv160708317K}
for Grassmannians and their Calabi--Yau complete intersections
defined by equivariant vector bundles.
It follows that the  twisted chiral ring of 2d A-twisted GLSM,
whose Higgs branch is the Grassmannian, agrees with the small quantum cohomology ring of the Grassmannian.

When a supersymmetric quantum field theory on a manifold $M$ is related to a cohomological object, an $S^1$ uplift of the theory, i.e.,  
a supersymmetric quantum field theory on  $S^1 \times M$, is often related to a $K$-theoretic object. A typical example is Nekrasov's instanton partition functions;
roughly speaking,
instanton partition functions on $\mathbb{R}^4$ \cite{Nekrasov:2002qd}
are integrals of cohomology classes
over instanton moduli spaces \cite{Nakajima:2003pg}, whereas
five-dimensional instanton partition functions on $S^1 \times \mathbb{R}^4$ \cite{Nekrasov:2004vw}
are indices of instanton moduli spaces ($K$-theoretic instanton partition functions)  \cite{Nakajima:2005fg}. 
Hence  it is natural to expect the existence of 3d (topologically twisted) supersymmetric gauge theories relevant to quantum $K$-theory \cite{MR1943747},
which is a $K$-theoretic analogue of quantum cohomology.   

In this paper,
we establish a relation
between certain 3d topologically twisted supersymmetric gauge theories
and the quantum $K$-rings of Grassmannians.
Although there are earlier works
in this direction,
starting with \cite{Dimofte:2010tz}
which pointed out a similarity
between the small $K$-theoretic $J$-function of $\mathbb{P}^{M-1}$
and the vortex partition function of a certain 3d $U(1)$ gauge theory
in the context of open BPS state counting,
to our best knowledge,
an identification of the small quantum $K$-ring of Grassmannians
(or even projective spaces)
with the  level of precision of this paper is new.

To find the precise relation between the 3d supersymmetric theories and the quantum $K$-ring of Grassmannians,
we focus on Bethe/Gauge correspondence. 
It was pointed out in  \cite{Nekrasov:2009uh, Nekrasov:2009ui} that the saddle point equations of the mass deformations of 
a 2d  $\mathcal{N}=(4,4)$ $U(N)$ GLSM 
 coincide with the Bethe ansatz equation of
the spin-$\frac{1}{2}$ XXX  spin chain. 
The relation between the twisted chiral ring and the quantum cohomology mentioned above implies that the Bethe ansatz equation is closely related to 
the relations of the equivariant quantum cohomology ring of the cotangent bundle of the Grassmannian, which appears as the Higgs branch  of the $\mathcal{N}=(4,4)$ GLSM. 
See e.g.~\cite{Maulik:2012wi} for a mathematical formulation of this observation.

For the quantum cohomology of the Grassmannian,
a description
in terms of the algebraic Bethe ansatz of a  five-vertex model
is given by Gorbounov--Korff \cite{Gorbounov:2014oxa}.
In a subsequent work \cite{Gorbounov:2014bra},
they introduced a finite-dimensional commutative Frobenius algebra
depending on a parameter $\beta$.
They showed that
the Frobenius algebra with  $\beta=-1$ is isomorphic to the small quantum $K$-ring of Grassmannian
\footnote{The equivariant case was originally conjectured in \cite{Gorbounov:2014bra}
and proved in \cite{MR3847206}.}
and the Frobenius algebra with $\beta=0$ reproduces their previous result
on the quantum cohomology in \cite{Gorbounov:2014oxa}. 

By using supersymmetric localization formulas,
we show that the saddle point equations of
the 3d $U(N)$ Chern--Simons-matter theory with $M$ chiral multiplets
coincide with the Bethe ansatz equations for $\beta=-1$.
This implies that
the operator product expansions (OPEs) of
Wilson loop operators $\mathcal{W}_\lambda$
can be identified with
the quantum products of the equivariant quantum $K$-theory ring
$QK_T (\mathrm{Gr}(N,M))$
of the Grassmannian;
\begin{align}
\mathcal{W}_{\mu} \mathcal{W}_{\nu}= \sum_{\lambda} C_{\mu \nu}^{\lambda} \mathcal{W}_{\lambda}
\Longleftrightarrow [ \mathcal{O}_{\mu} ] \star [ \mathcal{O}_{\nu} ]= \sum_{\lambda} C_{\mu \nu}^{\lambda} [ \mathcal{O}_{\lambda}] .
\label{eq:ope1}
\end{align}
Here $\star$ is the quantum product in $QK_T (\mathrm{Gr}(N,M))$
and $[\mathcal{O}_{\lambda}]$'s are 
the $K$-theory classes of Schubert varieties of the Grassmannian.
We also revisit the relation between A-twisted 2d GLSM and quantum cohomology
from the viewpoint of Bethe/Gauge correspondence, and
derive an isomorphism of the twisted chiral ring with the equivariant small quantum cohomology of the Grassmannians.

Generalizations of the supersymmetric localization formula
from $S^1 \times S^2 $ (resp. $S^2$)
to higher genus cases $S^1 \times \Sigma_g$ (resp. $\Sigma_g$)
were studied in \cite{Benini:2016hjo, Closset:2016arn}.
We show that the correlation functions at higher genus can be identified
with the correlation functions of the 2d topological quantum field theories (TQFTs) associated with the Frobenius algebra.

Besides the correspondence between the algebra of Wilson loops in the topologically twisted CS--matter theories and 
the quantum $K$-rings of Grassmannians, we discuss several other aspects of CS--matter theories;
 CS--matter theories with $\Omega$-background, 
$K$-theoretic $I$-functions of Grassmannians with level structures,  supersymmetric index on 
 $S^1 \times D^2$, and deformations of Verlinde algebra in terms of indices over moduli stacks of 
$G$-bundles on Riemann surfaces.

This article is organized as follows. 
In section \ref{sec:resultGK}, we recall results by Gorbounov--Korff
which will be used later to show the relation  between quantum $K$-ring (resp.~quantum cohomology) and CS--matter theory (resp.~GLSM). 
In section \ref{sec:Frob2dTQFT}, we compute the correlation functions of
2d TQFT on a genus $g$ surface associated with the Frobenius algebra. 
In section \ref{sec:norm}, we evaluate the inner product of the on-shell Bethe vectors
by the Izergin--Korepin method.
In section \ref{sec:GLSM},
we show that the genus $g$ correlation function of the 2d TQFT
 associated with the Frobenius algebra with $\beta=0$
agrees with 
the genus $g$ correlation functions of the A-twisted GLSM.
In section \ref{sec:CSM}, which is the main part of our paper, we study the correspondence between CS--matter theories and the quantum $K$-theory  of Grassmannians.  In section \ref{sec:CSMsub1},
 we show that the supersymmetric localization formulas of the correlation functions of Wilson loops on $S^1 \times \Sigma_g$ agree with
the  genus $g$ correlation functions of 2d TQFT associated with the Frobenius algebra
 with $\beta = -1$.
These equalities lead to the isomorphism between the algebras of Wilson loops and the quantum $K$-ring of Grassmannians.
In section \ref{sec:CSMsec2}, we turn on the $\Omega$-background and show that 
the partition function of topologically twisted CS--matter theory with specific choices of CS levels factorizes
to a pair of $K$-theoretic $I$-functions,
 and discuss the relation between CS levels and level structures of quantum $K$-theory.
In section \ref{sec:S1xD2}, we show that  $K$-theoretic $I$-function derived in section \ref{sec:CSMsec2}
can be derived from a supersymmetric index on $S^1 \times D^2$. In section \ref{sec:Verlinde},
we discuss connections between CS--matter theories and 
deformations of Verlinde algebra.

{\it Note added}: When our paper\footnote{Results of our paper were presented by the second author at Workshop on 3d mirror symmetry and AGT conjecture, Zhejiang University, China, October 21-25, 2019, organized by {Hans Jockers}, Yongbin Ruan and Yefeng Shen.}
 was  being completed, we became aware of  the paper \cite{Jockers:2019lwe} by {H.~Jockers} {\it et al.}~on the arXiv, which has partial overlap with some results in our paper.

\section{Quantum $K$-ring, quantum cohomology and  algebraic Bethe ansatz}
\label{sec:section2}
 Gorbounov--Korff  \cite{Gorbounov:2014bra} defined a Frobenius algebra depending on a parameter $\beta$ in terms of  algebraic Bethe ansatz.
 The elements of the Frobenius algebra are defined by  the $N$ excitations (particle states)  in the Hilbert space of the quantum integrable system with $M$-sites. 
The Frobenius bilinear form is defined in terms of the inner product of states and dual of states in the Hilbert spaces. 
They showed that the structure constants  in the spin basis agree with  the structure constants of   
the equivariant quantum $K$-ring  of Grassmannian ${QK}_T (\mathrm{Gr}(N, M))$ for $\beta=-1$ and the equivariant quantum cohomology  of Grassmannian ${QH}_T (\mathrm{Gr}(N, M))$ for $\beta=0$.

In section \ref{sec:resultGK}, we summarize results from \cite{Gorbounov:2014bra}
which will be useful later.
In particular, we recall the definition and properties of the Frobenius algebra
introduced in \cite{Gorbounov:2014bra}.
In  section \ref{sec:Frob2dTQFT},
we
describe the correlation functions
of the 2d TQFT associated with the Frobenius algebra.
In section \ref{sec:norm}, we evaluate the on-shell square norm of Bethe vectors and dual Bethe vectors.

\subsection{Results of Gorbounov--Korff }
\label{sec:resultGK}
First we introduce the R-matrix and L-operator which characterize the quantum integrable model in \cite{Gorbounov:2014bra}.
The R-matrix ${\sf R}({ x}, { y}) \in  \mathrm{End}[ \mathbb{C}^2 ({ x}) \otimes \mathbb{C}^2 ({ y}) ]$ is  defined by
\begin{align}
{\sf R}({ x}, { y}):=\left(
\begin{array}{cccc}
 1 & 0 & 0&0 \\
 0 & 0 & 1 & 0\\
 0 & 1 + \beta ({ y} \ominus { x} )& { y} \ominus { x} &0\\
0 & 0 & 0& 1 \\
\end{array}
\right) , 
\label{eq:Rmatrix}
\end{align}
where $\ominus$ is defined in Appendix \ref{appendix1}. 
The R-matrix satisfies the following Yang-Baxter equation:
\begin{align}
{\sf R}_{1 2}({ x, y}) {\sf R}_{1 3 }({ x, z}) { \sf  R}_{2 3 }({ y, z})= { \sf R}_{1 3 }({ x, z}) {\sf R}_{2 3 }({ y, z}) {\sf R}_{1 2}({ x, y}),
\label{eq:YBequation}
\end{align}
with
\begin{align}
&{\sf R}_{1 2}({ x, y}):={\sf R}({ x, y}) \otimes \mathbb{I}_{2}, \\
&{\sf R}_{2 3}({ x, y}):=\mathbb{I}_{2} \otimes {\sf R}({ x, y}), \\
&{\sf R}_{1 3}({ x, y}):=(\mathbb{I}_{2} \otimes {\sf P} ) {\sf R}_{12}({ x, y}) (\mathbb{I}_{2} \otimes {\sf P} ) .
\end{align}
Here  $\mathbb{I}_{2}$ is the  identity matrix of size two and ${\sf P} $ is a permutation matrix defined by ${\sf P} (v \otimes u )=u \otimes v$ for $u, v \in \mathbb{C}^2$. 
The L-operator ${\sf L}({ x | t}) \in \mathrm{End} [\mathbb{C}^2 ({ x}) \otimes \mathbb{C}^2 ({t})] $ is defined by  
\begin{align}
{\sf L} ({ x | t})=\left(
\begin{array}{cc}
 \sigma^+ \sigma^- + ({x}  \ominus { t} )\sigma^- \sigma^+ & (1+\beta ({ x} \ominus { t}) ) \sigma^+  \\
 \sigma^- & \sigma^- \sigma^+ \\
\end{array}
\right), 
\end{align}
with
\begin{align}
\sigma^+ =\left(
\begin{array}{cc}
 0 & 0  \\
 1 & 0 \\
\end{array}
\right), \quad 
\sigma^- =\left(
\begin{array}{cc}
 0 & 1  \\
 0 & 0 \\
\end{array}
\right) .
\end{align}
The R-matrix and the L-operator  satisfy the following Yang-Baxter type relation (RLL relation); 
\begin{align}
{\sf R}_{12} ({ x, y}) \left( {\sf L}({ x | t}) \otimes {\sf L} ({ y | t}) \right)=\left( {\sf L}({ y | t}) \otimes {\sf L}({ x | t}) \right)  {\sf R}_{12}({ x, y}) .
\label{eq:RLL}
\end{align}
From \eqref{eq:RLL} one can show that  the monodromy matrix ${\sf M} ({ x| t}) \in \mathrm{End}[\mathbb{C}^2 ({ x}) \otimes \otimes_{i=1}^M \mathbb{C}^2 ({ t}_i)]$  defined by
\begin{align}
{\sf M} ({ x| t}) =
\left(
\begin{array}{cc}
 {\sf A}({ x|t}) & {\sf B}({ x|t})  \\
 {\sf C}({ x|t}) & {\sf D}({ x|t}) \\
\end{array}
\right)
:= {\sf L} ({ x} | { t}_{M}) \cdots {\sf L}({ x} | { t}_{2}) {\sf L}({ x} | { t}_{1})
\end{align}
also satisfies the Yang-Baxter  type relation (RTT relation)
\begin{align}
{\sf R}_{12}(x, y) \left( {\sf M} (x | t) \otimes {\sf M} (y | t) \right)=\left( {\sf M} (y | t) \otimes {\sf M}(x| t) \right) {\sf R}_{12}(x, y) .
\label{eq:RTT}
\end{align}
Here ${\sf L}({ x} | { t}_{i})$ non-trivially acts on $ \mathbb{C}^2 ({ x}) \otimes \mathbb{C}^2 ({ t}_i)$.
\eqref{eq:RTT} gives the sixteen relations between four $2^{M} \times 2^{M}$ matrices $ {\sf A}(x|t) ,{\sf B}(x|t), {\sf C}(x|t)$ and   ${\sf D}(x|t) $  listed in appendix \ref{sec:RTT}.
The twisted transfer matrix is defined by the trace taken over the auxiliary space as
\begin{align}
{\sf T} (y|t):=\mathrm{Tr} \left( {\sf M}(y|t)  
\left(
\begin{array}{cc}
 1 & 0  \\
 0 & q \\
\end{array}
\right) \right)
=  {\sf A}({ y | t})+q {\sf D}({ y|t}) .
\label{eq:trans}
\end{align}
The periodic boundary condition is given by  $q=1$. Note that the  twist parameter $q$ will be identified with the 
quantum parameter of the quantum $K$-ring for $\beta=-1$ and the quantum parameter of  the quantum cohomology for $\beta=0$. 
From the relation \eqref{eq:RTT}, the transfer matrices commute with each other:
\begin{align}
[{\sf T} ({ y}|t), {\sf T}({ y^{\prime}|t})]=0 .
\end{align}
The  coefficients of ${y}^l$ for $l=0,1, \cdots$ in  ${\sf T}({y}|{ t})$ give  conserved charges.

The states and dual states on which ${\sf A,B,C,D}$ act are defined as follows.
First we introduce a basis of the $2^M$-dimensional space $\otimes_{i=1}^M \mathbb{C}^2({t}_i)$ by 
\begin{align}
\otimes_{i=1}^M {v}_{\ell_i} \, \text{ with } \, \ell_i \in \{ 0, 1 \} ,   
\end{align}
where $v_{0}$ and $v_1$ are defined by
\begin{align}
v_0:=\left(
\begin{array}{c}
 1  \\
 0 \\
\end{array}
\right),
v_1:=\left(
\begin{array}{c}
 0  \\
 1 \\
\end{array}
\right) ,
\end{align}
and we also introduce the dual space  spanned by 
\begin{align}
\otimes_{i=1}^M {v}_{\ell_i}^T  \, \text{ with } \, \ell_i \in \{0, 1 \}   .
\end{align}
The $2^{M}$-dimensional space is decomposed to the direct sum  $\bigoplus_{N=0}^{M} \mathcal{V}_{N, M}$, where   
 $\mathcal{V}_{N, M}$ is a $M !/ N ! (M-N)!$-dimensional vector space spanned by the vectors $\otimes_{i=1}^M {v}_{\ell_i}$ with $\sum_{i=1}^M \ell_i=N$. 
 The vectors $\otimes_{i=1}^M {v}_{\ell_i}$ with $\sum_{i=1}^M \ell_i=N$ 
are in one-to-one correspondence with the elements of a set of partitions $\mathcal{P}_{N,M}$ defined by
\begin{align}
\mathcal{P}_{N,M}:= \{ \lambda=(\lambda_1, \dots\lambda_N) |  M-N \ge \lambda_1 \ge \lambda_2 \ge \cdots \ge \lambda_N \ge 0  \} .
\end{align}
 We define a vector $| v_{\lambda} \rangle$ by $\otimes_{i=1}^M {v}_{\ell_i}$ which correspond to a partition $\lambda \in \mathcal{P}_{N,M}$ . We call $\{ | v_{\lambda} \rangle \}_{\lambda \in \mathcal{P}_{N, M}}$ as the spin basis of $\mathcal{V}_{N, M}$. 
The dual spin basis $\{ \langle v_{\lambda} | \}_{\lambda \in \mathcal{P}_{N, M}}$ for the dual space of 
$\mathcal{V}_{N, M}$  is defined in a similar way and the inner  products satisfy  $\langle v_{\lambda} | v_{\mu} \rangle=\delta_{\lambda \mu}$.
We define $\emptyset$ by $\emptyset:= (0, \cdots, 0) \in \mathcal{P}_{N, M}.$

The pseudo vacuum  $| 0 \rangle$ and the dual pseudo vacuum $\langle 0 |$ are defined by  
\begin{align}
|0 \rangle := \otimes_{i=1}^M v_1, \quad \langle 0 | := \otimes_{i=1}^M v_1^{T} .
\end{align}
The ${\sf A}, {\sf B}, {\sf C}, {\sf D}$ act on the pseudo vacuum as
\begin{align}
&{\sf A}({ y}| { t}) | 0 \rangle={\sf a}({ y},  { t}) | 0 \rangle, \, \langle  0 | {\sf B}({ y}| { t}) =0, \, {\sf C}({ y}|{ t}) | 0 \rangle=0, \,  {\sf  D}({ y}| { t}) | 0 \rangle={\sf d}({ y}, { t}) | 0 \rangle,
\label{eq:AaDd}
\end{align}
with
\begin{align}
{\sf a}({ y}, { t}):= \prod_{i=1}^M (y \ominus t_i),   \quad {\sf d}({ y}, { t}):= 1. 
\end{align}

Suppose  $\prod_{a=1}^N {\sf B}({ y}_a) | 0 \rangle$  is  an eigen vector of 
the transfer matrix \eqref{eq:trans}, then   
 $\{{ y}_1, \cdots, {y}_N\}$ have to satisfy the following system of equations  called the Bethe ansatz equation:
\begin{align}
 \frac{(-1)^{N-1} q (1+ \beta y_a)^N}{\prod_{b=1}^N (1+ \beta y_b) \prod_{i=1}^M (y_a \ominus t_i)}  =1, \quad  a=1, \cdots, N .
\label{eq:Betheansatz}
\end{align}
A root $\{y_a \}_{a=1}^N$ of the Bethe ansatz equation  is called  as a Bethe root.
When $\{y_a \}_{a=1}^N$ is a Bethe root, $\prod_{a=1}^N {\sf B}(y_a) | 0 \rangle$ $(\text{resp. } \langle 0 | \prod_{a=1}^N {\sf C} (y_a) $) is  called an on-shell Bethe vector, (resp. on-shell dual Bethe vector), vice versa.
When  $\{ y_a \}_{a=1}^N$ is indeterminate, Bethe vectors are called off-shell.  

The  Bethe ansatz equation has $\frac{M!}{N! (M-N)!}$ distinct Bethe roots.  The $\frac{M!}{N! (M-N)!}$ Bethe roots are characterized by the
partitions in $\mathcal{P}_{N,M}$ as
\begin{align}
{ y}_{\lambda}=(y_{\lambda_N+1}, \cdots, y_{\lambda_2+N-1}, y_{\lambda_1+N}) \, \text{ for } \lambda \in \mathcal{P}_{M,N},
\label{eq:BrootsP}
\end{align}
which has the following expansions:
\begin{align}
y_j=
\left\{
\begin{array}{cc}
t_j+q  (-1)^{N-1} \frac{(1+\beta t_j)^{N+1}}{\Pi(t_{\lambda}) \prod_{l  \neq j} t_j \ominus t_l }+ O(q^2),  &  (\text{inhomogeneous}), \\
q^{\frac{1}{M}} e^{\frac{2\pi {\rm i} j}{M}} + \beta (-1)^{N-1}  q^{\frac{2}{M}} e^{\frac{2 \pi {\rm i} j}{M}} \sum_{l \neq j} (e^{\frac{2 \pi {\rm i} l}{M}} -e^{\frac{2 \pi {\rm i} j}{M}})+ O(\beta^2), & (\text{homogeneous}). 
\end{array}
\right.
\end{align}
Here   homogeneous  means  $t_i=0$ for $i=1, \cdots, M$. Inhomogeneous means    $t_i \neq 0 $ for $i=1, \cdots, M$. 
 On-shell (dual) Bethe vectors  $| { y}_{\lambda} \rangle$ and $\langle { y}_{\lambda} |$ for $\lambda \in \mathcal{P}_{N,M}$ associated to the Bethe root \eqref{eq:BrootsP}, are defined by
\begin{align}
&| { y}_{\lambda} \rangle:=  \prod_{a=1}^N {\sf B}(y_{\lambda_a+N+1-a}) | 0 \rangle, \quad 
\langle { y}_{\lambda} |:= \langle 0 |  \prod_{a=1}^N {\sf C}(y_{\lambda_a+N+1-a}) .
\end{align}
  $\{ | { y}_{\lambda} \rangle \}_{\lambda \in \mathcal{P}_{N ,M}}$ forms a basis of $\mathcal{V}_{N, M}$. 
 From the orthogonality and completeness conditions of the on-shell Bethe vectors, the following 
relations hold:
\begin{align}
\delta_{\lambda \mu}&=\sum_{\alpha \in \mathcal{P}_{M,N}} \frac{\Pi(y_{\lambda})}{\Pi (t_{\alpha})}
\frac{G_{\alpha^{\vee} }(y_{\lambda} |\ominus t^{\prime}) G_{\alpha}(y_{\mu} | \ominus t) }{{\sf e} ( y_{\lambda} ,  y_{\lambda})  },
\label{eq:orth} \\
\delta_{\lambda \mu}&=\sum_{\alpha \in \mathcal{P}_{M,N} } \frac{\Pi(y_{\alpha})}{\Pi (t_{\lambda})}
\frac{G_{\lambda^{\vee} }(y_{\alpha} |\ominus t^{\prime}) G_{\mu}(y_{\alpha} | \ominus t) }{ {\sf e}(y_{\alpha},y_{\alpha}) } .
\label{eq:complete}
\end{align}
Here $G_{\lambda}(y | t)=G_{\lambda}(y_1, \cdots, y_N | t)$ is a factorial Grothendieck polynomial which is given by a  determinant \cite{MR3062739} as
\begin{align}
G_{\lambda}(y|t)= \frac{\det_{1 \le a,b \le N} [(y_b|t)^{\lambda_a+N-a} (1+ \beta y_b)] }{\det_{1 \le a,b \le N}[y_b^{N-a}] }.
\end{align}
$\Pi(y)$, $\Pi(t_{\lambda})$,  $\ominus t$, $\ominus t^{\prime}$ and $(y_b|t)^{k}$ are defined in Appendix \ref{appendix1} .
The partition $\lambda^{\vee}$ associated to $\lambda=(\lambda_1, \cdots,\lambda_N) \in \mathcal{P}_{N,M}$ is defined by  
\begin{align}
\lambda^{\vee}:=(M-N-\lambda_N,M-N-\lambda_{N-1}, \cdots, M-N-\lambda_1) .
\end{align}
Note that  $G_{\lambda}(y|t)$ for $\beta=0$ agrees with a factorial Schur polynomial given by
\begin{align}
s_{\lambda}(y|t)= \frac{\det_{1 \le a,b \le N} [ \prod_{i=1}^{{\lambda_a+N-a}} (y_b+t_i) ] }{\det_{1 \le a,b \le N}[ \prod_{i=1}^{{N-a}} (y_b+t_i)] }.
\end{align}
When $t_i=0$, a factorial Schur polynomial $s_{\lambda}(y|t)$ is reduced to a Schur polynomial $s_{\lambda}(y)$. 
${\sf e} ({ y}, { y})$ in \eqref{eq:orth} and \eqref{eq:complete} is  the on-shell square norm defined by the inner product of an on-shell Bethe vector  and an on-shell dual Bethe vector;
\begin{align}
{\sf e} ({y}, { y}):= \langle 0 | \prod_{a=1}^N {\sf C}(y_a) \prod_{a=1}^N {\sf B}(y_a)| 0 \rangle .
\end{align}
Here the inner product  is called   on-shell (resp.~off-shell), when $\{y_a \}_{a=1}^N$ are a Bethe root (resp. indeterminate).
As we will see  \eqref{eq:orth} and \eqref{eq:complete} are useful to show the equality of  the equivalence between the partition function of 2d TQFT on $\Sigma_g$ 
and the topologically twisted index (for simplicity we call partition function)  of a Chern--Simons--matter theory on $S^1 \times \Sigma_g$ for 
$\beta=-1$ and also the partition function of a 2d A-twisted GLSM on $\Sigma_g$ for $\beta=0$.

\subsubsection*{The definition and properties of the Frobenius algebra}
Now we explain the definition and properties of the Frobenius algebra in \cite{Gorbounov:2014bra}.
Let $\{ Y_{\lambda} \}_{\lambda \in  \mathcal{P}_{N, M} }$  be  a basis of   $\mathcal{V}_{N, M}$ defined by 
\begin{align}
Y_{\lambda}:=\frac{1}{{\sf e}(y_{\lambda}, y_{\lambda})} | y_{\lambda} \rangle.
\end{align}
The associative product $*: \mathcal{V}_{N,M} \otimes \mathcal{V}_{N, M} \to \mathcal{V}_{N, M}$ is defined by
\begin{align}
Y_{\mu} * Y_{\nu}:= \delta_{\mu \nu} Y_{\nu}.
\end{align}
The identity element is given by $| v_{\emptyset} \rangle =\sum_{\lambda \in \mathcal{P}_{N,M}} Y_{\lambda}$.
A Frobenius bilinear  form  $\eta: \mathcal{V}_{M,N} \otimes \mathcal{V}_{M,N} \to R$ 
is defined by 
\begin{align}
\eta (Y_{\mu}, Y_{\nu}) :
 = \frac{\delta_{\mu \nu }}{{\sf e}(y_{\mu}, y_{\mu})} .
\label{eq:innerprod}
\end{align}
Here the coefficient ring is defined by $R:=\mathbb{Z}[\![ q]\!]\otimes \mathcal{R}(t_1, t_2, \cdots, t_M)$ and 
 $\mathcal{R}$ is the ring of rational function of $\beta$  regular at $\beta=0,-1$. 
$\eta$ is non-degenerate, since $\{ | y_{\lambda} \rangle \}_{\lambda \in \mathcal{P}_{N, M}}$ is a basis.    
 $(\mathcal{V}_{N,M}, *, \eta)$ is  a finite dimensional commutative Frobenius algebra. For simplicity we call $\mathcal{V}_{N,M}$ as the Frobenius algebra. 

Next we explain the relations between the Frobenius algebra, quantum $K$-ring and quantum cohomology.
Let  $\{C_{\mu \nu}^{\lambda}(q,t, \beta) \}_{\lambda, \mu, \nu \in \mathcal{P}_{N, M} }$ be the structure constants of $\mathcal{V}_{N,M}$ in the spin basis $\{ | v_{\lambda} \rangle \}_{\lambda \in \mathcal{V}_{N,M}}$:
\begin{align}
 | v_{\mu} \rangle  *  | v_{\nu} \rangle= \sum_{\lambda \in \mathcal{P}_{N,M}} C_{\mu \nu}^{\lambda}(q,t, \beta)  | v_{\lambda} \rangle .
\end{align}
The transformation from the on-shell Bethe vectors to the spin basis,  $C_{\mu \nu}^{\lambda}(q,t, \beta)$  is written as
\begin{align}
C_{\mu \nu}^{\lambda}(q,t, \beta)=\sum_{\alpha \in \mathcal{P}_{M,N}} \frac{\Pi(y_{\alpha})}{\Pi(t_{\lambda})} 
\frac{G_{\mu}(y_{\alpha} | \ominus t)G_{\nu}(y_{\alpha} |\ominus t) G_{\lambda^{\vee}}(y_{\alpha} | \ominus t^{\prime}) }{{\sf e} ({ y}_{\alpha}, { y}_{\alpha})}. 
\label{eq:strconst}
\end{align} 
Gorbounov--Korff showed that the structure constants \eqref{eq:strconst}  for $\beta=-1$  agree with the structure constants  of the $K$-theory classes of structure sheaves  
of  the $T:=(\mathbb{C}^{\times})^{M}$ equivariant quantum $K$-ring of Grassmannian $QK_T(\mathrm{Gr}(N,M))$,  i.e., 
\begin{align}
 [\mathcal{O}_{\mu} ]  \star  [ \mathcal{O}_{\nu} ]= \sum_{\lambda \in \mathcal{P}_{N,M}}C_{\mu \nu}^{\lambda}(q,t, \beta=-1) [\mathcal{O}_{\lambda} ]  .
\end{align}
Here $\{ [\mathcal{O}_{\lambda} ]  \}_{\lambda \in \mathcal{P}_{N,M}}$ are the $K$-theory classes of the structure sheaves of  the Schubert varieties $\{ X_{\lambda} \}_{\lambda \in \mathcal{P}_{N,M}}$ of $\mathrm{Gr}(N,M)$. 
$\star$ is the quantum product of the small quantum $K$-ring of Grassmannian. Then the  isomorphism between $\mathcal{V}_{N,M}/( \beta+1 )$ and $QK_T(\mathrm{Gr}(N, M))$ is given by $| v_{\lambda} \rangle  \mapsto [\mathcal{O}_{\lambda}]$ and $t_i \mapsto 1-e^{\varepsilon_{M-i+1}}$.
Here $\varepsilon_i:T \to \mathbb{C}^{\times}$ is the character defined by $\varepsilon_i (\mathrm{\tau}_1,\cdots,\tau_M)=\tau_i$ and $e^{\varepsilon_i}:=[\mathbb{C}_{\varepsilon_i}] \in K_T(\mathrm{pt})$.

 When $\beta=0$, the structure constants  \eqref{eq:strconst}    agree with the structure constants of  the equivariant  Schubert classes $\{ [X_{\lambda} ] \}_{\lambda \in \mathcal{P}_{N,M}}$
in  the $T$-equivariant quantum cohomology of Grassmannian $QH_T(\mathrm{Gr}(N,M))$:
\begin{align}
 [X_{\mu} ]  \cup_q  [ X_{\nu} ]= \sum_{\lambda \in \mathcal{P}_{N,M}} C_{\mu \nu}^{ \lambda}(q,t, \beta=0) [ X_{\lambda} ] .
\end{align}
Therefore the  isomorphism between $\mathcal{V}_{N,M}/( \beta  )$ and $QH_T(\mathrm{Gr}(N, M))$ is given by $| v_{\lambda} \rangle  \mapsto [X_{\lambda}]$.

We define $\eta_{\mu \nu} \in R$ as the elements of  the Frobenius  bilinear form $\eta$ in the spin basis:
\begin{align}
\eta_{\mu \nu}(q,t, \beta):=\eta ( | v_{\mu} \rangle,  | v_{\nu} \rangle ) = \sum_{\lambda \in \mathcal{P}_{N, M}}  
\frac{G_{\mu}(y_{\lambda} | \ominus t)G_{\nu}(y_{\lambda} |\ominus t)  }{{\sf e}  ( {  y}_{\lambda}, { y}_{\lambda})} .
\label{eq:metric}
\end{align}
We define the genus zero three point function $C^{(0)}_{\lambda \mu \nu}(q,t, \beta)$ by
\begin{align}
C^{(0)}_{\lambda \mu \nu}(q,t, \beta)&:= \eta (| v_{\lambda } \rangle  ,| v_{\mu } \rangle * | v_{\nu} \rangle ) 
=\sum_{\rho \in \mathcal{P}_{N},M} \eta_{\lambda \rho } C^{\rho}_{\mu \nu}
\nonumber \\
&=\sum_{\alpha \in \mathcal{P}_{M,N}} 
\frac{G_{\mu}(y_{\alpha} | \ominus t)G_{\nu}(y_{\alpha} |\ominus t) G_{\lambda}(y_{\alpha} | \ominus t) }{{\sf e}  ( { y}_{\alpha}, { y}_{\alpha}) } .
\label{eq:3ptfunction}
\end{align}
Note that $\eta (| v_{\lambda } \rangle  ,| v_{\mu } \rangle * | v_{\nu} \rangle )=\eta (| v_{\lambda } \rangle * | v_{\mu } \rangle   , | v_{\nu} \rangle )$.
$C^{(0)}_{\lambda \mu \nu}$ is invariant under the permutations of the indices $\lambda, \mu, \nu$ and satisfies $C^{(0)}_{\emptyset \mu \nu}=\eta_{\mu \nu}$.

\subsubsection*{\underline{Level-rank duality}}
The level-rank duality is a ring isomorphism between  $\mathcal{V}_{N, M} \simeq \mathcal{V}_{M-N, M}$ which leads to 
 $QK_T (\mathrm{Gr}(N, M)) \simeq QK_T (\mathrm{Gr}(M-N, M))$ and  $QH_T (\mathrm{Gr}(N, M)) \simeq QH_T (\mathrm{Gr}(M-N, M))$.
In the spin basis, the level-rank duality  follows from the equalities between the structure constants and the Frobenius bilinear forms between 
$\mathcal{V}_{N, M}$ and $\mathcal{V}_{M-N, M}$:
\begin{align}
C^{\lambda}_{ \mu \nu}(q,t, \beta)&=C^{\lambda^{\prime}}_{ \mu^{\prime} \nu^{\prime}}(q,\ominus t^{\prime}, \beta), 
\label{eq:levelrank1}\\
\eta_{ \mu \nu}(q,t, \beta)&=\eta_{ \mu^{\prime} \nu^{\prime}}(q,\ominus t^{\prime}, \beta) .
\label{eq:levelrank2}
\end{align}
Here $\lambda^{\prime}$ is the conjugate (transpose) partition of $\lambda$.
We will see  the  level-rank duality leads to a Seiberg-like (level-rank) duality between correlation function of  $U(N)$ Chern--Simons--matter theory and a $U(M-N)$ Chern--Simons--matter theory on $S^1 \times \Sigma_g$ 
and also leads to  Seiberg-like duality of  two  A-twisted GLSMs on $\Sigma_g$.

\subsection{ Frobenius algebra and  genus $g$ $n$-point correlation functions in  2d TQFT}
\label{sec:Frob2dTQFT}

\begin{pdffig}
\begin{figure}[thb]
\centering
\subfigure[]{\label{fig:unit}
\includegraphics[width=2cm]{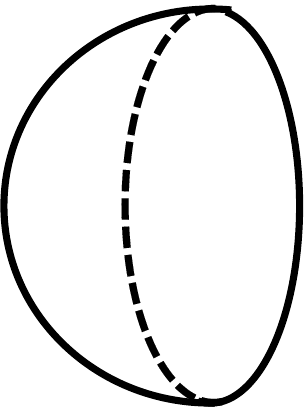}}
\hspace{1cm}
\subfigure[]{\label{fig:metric}
\includegraphics[width=1.6cm]{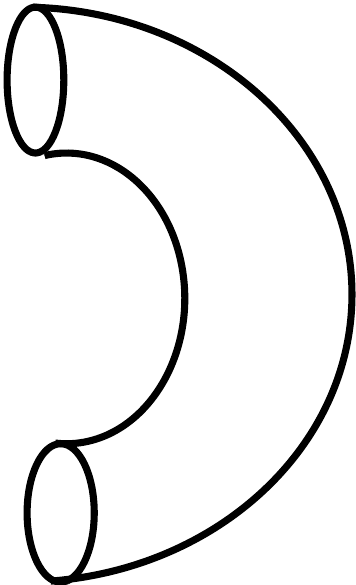}}
\hspace{1cm}
\subfigure[]{\label{fig:strconst}
\includegraphics[width=2cm]{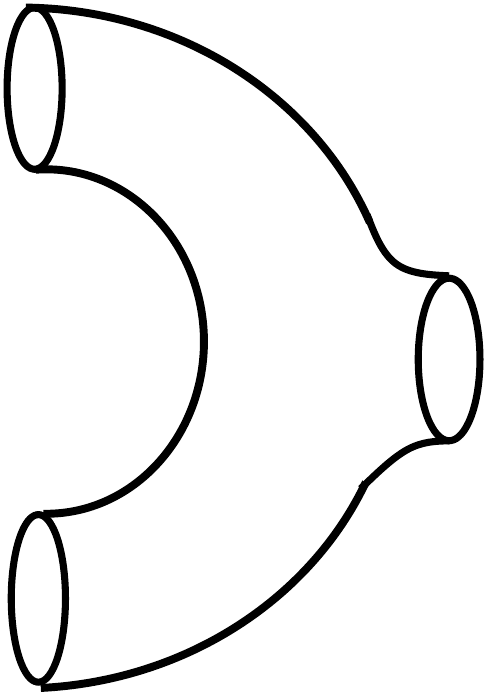}}
\caption{$(a)$ The identity of the Frobenius algebra $\mathcal{V}_{N,M}$ corresponds to a sphere with an out-boundary (right boundary).
$(b)$: The Frobenius form $\eta: \mathcal{V}_{N,M} \otimes \mathcal{V}_{N,M}  \to R$ corresponds to a annulus with two  in-boundaries (left boundaries). 
$(c)$: The product $*: \mathcal{V}_{N,M} \otimes \mathcal{V}_{N,M} \to \mathcal{V}_{N,M}$ corresponds to a pant with  two  in-boundaries and an out-boundary .}
\label{fig:TFTdata}
\end{figure}
\end{pdffig}
In order to show the correspondence between the quantum $K$-rings and   CS--matter theories on $S^1 \times \Sigma_g$ and also 
 the correspondence between the quantum cohomologies and  A-twisted GLSMs on $\Sigma_g$,
it is useful to introduce a pictorial description 
\cite{Dijkgraaf:1989} (see also \cite{MR2037238} for a more rigorous treatment) of the finite dimensional commutative Frobenius algebras in terms of 2d TQFTs. 
After explaining the 2d TQFT aspect of the Frobenius algebra, we will calculate  quantities  associated to the surface $\Sigma_{g,n}$ with genus $g$ and  $n$ in-boundaries which will be 
identified with  genus $g$ correlation function of topologically twisted gauge theories in two and three dimensions. 

\begin{pdffig}
\begin{figure}[thb]
\centering
\subfigure[]{\label{fig:Cijk}
\includegraphics[width=5cm]{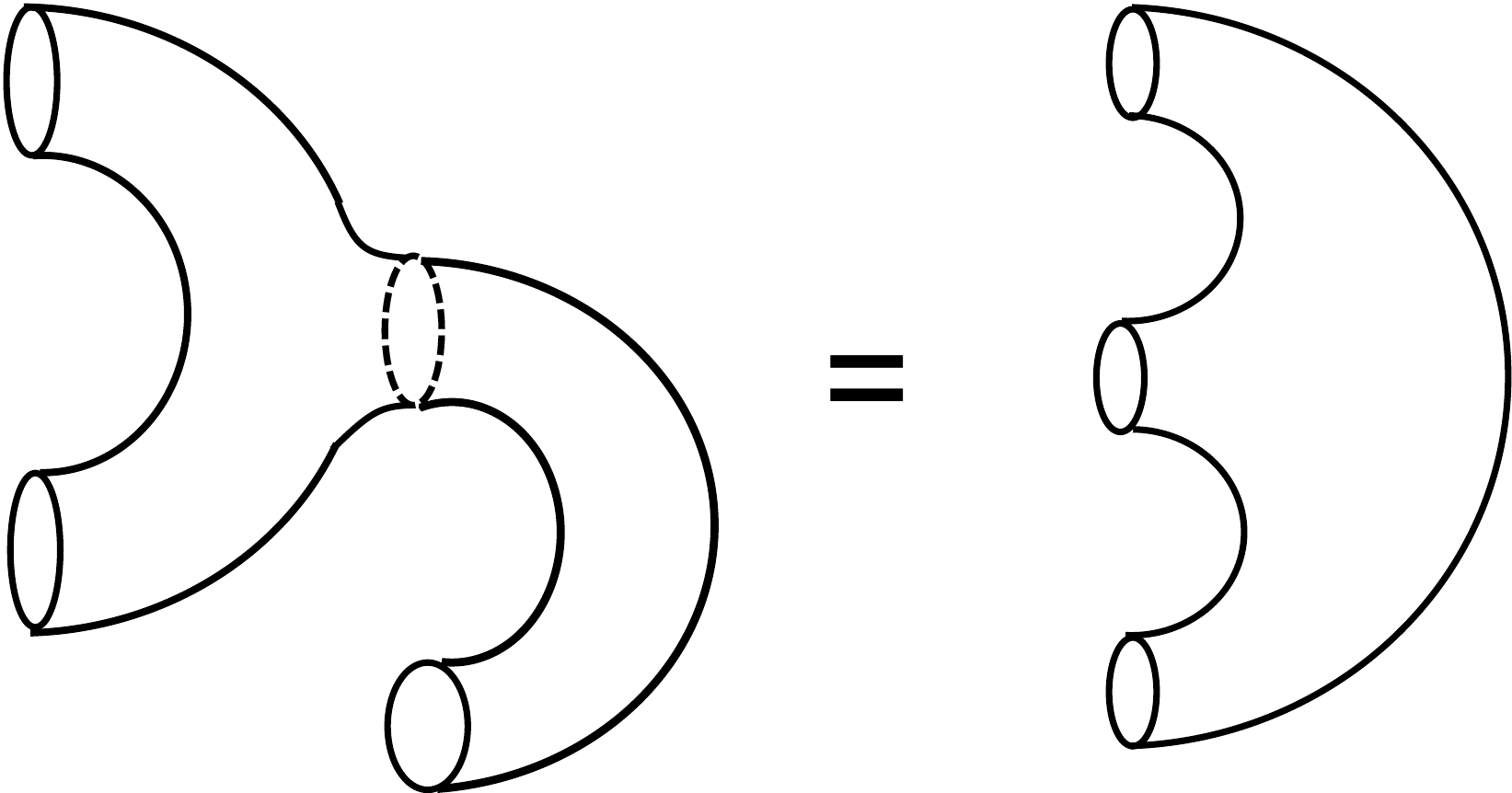}}
\hspace{1cm}
\subfigure[]{\label{fig:associative}
\includegraphics[width=6cm]{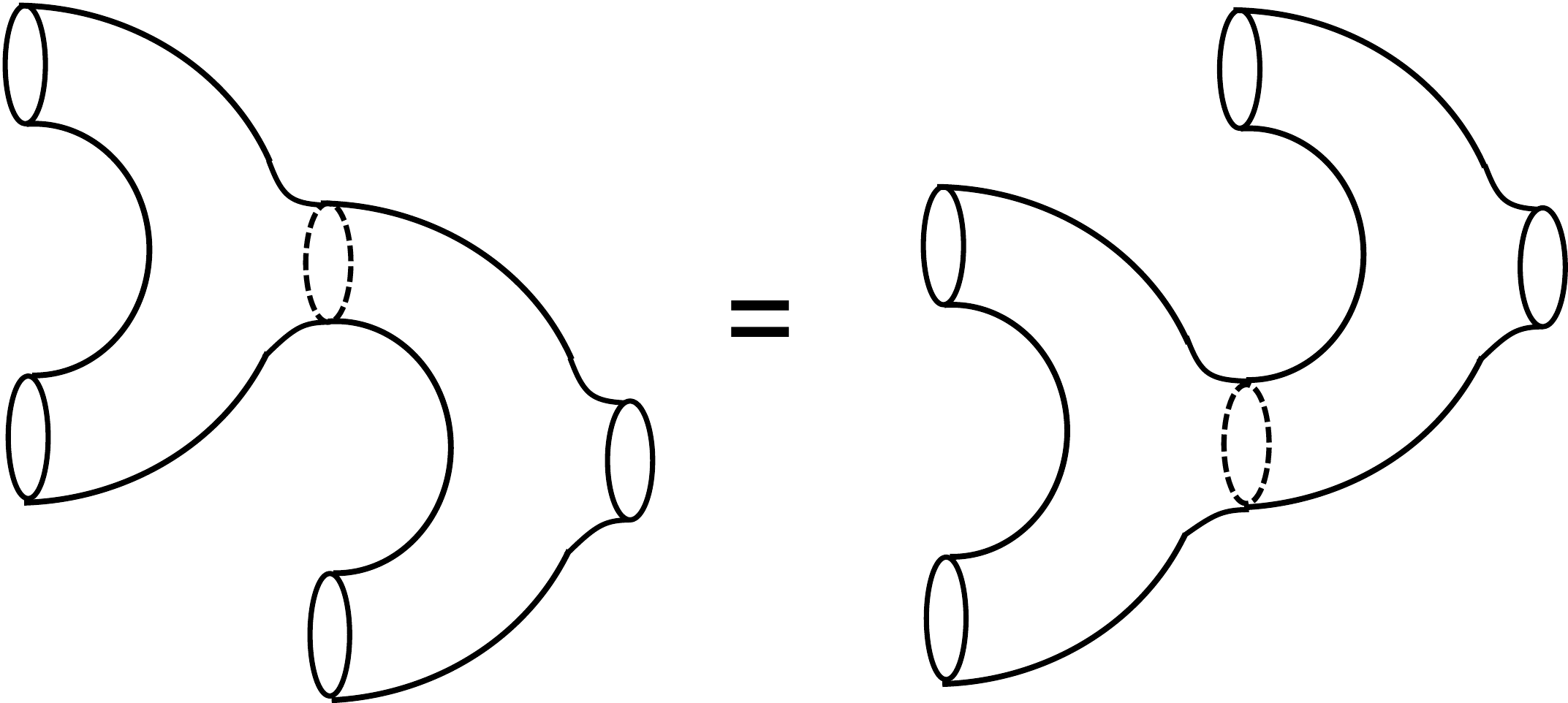}}
\caption{$(a)$: The composition $\eta \circ (* \otimes \mathrm{id} )= \eta \circ (\mathrm{id} \otimes *   ) : \mathcal{V}_{N,M}^{\otimes 3}   \to R$ corresponds to gluing an in-boundary of  an annulus and an out-boundary of a pant 
which is equivalent to a sphere with three in-boundaries.  $(b)$: The associativity condition
for $ \mathcal{V}_{N,M}^{\otimes 3}   \to \mathcal{V}_{N,M}$ means the equivalence of  two different gluing of two pants .   }
\label{fig:TFTdata1}
\end{figure}
\end{pdffig}
 First we introduce $\eta^{\mu \nu}$ defined by
\begin{align}
\eta^{ \mu \nu}:=\sum_{ \alpha \in \mathcal{P}_{N, M}} \frac{\Pi (y_{\alpha})^2 }{\Pi (t_{\mu}) \Pi (t_{\nu})}
\frac{G_{\mu^{\vee} }(y_{\alpha}| \ominus t^{\prime}) G_{\nu^{\vee}}(y_{\alpha}| \ominus t^{\prime}) }{{\sf e}(y_{\alpha},y_{\alpha})} .
\end{align}
Then $\eta^{\mu \nu}$ is the inverse of $\eta_{\mu \nu}$, since these satisfy
 the relation $\sum_{\nu \in \mathcal{P}_{N, M}} \eta_{\mu \nu} \eta^{\nu \rho}=\delta^{\rho}_{\mu}$ which follows from the  formula \eqref{eq:complete}.
We introduce the pictorial expression of the product $*$, the bilinear form $\eta$, and identity $| v_{\emptyset} \rangle$ as depicted in Figure \ref{fig:TFTdata}.
The inverse of $\eta$ is associated to an annulus with two out-boundaries.  A contraction of an upper index and a lower index corresponds to 
a gluing of an in-boundary and an out-boundary (See Figure \ref{fig:handle2}).  

\begin{pdffig}
\begin{figure}[thb]
\centering
\subfigure[]{\label{fig:handle}
\includegraphics[width=5cm]{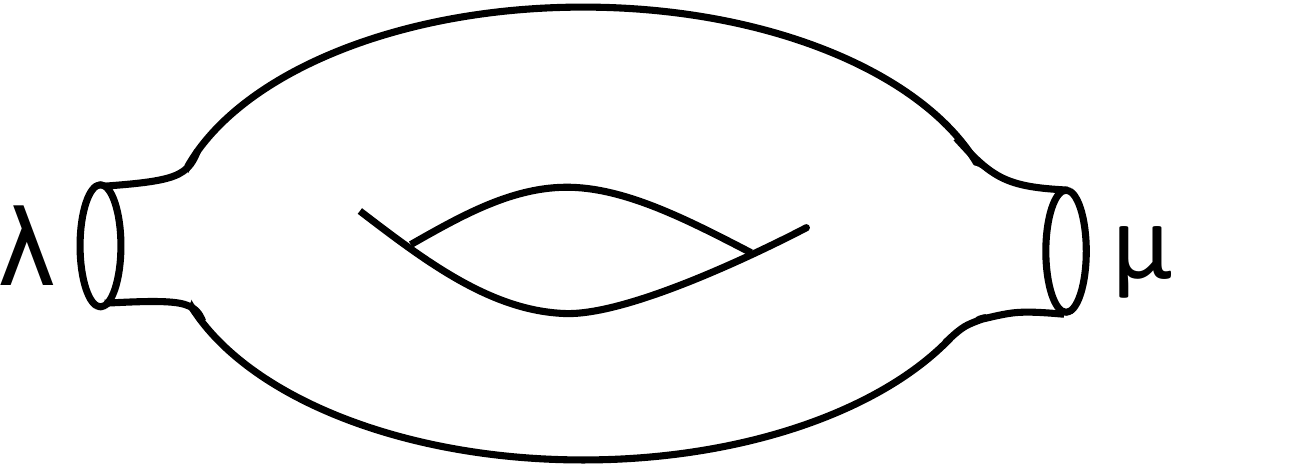}}
\hspace{0.5cm}
\subfigure[]{\label{fig:handle2}
\includegraphics[width=8cm]{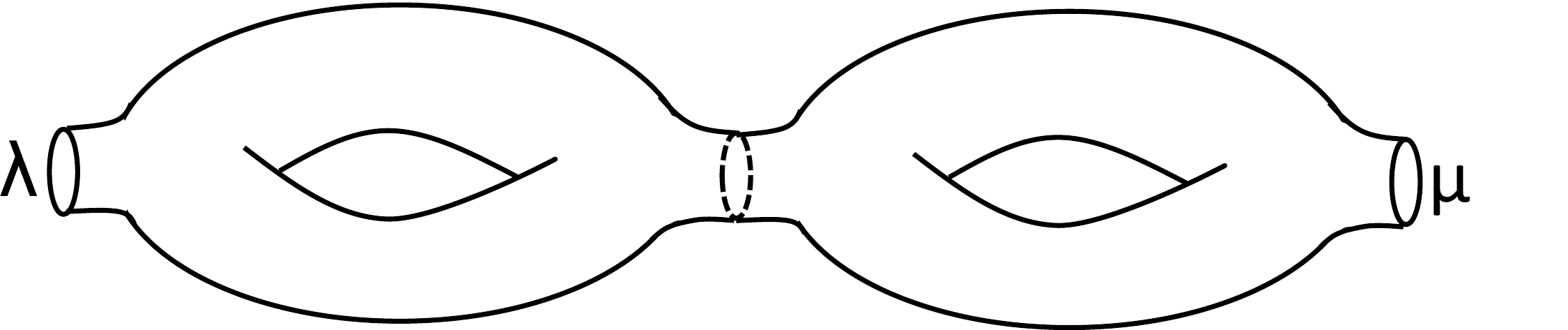}}
\caption{$(a)$: We associate $C^{(1) \mu}_{\quad \lambda}$ to a genus one surface with an in-boundary and out-boundary. 
$\lambda$ at the in-boundary and $\mu$ at  the out-boundary denote the lower and upper indices of $C^{(1) \mu}_{\quad \lambda}$, respectively. 
$(b)$: A contraction $\sum_{\nu} C^{(1) \mu}_{\quad \nu} C^{(1) \nu}_{\quad \lambda}$ corresponds to gluing two genus one surface with an in-boundary and out-boundary, which is
  a genus two surface with an  in-boundary and out-boundary and  is expressed  by $C^{(2) \mu}_{\quad \lambda}$.}
\label{fig:handleop}
\end{figure}
\end{pdffig}
We introduce $C^{(g) \mu_1 \cdots \mu_l}_{\quad \lambda_1 \cdots \lambda_n} (q,t , \beta)$ defined by contractions of  $C^{\lambda_1}_{\lambda_2 \lambda_3}$'s, $\eta_{\mu_1 \mu_2}$'s, and  $\eta^{\nu_1 \nu_2}$'s  which corresponds to the genus $g$ surface with  $n$ in-boundaries and  $l$ out-boundaries. $\lambda_i$ is assigned to an in-boundary and $\mu_i$ is assigned to an out-boundary. For example,  $g=1,2$ with $n=l=1$ are  depicted in Figure \ref{fig:handleop}.

Note that $C^{(g) \mu_1 \cdots \mu_l}_{\quad \lambda_1 \cdots \lambda_n}$ satisfies the following properties
\begin{align}
C^{(0) \lambda}_{\quad \mu \nu}&=C^{\lambda}_{\mu \nu}, \quad C^{(0) }_{\quad \mu \nu}=\eta_{\mu \nu}, \quad C^{(0) \mu \nu}=\eta^{\mu \nu}, \\
C^{(g) \mu_1 \cdots \mu_{l-1} \emptyset}_{\quad \lambda_1 \cdots \lambda_n}&=C^{(g) \mu_1 \cdots \mu_{l-1}}_{\quad \lambda_1 \cdots \lambda_n}, \\
C^{(g) \mu_1 \cdots \mu_{l}}_{\quad \lambda_1 \cdots \lambda_{n-1} \emptyset}&=C^{(g) \mu_1 \cdots \mu_{l}}_{\quad \lambda_1 \cdots \lambda_{n-1}}, \\
C^{(g_1+g_2) \mu_1 \cdots \mu_{l_1+l_2}}_{\quad \quad \lambda_1 \cdots \lambda_{n_1+n_2} }&=\sum_{\nu_1 \nu_2 } C^{(g_1) \mu_1 \cdots \mu_{l_1}  }_{\quad \lambda_1 \cdots \lambda_{n_1} \nu_1}  
\eta^{\nu_1 \nu_2} C^{(g_2) \mu_{l_1+1} \cdots \mu_{l_1+l_2}  }_{\quad \nu_2 \lambda_{n_1+1} \cdots \lambda_{n_1+n_2} },\\
C^{(g_1+g_2) \mu_1 \cdots \mu_{l_1+l_2}}_{\quad \quad \lambda_1 \cdots \lambda_{n_1+n_2} }&=\sum_{\nu_1 \nu_2 }  
C^{(g_1) \mu_1 \cdots \mu_{l_1} \nu_1  }_{\quad \lambda_1 \cdots \lambda_{n_1} }  
\eta_{\nu_1 \nu_2} C^{(g_2) \nu_2 \mu_{l_1+1} \cdots \mu_{l_1+l_2}  }_{\quad \lambda_{n_1+1} \cdots \lambda_{n_1+n_2} } .
\end{align}
$C^{(1) \mu}_{\quad \lambda}$ is written in terms of Grothendieck polynomial as
\begin{align}
C^{(1) \mu}_{\quad \lambda} (q,t, \beta) &= \sum_{\nu_1, \nu_2 \nu_3  \in \mathcal{P}_{N, M}} \eta^{\nu_2 \nu_3 } C^{\nu_1 }_{\lambda \nu_2  } C^{\mu }_{  \nu_1  \nu_3} \nonumber \\
&=\sum_{\alpha \in \mathcal{P}_{N,M} } \frac{\Pi (y_{\alpha})}{\Pi (t_{\mu})} 
{G_{\lambda}(y_{\alpha}| \ominus t)  G_{\mu^{\vee}}( y_{\alpha}| \ominus t^{\prime})} .
\end{align}
Here we used \eqref{eq:orth}.  
$C^{(g) \mu}_{\quad \lambda}$ is calculated by  contracting  upper and lower indices of $C^{(1) \mu}_{\quad \lambda}$ as 
\begin{align}
{C}^{(g) \mu}_{\quad \lambda} &=  \sum_{\nu_1, \cdots, \nu_{g-1} \in \mathcal{P}_{N,M}}
 C^{(1) \mu }_{\quad \nu_1 }   \left( \prod_{k=1}^{g-2} C^{(1) \nu_{k} }_{\quad \nu_{k+1} }  \right) C^{(1) \nu_{g-1} }_{\quad \lambda } \nonumber \\
&=\sum_{{\alpha} \in \mathcal{P}_{N,M} } \frac{\Pi (y_{\alpha})}{\Pi (t_{\mu})} 
\frac{G_{\lambda}(y_{\alpha}| \ominus t)  G_{\mu^{\vee}}( y_{\alpha}| \ominus t^{\prime})} {{\sf e}(y_{\alpha}, y_{\alpha})^{1-g} }.
\label{eq:g11}
\end{align}
The genus $g$ Riemann surface corresponds to the contraction $\sum_{\mu \in \mathcal{P}_{N,M}}C^{(g) \mu}_{\quad \mu}$ given by    
\begin{align}
Z( \Sigma_g )&:=\sum_{\mu \in \mathcal{P}_{N,M}}C^{(g) \mu}_{\quad \mu} =\sum_{\alpha \in \mathcal{P}_{M,N}} {\sf e} ( y_{\alpha}, y_{\alpha}) ^{g-1} .
\end{align}
Although we defined $Z( \Sigma_g )$ in a particular basis $\{ v_{\lambda} \}_{\lambda \in \mathcal{P}_{N, M}}$ of $\mathcal{V}_{N, M}$,  $Z( \Sigma_g )$ is independent of the choices  of bases. 
$Z(\Sigma_g)$ is sometimes called the genus $g$ partition function of  2d TQFT. 
In section \ref{sec:GLSM}, we will show  $Z(\Sigma_g)$ for $\beta=0$ agrees with a partition function of A-twisted GLSM on $\Sigma_g$ and, in section \ref{sec:CSMsub1}, 
we will show $Z(\Sigma_g)$ for $\beta=-1$ agrees with  the partition function (topologically twisted index) of a CS--matter theory on $S^1 \times \Sigma_g$.

Next we express $C^{(g)}_{\mu_1 \mu_2 \cdots \mu_n }$ in terms of Grothendieck polynomials and the on-shell square norms.  By using the  relation \eqref{eq:orth},  we obtain
\begin{align}
C^{(0)}_{\lambda_1 \lambda_2 \cdots \lambda_n }(q,t, \beta)&=\sum_{\nu_1 ,\cdots, \nu_{n-2} \in \mathcal{P}_{N,M}}C^{\nu_1}_{\lambda_1 \lambda_2} C^{\nu_2}_{\nu_1 \lambda_3} C^{\nu_3}_{\nu_2 \lambda_4} \cdots C^{\nu_{n-2}}_{\nu_{n-3} \lambda_{n-1}} 
C^{\emptyset}_{\nu_{n-2} \lambda_{n}}
\nonumber \\
&=\sum_{\alpha \in \mathcal{P}_{M,N}} 
\frac{ \prod_{l=1}^n G_{\lambda_l}(y_{\alpha} | \ominus t) }{ {\sf e}  ( {  y}_{\alpha}, { y}_{\alpha}) } .
\label{eq:genus0nptfunction}
\end{align}
From  \eqref{eq:g11} and \eqref{eq:genus0nptfunction}, 
$C^{(g)}_{\lambda_1 \lambda_2 \cdots \lambda_{n}  }$ is given by
\begin{align}
C^{(g)}_{\lambda_1 \lambda_2 \cdots \lambda_{n}  }(q,t, \beta)&=
\sum_{\nu \in \mathcal{P}_{N,M}} C^{(0)}_{\lambda_1 \lambda_2 \cdots \lambda_{n-1} \nu } C^{(g) \nu}_{\quad \lambda_n}
\nonumber \\
&=\sum_{\alpha \in \mathcal{P}_{M,N}} 
\frac{ \prod_{l=1}^n G_{\lambda_l}(y_{\alpha} | \ominus t) }{ {\sf e}  ( { y}_{\alpha}, {  y}_{\alpha})^{1-g} } .
\label{eq:genusgnptfunction}
\end{align}
We will show that $C^{(g)}_{\lambda_1 \lambda_2 \cdots \lambda_{n}  }(q,t, \beta)$ is same as 
an $n$-point correlation function in A-twisted GLSM on $\Sigma_g$ for $\beta=0$ and  an $n$-point correlation function in 
CS--matter theory on $S^1\times \Sigma_g$ for $\beta=-1$. We call \eqref{eq:genusgnptfunction} as a genus $g$ correlation function of 2d TQFT. 

The level-rank duality \eqref{eq:levelrank1} and \eqref{eq:levelrank2} leads to 
the equality  of $C^{(g)}_{\lambda_1 \lambda_2 \cdots \lambda_{n}  }$ between
 $\mathcal{V}_{N, M}$ and $\mathcal{V}_{M-N, M}$:
\begin{align}
C^{(g)}_{\lambda_1 \lambda_2 \cdots \lambda_{n}  }(q,t, \beta)&=
C^{(g)}_{\lambda^{\prime}_1 \lambda^{\prime}_2 \cdots \lambda^{\prime}_{n}  }(q, \ominus t^{\prime}, \beta) .
\label{eq:levelrankg}
\end{align}

\subsection{Determinant representation of the on-shell square norm}
\label{sec:norm}
In order to relate the  quantum $K$-ring with the inner product \eqref{eq:innerprod} to  correlation functions and the algebra of supersymmetric Wilson loops in a supersymmetric Chern--Simons--matter theory, 
we have to express   the  on-shell inner product ${\sf e}(y_{\lambda},y_{\lambda})$ in \eqref{eq:3ptfunction} as a nice function of the Bethe roots $\{  y_{\lambda}\}_{\lambda \in \mathcal{P}_{N, M}}$ and the inhomogeneous parameters $\{ t_i \}_{i=1}^M$.  
  It was shown in \cite{Okuda:2012nx, Okuda:2013fea} that 
the partition functions of $U(N)/U(N)$ gauged Wess-Zumino-Witten (WZW) model and its deformation (called gauged WZW--matter model) 
\footnote{$G/G$ gauged WZW model does not depend on the metric of the world sheet  \cite{Witten:1993xi} and possesses a nilpotent scalar charge $\mathcal{Q}$.
In \cite{Okuda:2013fea} a $\mathcal{Q}$-deformation (by adding $\mathcal{Q}$-exact terms ) of gauged WZW model  was studied. 
It is known that the partition function of $\mathcal{Q}$-deformation of gauged WZW model agrees with the partition function of topologically twisted Chern--Simons-matter theory   with an adjoint chiral multiplet of R-charge $r=0$.
} 
   are written in terms of  the determinant type formulas of the on-shell norms for the phase model and $q$-boson, respectively. Moreover the partition functions of the gauged WZW model and the gauged WZW-matter model  on the Riemann surfaces 
agree with the  partition functions of 2d TQFTs associated to the Frobenius algebras in \cite{MR2669352, MR3017067}.
Motivated by these agreements between gauge theories and 2d TQFTs observed 
 in \cite{Okuda:2012nx, Okuda:2013fea, Okuda:2015yea, Kanno:2018qbn}, we apply Izergin-Korepin method \cite{Korepin:1982gg} 
and  express the on-shell square norm by the  determinant type formula. 
See also  \cite{MR2358738} for a review of derivation of the norm for  the  spin-$\frac{1}{2}$ XXZ spin chain. 
The derivation of determinant type formula for  ${\sf e}(y_{\lambda},y_{\lambda})$ is parallel to that for the XXZ spin chain. 

First we study the  off-shell inner product between the Bethe vector and the dual Bethe vector 
for indeterminate variables $\{ x \}:= \{x_1, \cdots, x_N \}$ and $\{ y \} =\{y_1, \cdots, y_N \}$. 
Recall that the off-shell inner product is defined by 
\begin{align}
S_{N} (\{ x \}, \{ y \} ):= \langle 0 | \prod_{a=1}^N{\sf C}(x_a) \prod_{a=1}^N {\sf B}(y_a)  | 0 \rangle .
\label{eq:offinner}
\end{align}
From \eqref{eq:CB} and \eqref{eq:AaDd}, the most general form of \eqref{eq:offinner} is expressed as 
\begin{align}
S_{N} (\{ x\} | \{ y \})&=\sum_{\alpha \cup \overline{\alpha} \atop \gamma \cup \overline{\gamma} }
\prod_{j \in \gamma} {\sf a}(x_j) \prod_{k \in \overline{\gamma}} {\sf d}(x_k)
 \prod_{k \in \overline{\alpha}} {\sf a}(y_k) \prod_{j \in \alpha} {\sf d}(y_j) 
\nonumber \\
&\qquad \qquad \times
 {\sf K}_{N} (\{ x\}_{\gamma}, \{ x \}_{\overline{\gamma}} | \{ y \}_{\alpha},  \{ y \}_{\overline{\alpha}})  .
\label{eq:offshellnorm}
\end{align}
Here $\{ x \}_{\alpha}$ and $\{ x \}_{\overline{\alpha}}$ are two disjoint subsets of $\{ x\}$ with $\{ x \}_{\alpha} \cup \{ x \}_{\overline{\alpha}} =\{ x \}$.
 $\{ y \}_{\gamma}$ and $\{ y \}_{\overline{\gamma}}$ are two disjoint subsets of $\{ y \}$ with $\{ y \}_{\gamma} \cup \{ y \}_{\overline{\gamma}} =\{ y \}$.
The subsets $\{ x \}_{\alpha}$, $\{ x \}_{\overline{\alpha}}$,  $\{ y \}_{\gamma}$, and $ \{ y \}_{\overline{\gamma}}$
  are ordered as $\{ x \}_{\alpha}=\{x_{\alpha_1}, \cdots, x_{\alpha_n}\}$, if $\alpha_1 < \cdots < \alpha_n$ and so on.  
Note that ${\sf K}_{N} (\{ x\}_{\gamma}, \{ x \}_{\overline{\gamma}} | \{ y \}_{\alpha},  \{ y \}_{\overline{\alpha}})$ depends on
the R-matrix \eqref{eq:Rmatrix}, but is independent of the eigen values  of ${\sf A}$ and ${\sf D}$ in the monodromy matrix.  In the following evaluation of  
${\sf K}_{N} (\{ x \}_{\gamma}, \{ x \}_{\overline{\gamma}} | \{ y \}_{\alpha},  \{ y \}_{\overline{\alpha}})$, we fix 
the partition $\alpha, \overline{\alpha}$ and  $\gamma, \overline{\gamma}$.
 
The leading coefficient ${\sf K}_N(\{x \}, \{ y \})$ and the conjugate leading coefficient $\overline{{\sf K}}_N(\{x \}, \{ y \})$ are defined by
\begin{align}
& {\sf K}_{N} (\{ x\} | \{ y \} ):=
  {\sf K}_{N} (\{ x\}_{\gamma}, \{ x \}_{\overline{\gamma}}= \emptyset | \{ y \}_{\alpha},  \{ y \}_{\overline{\alpha}}= \emptyset), \\
& \overline{{\sf K}}_{N} (\{ x\} | \{ y \} ):=
  {\sf K}_{N} (\{ x\}_{\gamma}= \emptyset, \{ x \}_{\overline{\gamma}} | \{ y \}_{\alpha}= \emptyset,  \{ y \}_{\overline{\alpha}})  .
\end{align}
An important property  is that ${\sf K}_{N} (\{ x\}_{\gamma}, \{ x \}_{\overline{\gamma}} | \{ y \}_{\alpha},  \{ y \}_{\overline{\alpha}})$ 
is expressed in terms of the leading and conjugate leading coefficients as follows.
\begin{align}
 {\sf K}_{N} (\{ x \}_{\gamma}, \{ x \}_{\overline{\gamma}} | \{ y \}_{\alpha},  \{ y \}_{\overline{\alpha}}) 
&=\prod_{a \in \alpha} \prod_{b \in \overline{\alpha}} 
g (y_a, y_b ) \prod_{a \in \gamma} \prod_{b \in \overline{\gamma}}g ( x_a, x_b ) \nonumber \\
& \times {\sf K}_{n} (\{ x \}_{\gamma} | \{ y \}_{\alpha} ) 
 \overline{{\sf K}}_{N-n} (\{ x \}_{\overline{\gamma}} | \{ y \}_{\overline{\alpha}} ) .
\label{eq:coeffnon}
\end{align}
Here $g(x, y)$ is  a coefficient  in the RTT relation given by  
\begin{align}
{\sf C}(x) {\sf B}(y)=g(y, x) \left({\sf A}(x) {\sf D}(y) -{\sf A}(y) {\sf D}(x) \right) ,
\label{eq:cbad}
\end{align}
with
\begin{align}
g(y, x):=\frac{1}{x \ominus y} .
\end{align}
Then, it is enough to evaluate the leading and  the conjugate leading coefficients to  evaluate the off-shell inner product \eqref{eq:offshellnorm}.
Let us consider to evaluate the leading and the conjugate leading coefficients.
 From \eqref{eq:cbad}, we obtain
\begin{align}
{\sf C}(x_N) \prod_{j=1}^N {\sf B}(y_j)  | 0 \rangle
&=  
\sum_{l=1}^N  {\sf a}(x_N )   {\sf d}(y_{l})  g(y_l, x_N)   \left( \prod_{j=1 \atop j \neq l}^N g(y_j, x_N)  g(y_l, y_j)  \right)
\prod_{j=1 \atop j \neq l}^n  {\sf B}(y_{j}) 
| 0 \rangle 
+ \cdots
\label{eq:CBn}
\end{align}
Here the ellipses denote the terms which are irrelevant to the evaluation of the leading coefficient. 
Multiplying  \eqref{eq:CBn} by $\langle 0 | \prod_{i=1}^{N-1} {\sf C} (x_i)$ , we find the leading coefficients satisfy the following recurrence relation.
\begin{align}
{\sf K}_N (\{ x \} | \{ y \})
&=\sum_{l=1}^N  g(y_l, x_N)
 \left(\prod_{a=1 \atop a \neq l}^N g(y_a, x_N)  g(y_l, y_a)  \right) {\sf K}_{N-1} (\{ x \neq x_N \} | \{ y \neq y_l \})  .
\label{eq:recurrence1}
\end{align}
In a similar way,  the recurrence relation for the conjugate leading coefficients is given by
\begin{align}
\overline{{\sf K}}_N (\{ x \} | \{ y \})
&=-\sum_{l=1}^N  g(y_l, x_N)
 \left(\prod_{a=1 \atop a \neq l}^N g(x_N, y_a)  g(y_a, y_l)  \right) \overline{{\sf K}}_{N-1} (\{ x \neq x_N \} | \{ y \neq y_l \})  .
\label{eq:recurrence2}
\end{align}
With the following initial condition
\begin{align}
{\sf K}_1 (\{ x \} | \{ y \}) = \frac{1+ \beta y}{x - y}, \quad 
\overline{{\sf K}}_1 (\{ x \} | \{ y \}) = \frac{1+ \beta y}{y - x},
\end{align}
the recurrence relations \eqref{eq:recurrence1} and \eqref{eq:recurrence2} are  solved   as  
\begin{align}
{\sf K}_N (\{ x \} | \{ y \}) 
&=  
 \frac{ \prod_{b=1}^N (1+\beta y_b)^{N+1} }{\prod_{a > b}^N (x_b-x_a) ( y_a- y_b) } \det_{k,j} {\sf t}( x_k, y_j) , 
\label{eq:leadKN} \\
\overline{{\sf K}}_N (\{ x \} | \{ y \}) 
&= \frac{\prod_{a=1}^N   (1+ \beta y_a)  (1+\beta x_a)^N}{\prod_{a > b}^N (x_b-x_a) ( y_a- y_b) } \det_{j, k}{\sf t}(y_j , x_k) , 
\label{eq:leadKNbar} 
\end{align}
where ${\sf t}(x, y)$ is defined by
\begin{align}
{\sf t}(x, y):=\frac{1}{(x - y){(1+ \beta y)}} .
\end{align}
 $\mathrm{det}_{j,k}$ denotes the determinant with respect to indices $j, k$. 
We show that \eqref{eq:leadKN} satisfies the recurrence relation \eqref{eq:recurrence1} as follows.
By adding   the $j$-th row  of the matrix ${\sf t}(x_j, y_k)$ multiplied by $u_j/u_N $  with $j=1,\cdots, N-1$   to the $N$-th row of  ${\sf t}(x_j, y_k)$, 
we obtain the following relation between  determinants for ${\sf t}( x_j, y_k)$:
\begin{align}
\det_{j, k}   {\sf t}( x_j, y_k)
&=
 \frac{1}{u_N} \sum_{l=1}^N (-1)^{N+l} G_l
\det_{j, k \atop j \neq N k \neq l} { \sf t}( x_{j}, y_k) ,
\label{eq:recurt}
\end{align}
where $G_l$ is defined by $G_l:= \sum_{k=1}^N u_k {\sf t}(x_k, y_l)$.
If $u_k$ is chosen as
\begin{align}
u_{k}=\frac{\prod_{a=1}^N (x_k - y_a)}{\prod_{a=1 \atop a \neq k}^{N}( x_k - x_a) },
\end{align}
then  $G_l$ satisfies the following identity:
\begin{align}
G_l=\frac{1}{1+\beta y_l}.
\label{eq:idGl}
\end{align}
To show \eqref{eq:idGl}, we consider the integral:
\begin{align}
I=\frac{1}{1+\beta y_l}\oint_{\Gamma} \frac{dz}{2 \pi {\rm i}} \frac{\prod_{a=1 \atop a \neq l}^N (z-y_a)}{\prod_{a=1}^N (z-x_a)}\,.
\label{eq:residueaux}
\end{align}
Here the contour encloses  all the poles of the integrand; $z=\infty, x_1,\cdots, x_N$. The residues at $z=\infty$ and $z=x_1,\cdots, x_N$ are evaluated as
\begin{align}
\frac{1}{1+\beta y_l}\oint_{z=\infty} \frac{dz}{2 \pi {\rm i}} \frac{\prod_{a=1 \atop a \neq l}^N (z-y_a)}{\prod_{a=1}^N (z-x_a)}&=-\frac{1}{1+\beta y_l}\,, \nonumber \\
\frac{1}{1+\beta y_l}\sum_{k=1}^N \oint_{z=x_k} \frac{dz}{2 \pi {\rm i}} \frac{\prod_{a=1 \atop a \neq l}^N (z-y_a)}{\prod_{a=1}^N (z-x_a)}&=G_l \,.
\end{align}
Since $I=0$, we obtain \eqref{eq:idGl}

\rem{
Next we rewrite 
\begin{align}
f(z):=\frac{1}{(z-\lambda_l)(1+\beta \lambda_l)} \prod_{a=1}^N \frac{z-\lambda_a}{z-\mu_a} 
\end{align}
\begin{align}
&0=\oint_{z= \infty} \frac{dz }{2\pi i} f(z)+\sum_{k=1}^N \oint_{z= \mu_k } \frac{dz }{2\pi i}f(z) \\
&\leftrightarrow \frac{1}{(1+\beta \lambda_l)}
=\sum_{k=1}^N \frac{1}{(\mu_k-\lambda_l)} \frac{\prod_{a=1}^N (\mu_k-\lambda_a)}{\prod_{a=1 \atop a \neq k}^N (\mu_k-\mu_a)}
=\sum_{k=1}^N t(\mu_k,\lambda_l) u_k
\end{align}
}
From \eqref{eq:recurt} with the identity \eqref{eq:idGl},  we find that \eqref{eq:leadKN} satisfies the recurrence relation \eqref{eq:recurrence1}. Thus the right hand side of \eqref{eq:leadKN} is the leading coefficient.
In a similar way, we can show that \eqref{eq:leadKNbar} satisfies the recurrence relation for $\bar{K}_N$.

Substituting the solutions  of the recurrence relations  to \eqref{eq:offshellnorm}, we have 
\begin{align}
S_{N} (\{ x \} | \{ y\})
&=\left( \prod_{b=1}^N (1+ \beta y_b) \right) \left( \prod^N_{a>b} (x_b- x_a)^{-1} (y_a -y_b)^{-1} \right) \nonumber \\
 & \times \sum_{\alpha \cup \overline{\alpha} \atop \gamma \cup \overline{\gamma} } (-1)^{P_{\alpha}+P_{\gamma}}
\prod_{j \in \gamma} {\sf  a}(x_j) \prod_{k \in \overline{\gamma}} {\sf d}(x_k)
 \prod_{k \in \overline{\alpha}} {\sf a}(y_k) \prod_{j \in \alpha} {\sf d}(y_j) 
\det_{j \in \alpha \atop k \in \gamma  } {\sf t}( x_k, y_j)
\det_{j \in \overline{\alpha} \atop k \in \overline{\gamma}  } {\sf t}(  y_j, x_k) \nonumber \\
&\times 
\prod_{a \in {\alpha}} \prod_{b \in {\gamma}} h(y_a, x_b)
\prod_{a \in \overline{\alpha}} \prod_{b \in \overline{\gamma}} h(x_b, y_a)
\prod_{a \in \alpha} \prod_{b \in \overline{\alpha}} h(y_a, y_b) 
\prod_{a \in \gamma} \prod_{b \in \overline{\gamma}} h(x_b, x_a)
\label{eq:Betheoff2}
\end{align}
with
\begin{align}
h(x, y):=g(x, y)(y-x)=1+\beta x.
\end{align}
$P_{\alpha}$ is the parity of the partition $(\alpha_1,\cdots, \alpha_n, \bar{\alpha}_1,\cdots, \bar{\alpha}_{N-n})$,
defined as 0 (resp. 1)
when $(\alpha_1,\cdots, \bar{\alpha}_{N-n})$ is
an even (resp. odd) permutation of $(1,2, \cdots, N)$.
Now assume that $\{ y \}$ in \eqref{eq:Betheoff2} is a Bethe root.
By using the Bethe ansatz equation and the identity
\begin{align}
\det_{j,k} \left (U(y_j, x_k)+V(y_j, x_k) \right)=\sum_{\alpha \cup \bar{\alpha} \atop \gamma \cup \bar{\gamma} }(-1)^{P_{\alpha}+P_{\gamma}}
\det_{j \in \alpha \atop k  \in \gamma} U(y_j, x_k) \det_{j \in \bar{\alpha} \atop k  \in \bar{\gamma}} V(y_j, x_k)\,,
\end{align}
we obtain the following expression of the square norm  after some calculation:
\begin{align}
S_N ( \{ x \} | \{ y \}) 
 &= \frac{ \prod_{b=1}^N (1+ \beta y_b) } {  \prod_{a>b}^N (x_b -x_a) (y_a - y_b) }
   \det_{a, b  } \Omega_{q} (y_a, x_b) ,
\end{align}
where
\begin{align}
 \Omega_{q} (y_a, x_b )&:={\sf a}(x_b) {\sf t}( x_b, y_a) \prod_{c=1}^N (1+\beta y_c) -(-1)^N q  \, 
{\sf t}(y_a, x_b ) (1+\beta x_b)^N .
\end{align}
Finally, by taking limit $x_a \to y_a$, we obtain the determinant formula of   the on-shell square norm given by
\begin{align}
{\sf e} ( y, y )   
& =\frac{\prod_{i=1}^M\prod_{a=1}^N ( y_a-t_i )}{\prod_{i=1}^M (1+ \beta t_i)^N} \frac{  \prod_{a=1}^N (1+ \beta y_a)^{N-1} }{\prod_{a >b}^N ( y_a-y_b)^2}
 \det_{a,b} \left[ \delta_{a,b} \left(\sum_{i=1}^M \frac{1+\beta y_a}{y_a-t_i} -N \beta \right) +\beta\right] .
\end{align}
We are interseted in the expression of  ${\sf e} ( y, y )$ with slices $\beta=-1, 0$.  
When $\beta=0$, the inverse of ${\sf e}(y,y)$ is given by
\begin{align}
{\sf e} ( y, y )^{-1} =
&  \frac{ \prod_{a >b}^N ( y_a-y_b)^2 }{\prod_{i=1}^M\prod_{a=1}^N ( y_a-t_i )}
 \prod_{a=1}^N \left[  \left(\sum_{i=1}^M \frac{1}{y_a-t_i} \right) \right]^{-1} .
\label{eq:invdet0}
\end{align}
When $\beta=-1$, the inverse of ${\sf e}(y,y)$ is given by 
\begin{align}
{\sf e} ( y, y )^{-1}
& =\frac{ \prod_{i=1}^M (1- t_i)^N \prod_{a >b}^N ( y_a-y_b)^2 } { \prod_{a=1}^N (1- y_a)^{N-1}  \prod_{i=1}^M ( y_a-t_i )  }
 \det_{a,b} \left[ \delta_{a,b} \left(\sum_{i=1}^M \frac{1-y_a}{y_a-t_i} + N \right) -1\right]^{-1} .
\label{eq:invnorm3d}
\end{align}
%

\section{2d A-twisted GLSM and quantum coholomogy of Grassmannian}
\label{sec:GLSM}

In this section we study a 2d A-twisted GLSM on the genus $g$ Riemann surface $\Sigma_g$. 
 The gauge group of GLSM is taken to be $G=U(N)$ and the $M$  chiral multiplets are in the fundamental representation of $U(N)$ with $M >N$.  
If we put the GLSM on  $\mathbb{R}^2$, the Higgs branch vacua  is a Grassmannian $\text{Gr}(N,M)$ in  the positive Fayet--Iliopoulos (FI) parameter region. 

The motivation of this section is threefold.
The first is to establish Bethe/Gauge correspondence between the algebraic Bethe ansatz for $\beta=0$ and
 the  2d $U(N)$ GLSM with $M$ fundamental chiral multiplets.
As far as we know,
Bethe/Gauge correspondence for A-twisted GLSM  for $M$ fundamental chiral multiplet has not been studied. 
The second is to explain how the expression \eqref{eq:3ptfunction}  of three-point  Gromov--Witten invariants  appears in A-twisted 2d GLSM on $S^2$ 
\footnote{For  the cotangent bundle of Grassmannian $T^{*} \mathrm{Gr}(N,M)$,  it was observed in \cite{Chung:2016lrm} that correlation functions of a  mass deformation of $\mathcal{N}=(4,4)$ A-twisted GLSM on $S^2$
agrees with the   quantum cohomology. Also, Bethe/Gauge correspondence for the XXX spin chain and an A-twisted GLSM on $S^2$ was studied in \cite{Chung:2016lrm}.}.
The third is to show that the correlation functions of A-twisted 2d GLSM
on higher-genus Riemann surfaces
agree with those of 2d TQFT associated with the Frobenius algebra.
This suggests that these correlation functions are quasimap invariants with fixed domains.

\subsection{A-twisted GLSM on $S^2$ and genus zero quasimap invariants}
\label{sec:subAtwist}

First we consider the genus zero case. In terms of supersymmetric localization formula \cite{Closset:2015rna},
 the $n$-point correlation functions of gauge invariant functions $\widehat{\mathcal{O}}_{\text{2d}.l}$ of  the $U(N)$ vector multiplet scalar
 in the A-twisted GLSM on $S^2$ is written as the following form:
\begin{align}
\langle \prod_{l=1}^n  \widehat{\mathcal{O}}_{\text{2d}.l} \rangle_{S^2} 
&=\frac{1}{N!}\sum_{a=1}^N \sum_{k_a =0}^{\infty} \oint \prod_{a=1}^N \frac{d \sigma_a}{2\pi {\rm i} } \left( \prod_{l=1}^n \mathcal{O}_{\text{2d}.l}(\sigma, m) \right) 
Z_{\text{FI}} Z^{(0)}_{\text{2d.vec}} (\sigma, k) Z^{(0)}_{\text{2d.chi}} (\sigma, k) \nonumber \\
&=\frac{1}{N!} \sum_{a=1}^N \sum_{k_a =0}^{\infty} \oint \prod_{a=1}^N \frac{d \sigma_a}{2\pi {\rm i} } \left( \prod_{l=1}^n \mathcal{O}_{\text{2d}.l}(\sigma, m) \right)  \nonumber \\
& \qquad \qquad \times ( (-1)^{N-1} q )^{\sum_{a=1}^N k_a} \frac{  \prod_{1 \le a < b \le N}( \sigma_a-\sigma_b)^2} { \prod_{i=1}^M \prod_{a=1}^N  \left( \sigma_{a}-m_i \right)^{k_a+1-r} } .
\label{eq:residueformula2dS2}
\end{align}
Here the residues in \eqref{eq:residueformula2dS2} are 
evaluated at $\sigma_a=m_{i_a}$ with $a=1, \cdots, N$ and $ i_a=1, \cdots, M$. 
$\{ \sigma_{a} \}_{a=1}^N$ is  the saddle point value of $\mathfrak{u}(N) \otimes \mathbb{C}$ valued  scalar $\sigma$ in the $U(N)$ vector multiplet:
\begin{align}
\sigma |_{\text{saddle point}}=:\mathrm{diag} (\sigma_1, \cdots, \sigma_N). 
\end{align}
$\{ k_a \}_{a=1}^N$ is magnetic charges of gauge fields for the maximal torus of $U(N)$.
$Z_{\text{FI}}$ is the contribution from  the saddle point value of the FI term and theta term which is given by
\begin{align}
Z_{\text{FI}}=e^{-S_{\text{FI}}}|_{\text{saddle point}}=  q^{\sum_{a=1}^N k_a} ,
\end{align}
where $S_{\text{FI}}$ is the actions of FI-term and theta-term.  $\log q =  2 \pi {\rm i} \theta -\xi $ is the holomorphic  combination of the FI parameter $\xi$ and the theta angle $\theta$.
$q$ is identified with the quantum parameter or equivalently the parameter of twisted  boundary condition in the algebraic Bethe ansatz.   
 $Z^{(0)}_{\text{2d.vec}} (\sigma, k)$ and $Z^{(0)}_{\text{2d.chi}} (\sigma, k)$  are the one-loop determinant of the 2d A-twisted super Yang-Mills action on $S^2$  and the one-loop determinant of the $M$ chiral multiplets on $S^2$ given by
\begin{align}
Z^{(0)}_{\text{2d.vec}} (\sigma, k) &= (-1)^{(N-1) \sum_{a=1}^N k_a  } \prod_{1 \le a < b \le N}( \sigma_a-\sigma_b)^2, \\
Z^{(0)}_{\text{2d.chi}} (\sigma, k) &= \prod_{i=1}^M \prod_{a=1}^N  \left( \frac{1}{\sigma_{a}-m_i} \right)^{k_a+1-r}.
\end{align}
Here $m_i$  is the twisted mass  of the $i$-th chiral multiplet and $r \in \mathbb{Z}_{\ge 0}$ is a  vector type $U(1)$ R-charge. 

$\widehat{\mathcal{O}}_{\text{2d}.l}$ is a gauge invariant operator consisting of  $\sigma$.  
 $\mathcal{O}_{\text{2d}.l}(\sigma)$ in the right hand side in \eqref{eq:residueformula2dS2}  expresses the saddle point value of $\widehat{\mathcal{O}}_{\text{2d}.l}$
 which is a symmetric polynomial of $\{ \sigma_{a} \}_{a=1}^N$ and is a polynomial of the mass parameters $\{m_i \}_{i=1}^M$. 
For example,  let us consider the saddle point value of a gauge invariant operator $\mathrm{Tr}\sigma$. 
 Here $\mathrm{Tr}$ is  the  trace of a  $N \times N$ matrix. The saddle point value is given by 
\begin{align}
\mathrm{Tr} \sigma |_{\text{saddle point}} =\mathrm{Tr}  \, \mathrm{diag} (\sigma_1, \cdots, \sigma_N) =\sum_{a=1}^N  \sigma_{a} .
\end{align}
Note that the A-twisted GLSM is topologically twisted theory and the correlation function of vector multiplet scalar ($\mathcal{Q}$-closed operator) are  independent of the coordinates of $S^2$ on which 
the operators are inserted. Then an $n$-point correlation function  have the same value of  the expectation value of the  single operator consisting of the product of $n$-operators inserted on a  point on $S^2$.
 
For  any given  symmetric polynomial $f(\sigma_1, \cdots, \sigma_N)$ of $\sigma_1, \cdots, \sigma_N$, 
there exists  a gauge invariant operator of $\sigma$ whose saddle value is given by $f(\sigma_1, \cdots, \sigma_N)$.
We may define $\hat{s}_{\lambda}$ by the gauge invariant operator of $\sigma$ including $m_i$, whose saddle point is given by a  factorial Schur polynomial $s_{\lambda}(\sigma | m)$.
It is enough to consider the correlation functions of $\hat{s}_{\lambda}$, since $\{ s_{\lambda}(\sigma | m) \}_{\lambda \in \mathcal{P}_{N,M}}$ generate the symmetric polynomials  of $\sigma_1, \cdots, \sigma_N$. 

The algebra of scalars in the vector multiplet (twisted chiral ring) is given by
\begin{align}
\langle \cdots  \widehat{\mathcal{O}}_{\mathrm{2d.}i} \widehat{\mathcal{O}}_{\mathrm{2d.}j} \cdots \rangle_{S^2}=
\langle \cdots  \sum_{k } N^{\prime k}_{i j} \, \widehat{\mathcal{O}}_{\mathrm{2d.}k}  \cdots \rangle_{S^2} .
\end{align}
Here ellipses denote the arbitrary insertions of  gauge invariant combination of vector multiplet scalar and $N^{\prime k}_{ i j}$'s are the coefficients of  OPEs.

We remark on properties of  the algebra of scalars.
 The $\mathcal{Q}$-exact terms in the correlation functions vanish and do not contribute to the OPEs. 
 If there exist a basis of the algebra of vector multiplet scalars such that 
the coefficients $N^{\prime k}_{i j}$ agree with the structure constants $C^{\lambda}_{\mu \nu}$ of the quantum cohomology, then
the algebra of vector multiplet scalar is isomorphic to the quantum cohomology of Grassmannian $\mathrm{Gr}(N, M)$.  
When we regard A-twisted GLSM as a Frobenius algebra, Taking the expectation values of $\widehat{\mathcal{O}}_{\text{2d}.l}$ correspond to a Frobenius bilinear form and
a choice of R-charges $r$ corresponds to a choice of Frobenius bilinear form.    
Although the expectation values depend on the R-charge, The coefficients of OPE of $\widehat{\mathcal{O}}_{\text{2d}.l}$, i.e., the structure constants are  independent of $r$, we will show this later. 
We  will choose a particular integer of R-charge $r$ to give identification  with the Frobenius algebra with $\beta=0$.

It is shown in \cite{2016arXiv160708317K} that   \eqref{eq:residueformula2dS2} with $r=0$ 
is the generating function of  genus zero degree-$k$ ($k:=k_1+\cdots+k_N$) quasimap invariants, which imply that the three point function correlation functions of factorial Schur polynomials agrees with 
 the genus zero three points $T$-equivariant  Gromov--Witten invariants of  Schubert classes $[X_{\lambda}], [X_{\mu}],$ and $[X_{\nu}]$  of $\mathrm{Gr}(N,M)$:
\begin{align}
\langle \hat{s}_{\lambda}  \hat{s}_{\mu} \hat{s}_{\nu}  \rangle_{S^2}&= C^{(0)}_{\lambda \mu \nu  }(q,t, \beta=0).
\label{eq:3pt3pt}  
\end{align}

We show  \eqref{eq:3pt3pt} as follows.
We first perform sums of $\{ k_a\}_{a=1}^N$ in \eqref{eq:residueformula2dS2} and deform the integration contours. Then we obtain the following expression of 
$n$-point correlation function:
\begin{align}
\langle  \prod_{l=1}^n \hat{s}_{\lambda_l} \rangle_{S^2} 
&=\frac{1}{N!} \oint \prod_{a=1}^N \frac{d \sigma_a}{2\pi {\rm i} }  \left( \prod_{l=1}^n s_{\lambda_l}(\sigma | m) \right)   
\frac{Z^{(0)}_{\text{2d.vec}} (\sigma, k=0) Z^{(0)}_{\text{2d.chi}} (\sigma,k= 0)}{\prod_{a=1}^N (1- \exp \left( {\partial_{\sigma_a} W_{\text{2d.eff}} } \right) ) } \nonumber \\
&=
\sum_{\sigma} \left( \prod_{l=1}^n s_{\lambda_l} (\sigma | m) \right) 
 Z^{(0)}_{\text{2d.vec}} (\sigma, 0) Z^{(0)}_{\text{2d.chi}} (\sigma, 0) \left[ \det_{a, b} \left( \frac{\partial^2 W_{\text{2d.eff}}}{\partial \sigma_a \partial \sigma_b} \right) \right]^{-1}  \nonumber \\
&= 
\sum_{\sigma} \left( \prod_{l=1}^n s_{\lambda_l} (\sigma | m) \right) 
 \frac{\prod_{ a < b} ( \sigma_a-\sigma_b)^{2 } } {\prod_{i=1}^M \prod_{a=1}^N (\sigma_{a}-m_i )^{ 1-r  } }   \prod_{a=1}^N \left(\sum_{i=1}^M \frac{1}{\sigma_a-m_i} \right)^{-1} .
\label{eq:genus0corr}
\end{align}
Here $W_{\text{2d.eff}}$ is the effective twisted superpotential defined by 
\begin{align}
 W_{\text{2d.eff}}&:=
 \left ( \log q +\pi {\rm i} (N-1) \right)\sum_{a=1}^N  \sigma_a-\sum_{i=1}^M \sum_{a=1}^N (\sigma_a -m_i) \left[  \log  (\sigma_a -m_i)-1 \right] .
\end{align}
From the first to the second line in \eqref{eq:genus0corr}, the residues are evaluated at roots of  the saddle 
point equations of the twisted superpotential; $\exp \left( \partial_{\sigma_a} W_{\mathrm{2d.eff}} \right)=1, a=1, \cdots, N$.  
\begin{align}
 \exp \left( {\partial_{\sigma_a} W_{\text{2d.eff}} } \right) =1 
\Leftrightarrow  \prod_{i=1}^N (\sigma_a-m_i)= (-1)^{N-1}q, \quad a=1, \cdots, N. 
\label{eq:2dsaddle}
\end{align}
with $\sigma_a \neq \sigma_b$ for $a \neq b$ and up to the permutations of $\sigma_1, \cdots, \sigma_N$. When $N=1$, \eqref{eq:2dsaddle} agrees with the ring relation of $QH_T (\mathbb{P}^{M-1})$.   
The sum  $\sum_{\sigma}$  runs over the solutions of  the saddle point equation of twisted superpotential.
We find that  the saddle point equation \eqref{eq:2dsaddle} is  same as the Bethe ansatz equation \eqref{eq:Betheansatz} for $\beta=0$
with the following identification of variables between the GLSM and the Bethe ansatz \eqref{eq:Betheansatz}.
\begin{align}
\sigma_a \equiv y_a, \quad m_i \equiv t_i, \quad \text{for } \beta=0. 
\end{align}
In this article a symbol ''$\equiv$'' is used to express  identifications of variables between two different theories.

From  the determinant formula of on-shell square norm \eqref{eq:invdet0}, if we choose $r=0$, \eqref{eq:genus0corr} perfectly agrees with  the genus zero correlation functions  \eqref{eq:genus0nptfunction}
 for $\beta=0$  obtained by gluing the genus zero three point functions, i.e., the axiom of 2d TQFT. 
\begin{align}
\langle  \prod_{l=1}^n \hat{s}_{\lambda_l} \rangle_{S^2} 
&=C^{(0)}_{\lambda_1 \lambda_2 \cdots \lambda_{n}  }(q,t, \beta=0)  \quad \text{for } r=0.
\end{align}

Then  we obtained the two equivalent expressions of  the genus zero three point Gromov--Witten invariants of $\mathrm{Gr}(N,M)$ given by 
\begin{align}
\langle \prod_{l=1}^3 \hat{s}_{\lambda_i} \rangle_{S^2}  
&=\frac{1}{N!}\sum_{a=1}^N \sum_{k_a =0}^{\infty} \oint \prod_{a=1}^N \frac{d \sigma_a}{2\pi {\rm i} }  \left( \prod_{l=1}^3   s_{\lambda_l} (\sigma| t)  \right)  
((-1)^{N-1} q)^{\sum_{a=1}^N k_a}  \frac{\prod_{1 \le a < b \le N} (\sigma_a -\sigma_b)^2}{\prod_{i=1}^M  \prod_{a=1}^N (\sigma_a-m_i)^{k_a+1}} 
\label{eq:zeroquasi1}
 \\
&=\sum_{\sigma} \left( \prod_{l=1}^3   s_{\lambda_l} (\sigma| t)  \right)
 \frac{\prod_{ a < b} ( \sigma_a-\sigma_b)^{2 } } {\prod_{i=1}^M \prod_{a=1}^N (\sigma_{a}-m_i ) }  \prod_{a=1}^N \left(\sum_{i=1}^M \frac{1}{\sigma_a-m_i} \right)^{-1} .
\label{eq:zeroquasi2}
\end{align}
Here the expression of  the right hand side in the first line is same as the generating function of  quasimap invariants and the expression in the second line is the equivariant version of
genus zero Intriligator--Vafa formula.  Note that three point functions in the quasimaps gives genus zero three point Gromov--Witten invariants.
When $r=0$,   the residues for  $q^k$ with $k \ge 1$  in the two point functions $\langle \hat{s}_{\mu}  \hat{s}_{\nu} \rangle_{S^2}$ in the GLSM
with $\mu, \nu \in \mathcal{P}_{N, M}$  vanish and the two point functions agree with
 the equivariant integral of  Schubert classes  $[X_{\mu}], [X_{\nu}] \in H^*_{T}(\mathrm{Gr}(N,M))$ as
\begin{align}
\langle \hat{s}_{\mu}  \hat{s}_{\nu} \rangle_{S^2}&= \frac{1}{N!} \sum_{i_1, \cdots, i_N=1}^{M} \oint_{\sigma_a=m_{i_a}} \prod_{a=1}^N \frac{d \sigma_a}{2\pi {\rm i} }     s_{\mu} (\sigma| t)    s_{\nu} (\sigma| t) 
 \frac{\prod_{1 \le a < b \le N} (\sigma_a -\sigma_b)^2}{\prod_{i=1}^M  \prod_{a=1}^N (\sigma_a-m_i)} \nonumber \\
& =\int_{\mathrm{Gr}(N,M)}^T [X_{\mu}] \cup [X_{\nu}] .
\label{eq;two2d}
\end{align}

Since the saddle point equations is independent of the R-charge $r$, the structure constants of the algebra of $\hat{\mathcal{O}}_{\text{2d}}$ in the correlation functions 
are independent of the R-charge. This can been seen as follows. 
From the saddle point equation \eqref{eq:2dsaddle}, the  genus zero $n$-correlation function with $r$ is related to that with $r=0$ as
\begin{align}
\langle  \prod_{l=1}^n \hat{s}_{\lambda_l} \rangle_{S^2, r} 
=\langle \prod_{i=1}^M \prod_{a=1}^N (\sigma_{a}-m_i )^{ r  }  \prod_{l=1}^n \hat{s}_{\lambda_l}  \rangle_{S^2, r=0}
=((-1)^{N-1}q)^{N r} \langle  \prod_{l=1}^n \hat{s}_{\lambda_l} \rangle_{S^2, r=0}
\end{align}
Here $\langle  \cdots \rangle_{S^2, r}$ denotes the correlation functions on $S^2$ with  R-charge $r$.
Then we find that the R-charge dependence is canceled out in  the structure constants $C^{\lambda}_{ \mu \nu}=\eta^{\lambda \rho} C^{(0)}_{\rho \mu \nu}$.

\subsection{A-twisted GLSM on the general Riemann surfaces and 2d TQFT}
\label{sec:GLSMhighergenus}
Next we consider A-twisted GLSM on $\Sigma_g$ and show that the correlation  functions of A-twisted GLSM agree with \eqref{eq:genusgnptfunction}. 
In terms of the supersymmetric localization computation \cite{Benini:2016hjo}, 
an $n$-point correlation function  on $\Sigma_g$ is  written as 
\begin{align}  \label{eq:residueformula2dg} 
\langle \prod_{l=1}^n  \widehat{\mathcal{O}}_{\text{2d}.l} \rangle_{\Sigma_g} 
&=\frac{1}{N!} \oint \prod_{a=1}^N \frac{d \sigma_a}{2\pi {\rm i} } \sum_{a=1}^N \sum_{k_a =0}^{\infty}  \left( \prod_{l=1}^n  \mathcal{O}_{\text{2d}.l} (\sigma, m) \right) 
\nonumber \\
&\qquad \times Z_{\text{FI}} Z^{(g)}_{\text{2d.vec}} (\sigma, k) Z^{(g)}_{\text{2d.chi}} (\sigma, k) \left[ \det_{a, b} \left( \frac{\partial^2 W_{\text{2d.eff}}}{\partial \sigma_a \partial \sigma_b} \right) \right]^g .
\end{align}
Here the one-loop determinants for genus $g$ case are given by
\begin{align}
Z^{(g)}_{\text{2d.vec}} (\sigma, k) &= (-1)^{(N-1) \sum_{a=1}^N k_a  } \prod_{1 \le a < b \le N}( \sigma_a-\sigma_b)^{2-2g}, \\
Z^{(g)}_{\text{2d.chi}} (\sigma, k) &= \prod_{i=1}^M \prod_{a=1}^N  \left( \frac{1}{\sigma_{a}-m_i} \right)^{k_a+(1-r)(1-g)}.
\end{align}
When $g=0$, the formula is same as the localization formula on $S^2$ \eqref{eq:residueformula2dS2}.
We comment on the contour integrals in \eqref{eq:residueformula2dS2}, \eqref{eq:residueformula2dg}. When $g=0$,  in  the order by order residue computations in the power of $q$, 
 the contour integrals are determined by Jeffrey--Kirwan (JK) residues \cite{MR1318878}.  
As we have seen in the previous section,  we have two  expressions  \eqref{eq:zeroquasi1} and \eqref{eq:zeroquasi2} of the genus zero correlation functions in the A-twisted GLSM, 
both of them give the same result. 
In the derivation of the JK residue  in supersymmetric localization, it is required that the singular hyperplane arrangement is projective, for example see  \cite{Benini:2013xpa}.  
When $g >1$, the one-loop determinant of the vector multiplet of non-abelian gauge group ($N \ge 2$) violates the projective condition and 
 the order by order residue operations are not well-defined\footnote{The violation of  the projective condition  implies that
the order of the sums and the integrals do not commute.  
A prescription to make sense of JK residue operation is proposed in  \cite{Benini:2013xpa} that  regulator mass for vector multiplet is introduced to  satisfy the condition and
the regular mass is taken to zero after the JK residue operations.  A different choice of contour was proposed in \cite{Closset:2016arn}, but we do not have any agreement between the contour prescription in 
 \cite{Closset:2016arn}  and the higher genus partition functions constructed by the axiom of 2d TQFT.}.
We first perform the sums and obtain the following expression of the correlation function
\begin{align}
\langle  \prod_{l=1}^n \hat{s}_{\lambda_l} \rangle_{\Sigma_g} 
&=\frac{1}{N!} \oint \prod_{a=1}^N \frac{d \sigma_a}{2\pi {\rm i} }  \left( \prod_{l=1}^n {s}_{\lambda_l}(\sigma |m) \right) 
 Z^{(g)}_{\text{2d.vec}} (\sigma, 0) Z^{(g)}_{\text{2d.chi}} (\sigma, 0)  \frac{ \left[ \det_{a,b} \left( \partial_{\sigma_a} \partial_{\sigma_b} W_{\text{2d.eff}} \right) \right]^g  }{\prod_{a=1}^N (1- \exp \left( {\partial_{\sigma_a} W_{\text{2d.eff}} } \right) ) } \\
&=
\sum_{\sigma} \left( \prod_{l=1}^n {s}_{\lambda_l}(\sigma |m) \right) 
 Z^{(g)}_{\text{2d.vec}} (\sigma, 0) Z^{(g)}_{\text{2d.chi}} (\sigma, 0) \left[ \det_{a, b} \left( \frac{\partial^2 W_{\text{2d.eff}}}{\partial \sigma_a \partial \sigma_b} \right) \right]^{g-1}  \\
&= 
\sum_{\sigma} \left( \prod_{l=1}^n  {s}_{\lambda_l}(\sigma |m) \right) 
 \left( \frac{\prod_{ a < b} ( \sigma_a-\sigma_b)^{2 } } {\prod_{i=1}^M \prod_{a=1}^N (\sigma_{a}-m_i )^{ 1-r  } }   \prod_{a=1}^N \left(\sum_{i=1}^M \frac{1}{\sigma_a-m_i} \right)^{-1} \right)^{(1-g)} .
\label{eq:genusgcorr}
\end{align}
Again, if we choose $r=0$, \eqref{eq:genusgcorr} perfectly agrees with the correlation function on a genus $g$ correlation function obtained from the axiom of 2d TQFT;
\begin{align}
\langle  \prod_{l=1}^n \hat{s}_{\lambda_l} \rangle_{\Sigma_g} 
=C^{(g)}_{\lambda_1 \lambda_2 \cdots \lambda_{n}  }(q,t, \beta=0) \quad \text{for } r=0. 
\label{eq:gnpointc}
\end{align}
In particular the partition  function of the A-twisted  GLSM with  $r=0$  on $\Sigma_g$ agrees with the genus $g$ partition function of 2 TQFT with $\beta=0$;
\begin{align}
\langle 1 \rangle_{\Sigma_g}=\sum_{\lambda \in \mathcal{P}_{N,M }}{\sf e}(y_{\lambda}, y_{\lambda})^{g-1}=Z(\Sigma_g) \quad \text{for } \beta=0.
\end{align}
When $r=0$ and $m=0$, \eqref{eq:genusgcorr} reproduces the genus $g$ Intriligator--Vafa formula  \cite{Vafa:1990mu, Intriligator:1991an, MR1621570}.
 
\subsubsection*{\underline{Seiberg-like duality for A-twisted GLSMs}}
We explain Seiberg-like duality of a $U(N)$ A-twisted GLSM and a $U(M-N)$ A-twisted GLSM.
Let $\{m_i \}_{i=1}^{M}$, (resp. $\{m^{\prime}_{i} \}_{i=1}^{M}$ ) be the twisted masses of $M$ chiral multiplets in $U(N)$ GLSM, (resp. $U(M-N)$ GLSM). 
When $\beta=0$, $\ominus t^{\prime}=(-t_M, -t_{M-1}, \cdots, -t_1)$. Therefore \eqref{eq:levelrankg} and \eqref{eq:gnpointc} lead to the following equality of 
correlation functions between two A-twisted GLSMs.
\begin{align}
\langle  \prod_{l=1}^n \hat{s}_{\lambda_l} \rangle_{U(N), \Sigma_g} 
=\langle  \prod_{l=1}^n \hat{s}_{\lambda^{\prime}_l} \rangle_{U(M-N), \Sigma_g} 
\end{align}
under the identification of mass parameters $m^{\prime}_{i}=-m_{M+1-i}$ for $i=1, \cdots, M$.
Here $\langle \cdots \rangle_{U(N), \Sigma_g}$ denotes the correlation functions of A-twisted GLSM with gauge group $U(N)$ on $\Sigma_g$.

\section{3d $\mathcal{N}=2$  Chern--Simons--matter theory and quantum $K$-theory }
\label{sec:CSM}
\subsection{Algebra of Wilson loops and quantum $K$-ring of Grassmannian}
\label{sec:CSMsub1}
In this section we  introduce topologically twisted 3d $\mathcal{N}=2$  Chern--Simons (CS) theories coupled to chiral multiplets (CS--matter theory) on $S^1 \times \Sigma_g$. We consider the gauge group  $G=U(N)$. 
The $M$ chiral multiplets are in the fundamental representation of the gauge group $U(N)$. 
We will  show the correspondence  between the algebra of supersymmetric Wilson loops  and  the equivariant small quantum $K$-theory ring of Grassmannian $\mathrm{Gr}(N, M)$.
For the moment we take  supersymmetric gauge, flavor, R-symmetry mixed   Chern--Simons  terms with the generic CS levels: 
\begin{align}
{\rm S}_{\text{CS}}&= \int_{S^1 \times \Sigma_g} \Bigl( \frac{\kappa^{(1)}}{4 \pi {\rm i}} \mathrm{Tr} \left(A \wedge d A - \frac{2 {\rm i}}{3} A^3 \right) +\frac{\kappa^{(2)} }{4 \pi {\rm i}} \mathrm{Tr} A \wedge  d \mathrm{Tr} A 
+ \sum_{i=1}^M \frac{\kappa^{(3)}_i}{4 \pi {\rm i}}  C^{(i)} \wedge d \mathrm{Tr} A  
  \nonumber \\
&+\frac{1}{2 \pi {\rm i}}  B \wedge d \mathrm{Tr} A + \frac{\kappa^{(4)} }{4 \pi {\rm i}} \mathrm{Tr} A \wedge d A^{(R)} + \sum_{i=1}^M\frac{\kappa^{(5)}_i}{4 \pi {\rm i}} C^{(i)} \wedge d A^{(R)} + \sum_{i,j=1}^M\frac{\kappa^{(6)}_{i j}}{4 \pi {\rm i}} C^{(i)} \wedge d C^{(j)}+\cdots  \Bigr).
\label{eq:CSterm}
\end{align}
Here $A$ is the $U(N)$ dynamical gauge field. $B$ is the background gauge field which couples to the topological $U(1)$-current. $A^{(R)}$ is the background gauge field for $U(1)$ R-symmetry. 
$C^{(i)}$  is the  background gauge field for a $U(1)$ flavor symmetry non-trivially acting on the $i$-th chiral multiplet for $i=1,\cdots, M$.  The ellipses denote supersymmetric completion of the CS-terms which includes  3d $\mathcal{N}=2$ super partners of the dynamical and background gauge fields.  
$\kappa^{(1)}, \kappa^{(2)}, \kappa^{(3)}_i \kappa^{(4)}, \kappa^{(5)}_i, \kappa^{(6)}_{i j}$ denote integer or half integer valued  CS levels
$\kappa^{(1)}, \kappa^{(2)}$ are the gauge CS levels, $\kappa^{(3)}_i$'s are  the gauge-flavor mixed CS levels, $\kappa^{(4)}$ is the gauge-R-symmetry mixed CS level, $\kappa^{(5)}_i$'s are 
the flavor-R-symmetry mixed CS levels, and  $\kappa^{(6)}_{i j}$'s are the flavor CS levels. 
The CS levels in the Lagrangian have to satisfy  the condition for the absence of  anomaly.  The values of CS levels will be determined later.

By supersymmetric localization formula \cite{Benini:2015noa, Benini:2016hjo, Closset:2016arn} (see also  an earlier work \cite{Ohta:2012ev}), the path integrals of  the topologically twisted 3d $\mathcal{N}=2$ CS--matter theory on $S^1 \times \Sigma_g$ is reduced to $N$-dimensional multi-contour integrals and infinite magnetic sums. An $n$-point correlation function of supersymmetric gauge and flavor Wilson loops is given by
\begin{align}
\langle \prod_{l=1}^n \widehat{\mathcal{O}}_{\text{3d}.l} \rangle_{S^1 \times \Sigma_g} 
&=\frac{1}{N!} \oint \prod_{a=1}^N \frac{d x_a}{2\pi {\rm i} x_a} \sum_{a=1}^N \sum_{k_a =0}^{\infty} \left( \prod_{l=1}^n \mathcal{O}_{\text{3d}.l} (x,z)  \right) \nonumber \\
& \times Z^{(g)}_{\text{CS}}  Z^{(g)}_{\text{3d.vec}} (x) Z^{(g)}_{\text{3d.chi}} (x) 
\left[ \det_{a, b} \left( \frac{\partial^2 W_{\text{3d.eff}}}{\partial \log x_a \partial \log x_b} \right) \right]^{g} .
\label{eq:residueformula}
\end{align}
Here the integrand is given as follows. $x_a$ for $ a=1, \cdots, N$ are the saddle point value of supersymmetric Wilson loop for the $a$-th diagonal $U(1) \subset U(N)$. 
$z_i:=e^{\oint_{S^1} ( {\rm i} C^{(i)}_{\tau} - \sigma^{(i)}) d \tau}$ is the supersymmetric Wilson loop for  background  flavor $U(1)$ vector multiplet.  
$\sigma^{(i)}$  is the $U(1)$  scalar (real mass) belonging to the same background vector multiplet of $C^{(i)}$. $\tau$  is a coordinate of $S^1$. All the gauge and flavor Wilson loops we consider wrap on the $S^1$-direction  of $S^1 \times \Sigma_g$ and are located on  points on $\Sigma_g$.

$\widehat{\mathcal{O}}_{\text{3d}.l}$ in the left hand side is a polynomial function of supersymmetric gauge Wilson loops and background flavor Wilson  loops. 
 $\mathcal{O}_{\text{3d}.l}(x,z)$ is the saddle point value of $\widehat{\mathcal{O}}_{\text{3d}.l}$, which is symmetric Laurent polynomial of $\{ x_a \}_{a=1}^N$ and a Laurent polynomial of $\{ z_i \}_{i=1}^M$. 
Note that for any Laurent  polynomial $f$,
 there exists a polynomial function of Wilson loops whose saddle point value is given by $f$. 
For example,  the saddle point value of the Wilson loop $W_{\wedge^{\ell} {\Box}}$ (resp. $W_{\wedge^{\ell} \overline{\Box}}$)  in the $\ell$-th anti-symmetric products of fundamental (resp. anti-fundamental) representation of $U(N)$  is written as 
\begin{align}
W_{\wedge^{\ell} {\Box}} |_{\text{saddle point}}&=\mathrm{Tr}_{\wedge^{\ell} {\Box}} \mathrm{Pexp} \left(  \oint  ({\rm i} A_{\tau}  - \sigma^{\prime}) d \tau \right) \Big|_{\text{saddle point}}
 = \sum_{1 \le a_1 < \cdots < a_{\ell} \le N} \prod_{i=1}^{\ell} x_{a_i} , \\
W_{\wedge^{\ell} \overline{\Box}} |_{\text{saddle point}}&=\mathrm{Tr}_{\wedge^{\ell} \overline{\Box}} \mathrm{Pexp} \left(  \oint ({\rm i} A_{\tau}  - \sigma^{\prime}) d \tau \right) \Big|_{\text{saddle point}}
= \sum_{1 \le a_1 < \cdots < a_{\ell} \le N} \prod_{i=1}^{\ell} x^{-1}_{a_i} .
\end{align}
 $\sigma^{\prime}$ is the adjoint scalar in the $U(N)$ vector multiplet. 

The algebra of Wilson loop is defined by the following OPE 
\begin{align}
\langle \cdots  \widehat{\mathcal{O}}_{\mathrm{3d.}i} \widehat{\mathcal{O}}_{\mathrm{3d.}j} \cdots \rangle_{S^1 \times \Sigma_g}=
\langle \cdots  \sum_{k } N^{k}_{i j} \, \widehat{\mathcal{O}}_{\mathrm{3d.}k}  \cdots \rangle_{S^1 \times \Sigma_g} .
\end{align}
Here ellipses denote the arbitrary insertions of Wilson loop operators and $N^k_{i j}$ is the coefficient of OPE (structure constants).
 If there exist a basis of the algebra of Wilson loops such that 
the coefficients $N^k_{i j}$ agree with the structure constants $C^{\lambda}_{\mu \nu}$ of the quantum $K$-ring, then
the algebra of Wilson loops is isomorphic to the quantum $K$-ring. 
Note that the correlation functions and the coefficients of OPE do not depend on the coordinates of $S^1 \times \Sigma_g$, 
since  topologically twisted CS--matter theories  are topological field theories.

$Z^{(g)}_{\text{CS}}$ is  the saddle point value of  the mixed CS terms \eqref{eq:CSterm}  given by
\begin{align}
Z^{(g)}_{\text{CS}}&=e^{-{\rm S}_{\text{CS}}} \Big|_{\text{saddle point}} \nonumber \\
&=q^{\sum_{a=1}^N k_a } \left( \prod_{a=1}^N x_{a}^{\kappa^{(1)} k_a +\kappa^{(2)} \sum_{b=1}^N k_b+ \sum_{i=1}^M \kappa^{(3)}_i n_i+(g-1)\kappa^{(4)}} \right)  \nonumber \\
&\quad \quad \qquad  \times \left( \prod_{i=1}^M  z_{i}^{\kappa^{(3)}_{ i} \sum_{a=1}^N k_a+\sum_{j=1}^M \kappa^{(6)}_{ i j} n_j+(g-1) \kappa^{(5)}_i} \right) .
\label{eq:CSterms}
\end{align}
Here  $q=e^{\oint_{S^1} ( {\rm i} B_{\tau} - \sigma_B ) d \tau }$ is the  supersymmetric Wilson loop  for the background gauge field $B$. $\sigma_B$ is $U(1)$ scalar belonging to the same background supermultiplet of $B$, which is  the FI-parameter 
in three dimensions.  
 $k_a$ and $n_i$ are  magnetic charges  (the first Chern numbers) of the gauge field of the $a$-th diagonal $U(1) \subset U(N)$ and a $U(1)$ background gauge field 
$C^{(i)}$ in the Riemann surface direction, respectively.

$Z^{(g)}_{\mathrm{3d. vec}}$ and $Z^{(g)}_{\text{3d.chi}} $  are  the one-loop determinants of the 3d $U(N)$ super Yang-Mills action and 
 the one-loop determinant of  $M$-tuple chiral multiplets in the fundamental representation of $U(N)$ given by
\begin{align}
Z^{(g)}_{\mathrm{3d. vec}} (x,z, k)&= (-1)^{(N-1){\sum_{a=1}^N k_a} }\prod_{1 \le a \neq b \le N} (1-x_a x^{-1}_{b})^{1-g}, 
\label{eq:z1loopvec} \\
Z^{(g)}_{\text{3d.chi}} (x,z, k, n, r)&=\prod_{i=1}^M \prod_{a=1}^N\left( \frac{ (x_a z^{-1}_i)^{\frac{1}{2}} }{1-x_a z^{-1}_i  } \right)^{k_a+n_i +(1-g)(1-r)} .
\label{eq:z1loopvchi3}
\end{align}
$r \in \mathbb{Z}_{\ge 0}$ is an integer valued $U(1)$ R-charge of the 3d chiral multiplets. 
$W_{\text{3d.eff}}$ is the   effective twisted superpotential  given  by
\begin{align}
 W_{\text{3d.eff}}
&=
\frac{\kappa^{(1)}+\frac{M}{2} }{2} \sum_{a=1}^N  (\log x_a)^{2}+\frac{\kappa^{(2)}}{2} \left(\sum_{a=1}^N \log x_a\right)^{2}
+\sum_{i=1}^M \sum_{a=1}^N \left( \kappa^{(3)}_i +\frac{1}{2} \right)\log z_i \log x_a  \nonumber \\
&\quad
 +\left( \log q +\pi {\rm i} (N-1) \right)\sum_{a=1}^N  \log x_a+\sum_{i=1}^M \sum_{a=1}^N  \mathrm{Li}_2 (x_a z^{-1}_i) \nonumber \\
&\qquad \qquad \qquad \qquad  +\sum_{i=1}^M\frac{N}{4} (\log  z_i)^2 +\sum_{i, j=1}^M\frac{\kappa^{(6)}_{i j} }{2} (\log  z_i) (\log  z_j).
\label{eq:3deffsp}
\end{align}
The last two terms  in \eqref{eq:3deffsp} do not contribute to the computation of the correlation functions on $S^1 \times \Sigma_g$.
As explained in the section \ref{sec:GLSMhighergenus}, when $g=0,1$,  the singular hyperplane arrangements are projective and 
 the integration contours in \eqref{eq:residueformula}, order by order in powers of $q$  are  determined by the JK residue operations.
 The contours are  chosen to enclose $x_a=z_i$ for all $a=1, \cdots, N$ and $i=1,\cdots, M$.   
On the other hand, when $g\ge 2$ and $N \ge 2$, the singular hyperplane arrangements including the  one-loop determinant of  the super Yang-Mills action are not projective. 
We expect the ordering  of the sums and the contour integrals  does not commute in general.  
When  the sums of $k_a$ for  $a=1,\cdots, N$ are performed before the integration, we obtain the following 
expression of  the correlation function.
\begin{align}
&\langle \prod_{l=1}^n \widehat{\mathcal{O}}_{\text{3d}.l} \rangle_{S^1 \times \Sigma_g}  \nonumber \\
&=\frac{1}{N!} 
\oint \prod_{a=1}^N \frac{d x_a}{2\pi {\rm i} x_a}   \left( \prod_{l=1}^n {\mathcal{O}}_{\text{3d}.l} (x,z)  \right) 
\left( \prod_{a=1}^N x_{a}^{\sum_{i=1}^M \kappa^{(3)}_i n_i+(g-1)\kappa^{(4)} } \right) 
 \left( \prod_{i=1}^M  z_{i}^{(g-1) \kappa^{(5)}_i +\sum_{j=1}^M \kappa^{(6)}_{ i j} n_j} \right)
 \nonumber \\
& \qquad \qquad \times  
   \frac{Z^{(g)}_{\text{3d.vec}} (k=0) Z^{(g)}_{\text{3d.chi}} (k=0) }{\prod_{a=1}^N(1-e^{  \partial_{\log x_a} W_{\text{eff}} })} 
\left[ \det_{a, b} \left( \frac{\partial^2 W_{\text{3d.eff}}}{\partial \log x_a \partial \log x_b} \right) \right]^{g} 
\label{eq:CSresum12}
 \\
&= \left( \prod_{i=1}^M  z_{i}^{\sum_{j=1}^M \kappa^{(6)}_{ i j} n_j} \right) 
 \sum_{x} 
 \left( \prod_{l=1}^n {\mathcal{O}}_{\text{3d}.l} (x,z) \right) 
\left( \prod_{a=1}^N x_{a}^{\sum_{i=1}^M \kappa^{(3)}_i n_i } \right) 
\mathcal{H}^{1-g} (x, z, \kappa, r) .
\label{eq:CSresum2}
\end{align}
Here  we introduced $\mathcal{H} (x, z, \kappa, r)$ to shorten the formulas by
\begin{align}
\mathcal{H}(x, z, \kappa, r)&:=\left( \prod_{i=1}^M z_i^{  \frac{N(1-r)}{2}-\kappa^{(5)}_i } \right) 
 \left( \prod_{a > b} (x_a -x_b )^{2} \right)   
 \left(  \prod_{a=1}^N x_a^{ \frac{M(1-r)}{2} -\kappa^{(4)}-(N-1)  }\right)   \nonumber \\
& \qquad \qquad  \times \prod_{a=1}^N \prod_{i=1}^M  \left( \frac{ 1 }{z_i-x_a } \right)^{1-r} 
\left[ \det_{a, b} \left( \frac{\partial^2 W_{\text{3d.eff}}}{\partial \log x_a \partial \log x_b} \right) \right]^{-1} .
\end{align}
The Hessian of the 3d effective twisted superpotential is written as
\begin{align}
   \frac{\partial W_{\text{3d.eff}}}{\partial \log x_a \partial \log x_b} = \delta_{a b }\left(\kappa^{(1)}+\frac{M}{2} \right)+\kappa^{(2)} +\delta_{ab } \sum_{i=1}^M \frac{ x_a }{z_i-x_a} .
\label{eq:hess}
\end{align}
In \eqref{eq:CSresum12}, the residues are evaluated at the roots of  the saddle point equations of the 3d twisted superpotential 
$\exp \left( \partial_{\log x_a} W_{\mathrm{3d.eff}} \right)=1, a=1, \cdots, N$ 
with $x_a \neq x_b$ for $ 1 \le a \neq b \le N$.
The sum  $\sum_{x}$  runs over the solution of  the saddle point equation of twisted superpotential;
\begin{align}
1=\exp \left(   \frac{\partial W_{\text{3d.eff}}}{\partial \log x_a} \right)=
\frac{(-1)^{N-1} q x^{\kappa^{(1)} +\frac{M}{2}}_a }
{ \prod_{b=1}^N x^{-\kappa^{(2)} }_b}  \prod_{i=1}^M   \frac{ z_i^{\kappa^{(3)}_{i}+\frac{1}{2} }}{z_i-x_a } .
\label{eq:saddleeq}
\end{align}

Note that the expression of the correlation functions in terms of the roots of the saddle point equation (the Bethe roots) \eqref{eq:CSresum2} is useful to  show the isomorphism  between the Wilson loop algebra and the quantum $K$-ring $QK_T(\mathrm{Gr}(N,M))$ for general $N, M$, but it is not easy to compute correlation functions explicitly  from it, since 
 we do not have simple closed expressions of the Bethe roots for $\beta=-1$ and $N \ge 2$.  
\footnote{When $\beta=0$, the Bethe ansatz equation (the saddle point equation of $W_{\text{2d}}$)  factorized to $N$  saddle point equations for the  $U(1)$ GLSM. Then we have simple analytic expressions of Bethe roots for $\beta=0, m_i=0$ and can compute genus $g$ correlation functions directly from \eqref{eq:genusgcorr} by substituting the Bethe roots.  
}, 
On the 
other hand  \eqref{eq:residueformula} for $g=0$ gives an explicit procedure to compute genus zero 
correlation functions, order by order in the $q$-expansions in terms of JK residues, which will be useful to compute structure constants of quantum $K$-rings.

Now we verify the correspondence  between the algebra of Wilson loops and the quantum $K$-ring. 
We choose the gauge CS levels and the gauge-flavor mixed  CS levels as
\begin{align}
\kappa^{(1)} =N-\frac{M}{2}, \, \, \kappa^{(2)}=-1, \, \, \kappa^{(3)}_{i}=\frac{1}{2} .
\label{eq:pareid}
\end{align}
In particular, when  $N=1$, \eqref{eq:pareid} means that the effective level $\kappa^{(1)}+\kappa^{(2)}+M/2$ for $U(1)$ gauge CS term  equals to zero. 
 If we decompose the $U(N)$ gauge field to $SU(N)$ part and $U(1)$ part, we find  that the choice \eqref{eq:pareid} gives the zero effective CS level for  $U(1)$ part and 
$U(1)$ gauge CS term vanishes in the quantum level. The gauge-flavor mixed CS terms  also vanish, since the $U(1)$ flavor symmetries are  the Cartan part of $SU(M)$ global symmetry.
We will  see in section \ref{sec:S1xD2} that the choice in \eqref{eq:pareid} corresponds to the zero effective gauge CS levels and gauge-flavor mixed  CS levels on $S^1 \times D^2$.

When we choose gauge, gauge-flavor mixed CS levels as  \eqref{eq:pareid}, the saddle point equation \eqref{eq:saddleeq} becomes 
\begin{align}
1=\frac{(-1)^{N-1} q x^{N }_a}
{ \prod_{b=1}^N x_b}  \prod_{i=1}^M   \frac{ z_i}{z_i-x_a }, \quad a=1, \cdots, N.
\label{eq:saddleeq2}
\end{align}
We find that  the saddle point equation \eqref{eq:saddleeq2} equals to  the Bethe ansatz equation \eqref{eq:Betheansatz}  for $\beta=-1$
with the following   identification of variables: 
\begin{align}
x_a \equiv 1-y_a, \, \, z_i \equiv 1-t_i ,
\label{eq:monowilson}
\end{align}
where the $x_a$ and $z_i$ are a root of the twisted superpotentials and a flavor Wilson loop in the CS--matter theory and 
$y_a$ and $t_i$ are a Bethe root and an inhomogeneous parameter in the algebraic Bethe ansatz.

For $U(1)$ gauge theory, it is easy to see the agreement between the algebra of Wilson loops  and the quantum $K$-ring of projective space as follows.
In this case, the  saddle point value of a $U(1)$ gauge Wilson loop is an element of $\mathbb{C}[ x, x^{-1}]$, where $x:=x_1$   and  the saddle point equation is written as
\begin{align}
\exp \left( \frac{\partial W_{\text{3d.eff}} } {\partial \log x } \right) =1 \Leftrightarrow \prod_{i=1}^M (1-x z^{-1}_{i}) -q =0.
\end{align}
From \eqref{eq:CSresum2}, the following relation holds in the correlation function 
\begin{align}
\left\langle \widehat{\mathcal{O}}_{\mathrm{3d}}(x,z_i) \left(\prod_{i=1}^M (1-\hat{x} z^{-1}_{i}) -q \right) \right\rangle_{S^1 \times \Sigma_g}  =0 .
\end{align}
Here we express (polynomials of)  $U(1)$  Wilson loops as $\hat{x}$ and $\widehat{\mathcal{O}}_{\mathrm{3d}}(x,z_i)$ such that the saddle point 
values are given by $x$ and $\mathcal{O}_{\mathrm{3d}}(x,z_i)$, respectively.  ${\mathcal{O}}_{\mathrm{3d}}(x,z_i)$ is an arbitrary element of $\mathbb{C}[q, \{z^{\pm}_i\}_{i=1}^M ,x^{\pm1}]$. 
Then the algebra of Wilson loop for $U(1)$ theory with $M$ chiral multiplets is identified with
\footnote{The precise form of the  coefficient ring $R$ of Wilson loop algebra  depends on gauge-R-symmetry, flavor-R-symmetry mixed CS levels and R-charge $r$ and the flavor charges $n_i$. If we choose CS levels and the charges to agree with the Frobenius bilinear form   \eqref{eq:innerprod}, the correlation functions  are not polynomial of $q$. 
On the other hand, if we consider  parameters to agree with a Frobenius bilinear form defined in \cite{MR2772069}, the correlation functions are  polynomials with respect to $q$, for example see \eqref{eq:twopoint1}
 and 
\eqref{eq:twopoint2}.}
\begin{align}
R [ x, x^{-1}] / (  \prod_{i=1}^M(1-x z^{-1}_{i})-q ).
\label{eq:algU1Wilson}
\end{align}
If $x$ is identified with $K$-theory class of a line bundle  $\mathcal{O}(-1)$ of $\mathbb{P}^{M-1}$, we find that the algebra of Wilson loops \eqref{eq:algU1Wilson} reproduces the equivariant quantum $K$-ring $Q K_T (\mathbb{P}^{M-1})$. Note that the coefficients of OPEs of Wilson loops or equivalently the structure constants are independent of  
choices of $\kappa^{(4)}, \kappa^{(5)}_i, \kappa^{(6)}_{i j}, n_i$ and $r$.  
A choice of  $\kappa^{(4)}, \kappa^{(5)}_i, \kappa^{(6)}_{i j},  n_i$ and $r$ fixes the value of the Wilson loop correlation functions \eqref{eq:CSresum2} which 
 corresponds to  a choice of Frobenius bilinear form.

For general $N$, it is not easy to find 
the ring relation of the quotient of Laurent polynomial ring for the Wilson loop algebra. But, to 
show the isomorphism between the Wilson loop algebra and the quantum $K$-ring, 
it is enough to see the equality between 
the three point correlation functions of basis of Wilson loop algebra on $S^1 \times S^2$ and $C^{(0)}_{\lambda \mu \nu}$ for $\beta=-1$ since genus zero three point functions  fix the ring structure.

To show the agreement  between the quantum $K$-ring  and the CS--matter theory,
we define an operator $ \mathcal{W}_{\lambda }$ in the  algebra of the Wilson loops  such that 
 the saddle point value of ${\mathcal{W}}_{\lambda}$   equals to
 a Grothendieck polynomial $G_{\lambda} (y | \ominus t)$ for $\beta=-1$ with the identification of variables \eqref{eq:monowilson}:
\begin{align}
\mathcal{W}_{\lambda} |_{\text{saddle point}}= {G}_{\lambda} (y |\ominus t),  \quad \beta=-1, \,  x_i \equiv 1-y_i \text{ and } z_i \equiv 1-t_i.
\end{align}
For example, an operator $\mathcal{W}_{(1,0^{N-1})}$ with $(1,0^{N-1}):=(1,0 , \cdots,0) \in \mathcal{P}_{N,M}$  corresponds to the following combination gauge and flavor Wilson loops
\begin{align}
{\mathcal{W}}_{(1,0^{N-1})}=1- \left( \prod_{i=1}^N z^{-1}_i \right) W_{\wedge^N \Box } \Leftrightarrow G_{(1,0^{N-1})}(y| \ominus t)=1-\prod_{a=1}^N \frac{1-y_a}{1-t_a}  ,
\end{align}
where 
 $ W_{\wedge^N \Box }$ is the gauge Wilson loop in the $N$-th anti-symmetric products of the fundamental  representation  of 
$U(N)$. Since Grothendieck polynomials $\{ G_{\lambda}(y| \ominus t) \}_{\lambda \in \mathcal{P}_{N,M}}$ generate the symmetric polynomials of  a Bethe root $y_1, \cdots, y_N$ \cite{Gorbounov:2014bra}, 
the arbitrary functions of Wilson loops are generated by $\{ \mathcal{W}_{\lambda} \}_{\lambda \in \mathcal{P}_{N,M}}$.  Here the structure constants is determined by the Bethe ansatz equation.

We choose the R-charge, flavor charges, the gauge--R-symmetry mixed CS level, flavor--R-symmetry mixed CS levels as
\begin{align}
r=0, \quad n_i=0, \quad  \kappa^{(4)}=\frac{M}{2}, \quad \kappa^{(5)}_i=-\frac{N}{2}.
\label{eq:parameter2}
\end{align}
From the determinant formula of the on-shell square norm evaluated in the section \ref{sec:norm}, we find that $\mathcal{H}$ equals to  the inverse of the  on-shell square  norm of Bethe vectors  \eqref{eq:invnorm3d} for $\beta=-1$ under the choice 
\eqref{eq:pareid}, \eqref{eq:parameter2} and the identification of variables \eqref{eq:monowilson}:
\begin{align}
\mathcal{H}&= \prod_{a > b} (x_a -x_b )^{2}  \left(  \prod_{a=1}^N x_a^{1-N  }\right)   
\left( \prod_{a=1}^N \prod_{i=1}^M   \frac{ z_i }{z_i-x_a } \right) \det_{a,b} \left[ \delta_{a,b} \left(\sum_{i=1}^M \frac{x_a}{z_i-x_a} +N  \right) -1\right]^{-1} \nonumber \\
&={\sf e}(y,y)^{-1} .
\end{align}
Therefore the arbitrary $n$-point correlation functions of $\mathcal{W}_{\lambda_l}$ for $l=1,\cdots, n$  on $S^1 \times \Sigma_g$ perfectly agree with   $C^{(g)}_{\lambda_1, \cdots, \lambda_n}$
 obtained by the contractions of structure constants of $QK_T(\mathrm{Gr}(N,M))$,  $\eta_{\mu \nu}$ and $\eta^{\mu \nu}$:
\begin{align}
\langle \prod_{l=1}^n \mathcal{W}_{\lambda_l} \rangle_{S^1 \times \Sigma_g} =\sum_{\alpha \in \mathcal{P}_{M,N}} 
\frac{ \prod_{l=1}^n G_{\lambda_l}(y_{\alpha} | \ominus t) }{ {\sf e}(y_{\alpha}, y_{\alpha})^{1-g} }=C^{(g)}_{\lambda_1, \cdots, \lambda_n} (q,t, \beta=-1) .
\label{eq:npointg}
\end{align}
In particular,  \eqref{eq:npointg}  includes the agreement between the genus zero three point functions.
Therefore  the algebra of Wilson loops  is isomorphic to the  quantum $K$-ring of Grassmannian $QK_T(\mathrm{Gr}(N,M))$ with 
the Frobenius bilinear form in \cite{Gorbounov:2014bra}.   

We comment on properties of the algebra of Wilson loops.
If we choose different values of  $\kappa^{(4)}$, $\kappa^{(5)}$, $\kappa^{(6)}_{i j }$, $n_i$, and $r$, then correlation functions  correspond to  different Frobenius bilinear forms. 
For example, a choice of parameters
\begin{align}
r=0, \quad n_i=0, \quad  \kappa^{(4)}=1+\frac{M}{2}, \quad \kappa^{(5)}_i=-\frac{N}{2}.
\label{eq:parameter3}
\end{align}
agrees with a Frobenius bilinear form of the quantum $K$-ring in \cite{MR2772069} (see also sentences below (5.42)  in \cite{Gorbounov:2014bra}). If \eqref{eq:pareid} is satisfied,
both of  \eqref{eq:parameter2} and \eqref{eq:parameter3} gives the  structure constants of quantum $K$-ring. Namely  the OPEs of Wilson loop
 have the  same form of the quantum products in the quantum $K$-ring;
\begin{align}
\mathcal{W}_{\mu } \mathcal{W}_{\nu }=\sum_{\lambda \in \mathcal{P}_{N, M}} C^{\lambda}_{\mu \nu} \, \mathcal{W}_{\lambda}.
\end{align}
Here $C^{\lambda}_{\mu \nu}$ are the structure constants of $QK_T(\mathrm{Gr}(N,M))$.

If we choose as \eqref{eq:parameter2},  $\langle \mathcal{W}_{\lambda} \mathcal{W}_{\mu} \mathcal{W}_{\nu} \rangle_{S^1 \times S^2}$ is not polynomial with respect  to 
the quantum parameter $q$. On the other hand, if we choose \eqref{eq:parameter3},  $\langle \mathcal{W}_{\lambda} \mathcal{W}_{\mu} \mathcal{W}_{\nu} \rangle_{S^1 \times S^2}$ is 
polynomials with respect  to the quantum parameter $q$. 
For example  we explicitly calculate the genus zero two point functions and the structure constants for $z_i=1$, $N=2$ and $M=4$ .
Let us define operators $\{ \widehat{\mathcal{O}}_{1}, \widehat{\mathcal{O}}_{2}, \widehat{\mathcal{O}}_{3}, \widehat{\mathcal{O}}_{4}, \widehat{\mathcal{O}}_{5}, \widehat{\mathcal{O}}_{6} \}:=\{ \mathcal{W}_{\emptyset}, 
\mathcal{W}_{(1)}, \mathcal{W}_{(1,1)},  \mathcal{W}_{(2)},  \mathcal{W}_{(2,1)},  \mathcal{W}_{(2,2)} \}$ with fixed ordering; $\widehat{\mathcal{O}}_1=\mathcal{W}_{\emptyset}=1$, 
$\widehat{\mathcal{O}}_2=\mathcal{W}_{(1)}$ and so on.
We also define $C_{a b c}$ by  $\langle \widehat{\mathcal{O}}_{a} \widehat{\mathcal{O}}_{b} \widehat{\mathcal{O}}_{c} \rangle_{S^1 \times S^2}$ with the choices \eqref{eq:pareid} and \eqref{eq:parameter2}
and define $C^{\prime}_{a b c}$ by $ \langle \widehat{\mathcal{O}}_{a} \widehat{\mathcal{O}}_{b} \widehat{\mathcal{O}}_{c} \rangle_{S^1 \times S^2}$ with the choices \eqref{eq:pareid} and \eqref{eq:parameter3}.
From \eqref{eq:residueformula} for $g=0$,  we can compute the two point correlation functions $C_{1 a b}$ and $C^{\prime}_{1 a b}$. 
In matrix notation $(C_1)_{a b}:=C_{1 a b }=\langle \widehat{\mathcal{O}}_{a} \widehat{\mathcal{O}}_{b}  \rangle_{S^1 \times S^2}$, the two point functions are given by
\begin{align}
C_{1}=\frac{1}{1-q}\left(
\begin{array}{cccccc}
 1 & 1 & 1 & 1 & 1 & 1 \\
 1 & 1 & 1 & 1 & 1 & 0 \\
 1 & 1 & 1 & 0 & 0 & 0 \\
 1 & 1 & 0 & 1 & 0 & 0 \\
 1 & 1 & 0 & 0 & 0 & 0 \\
 1 & 0 & 0 & 0 & 0 & 0 \\
\end{array}
\right)
+
\frac{1}{1-q}\left(
\begin{array}{cccccc}
 0 & 0 & 0 & 0 & 0 & 0 \\
 0 & 0 & 0 & 0 & 0 & q \\
 0 & 0 & 0 & q & q & q \\
 0 & 0 & q & 0 & q & q \\
 0 & 0 & q & q & q & q \\
 0 & q & q & q & q & q^2 \\
\end{array}
\right) .
\label{eq:twopoint1}
\end{align}
The matrix notation of two point functions $(C^{\prime}_1)_{ab}:=C^{\prime}_{1 a b}$ is given by
\begin{align}
C^{\prime}_{1}=\left(
\begin{array}{cccccc}
 0 & 0 & 0 & 0 & 0 & 1 \\
 0 & 0 & 0 & 0 & 1 & 0 \\
 0 & 0 & 1 & 0 & 0 & 0 \\
 0 & 0 & 0 & 1 & 0 & 0 \\
 0 & 1 & 0 & 0 & 0 & 0 \\
 1 & 0 & 0 & 0 & 0 & 0 \\
\end{array}
\right) .
\label{eq:twopoint2}
\end{align}
In a similar way we  compute all the $C_{a b c}$ and $C^{\prime}_{a b c}$ and obtain the same structure constants  
$C^a_{ b c} =\sum_{d=1}^6 (C_{1}^{-1})_{a d} C_{d b c}=\sum_{d=1}^6 (C^{\prime \,-1}_{1})_{a d} C^{\prime}_{d b c}$ which 
reproduce the structure constants of $QK(\mathrm{Gr(2,4)})$ in \cite{MR2772069}:
\begin{equation}
\begin{aligned}
\mathcal{O}_{(1)} \star \mathcal{O}_{(1)}&=\mathcal{O}_{(1,1)}+\mathcal{O}_{(2)}-\mathcal{O}_{(2,1)}, &
\mathcal{O}_{(1)} \star \mathcal{O}_{(1,1)}&=\mathcal{O}_{(2,1)},  \\
\mathcal{O}_{(1)} \star \mathcal{O}_{(2)}&=\mathcal{O}_{(2,1)}, &
\mathcal{O}_{(1)} \star \mathcal{O}_{(2,1)}&=q -q\mathcal{O}_{(1)}+\mathcal{O}_{(2,2)},  \\
\mathcal{O}_{(1)} \star \mathcal{O}_{(2,2)}&=q \mathcal{O}_{(1)}, & 
\mathcal{O}_{(1,1)} \star \mathcal{O}_{(1,1)}&=\mathcal{O}_{(2,2)},  \\
\mathcal{O}_{(1,1)} \star \mathcal{O}_{(2)}&=q , &
 \mathcal{O}_{(1,1)} \star \mathcal{O}_{(2,1)}&=q \mathcal{O}_{(1)},  \\
 \mathcal{O}_{(1,1)} \star \mathcal{O}_{(2,2)}&=q \mathcal{O}_{(2)}, &
  \mathcal{O}_{(2)} \star \mathcal{O}_{(2)}&= \mathcal{O}_{(2,2)},  \\
 \mathcal{O}_{(2)} \star \mathcal{O}_{(2,1)}&=q \mathcal{O}_{(1)}, & \mathcal{O}_{(2)} \star \mathcal{O}_{(2,2)}&=q \mathcal{O}_{(1,1)}, \\
\mathcal{O}_{(2,1)} \star \mathcal{O}_{(2,1)}&=q \mathcal{O}_{(1,1)}+q \mathcal{O}_{(2)}-q \mathcal{O}_{(2,1)}, & \mathcal{O}_{(2,1)} \star \mathcal{O}_{(2,2)}&=q \mathcal{O}_{(2,1)}  ,  \\
\mathcal{O}_{(2,2)} \star \mathcal{O}_{(2,2)}&=q^2 . & 
\end{aligned}
\end{equation}
Here we write $[\mathcal{O}_{\lambda}]$ as $\mathcal{O}_{\lambda}$ to shorten the expression.
We explicitly evaluate the structure constants for $QK_T(\mathrm{Gr}(N,M))$ for small $N,M$ in terms of  supersymmetric localization formula. 
 The values of structure constants for $QK_T(\mathbb{P}^1)$, $QK_T(\mathbb{P}^{2}) \simeq QK_T(\mathrm{Gr(2,3)})$, $QK_T(\mathbb{P}^{3}) \simeq QK_T(\mathrm{Gr(3,4)})$, and  also a non-equivariant case  $QK(\mathrm{Gr(2,5)}) \simeq QK(\mathrm{Gr(3,5)})$ are listed in Appendix \ref{sec:Appstr}.

The correlation function \eqref{eq:residueformula2dS2}
of the 2d A-twisted GLSM on $S^2$ gives
the generating function of genus-zero quasimap invariants on $S^2$.
Similarly, in three dimensions,
the correlation functions 
\begin{align}
\langle \prod_{l=1}^n \widehat{\mathcal{O}}_{\text{3d}.l} \rangle_{S^1 \times S^2} 
&= \sum_{k=0}^{\infty} \frac{((-1)^{(N-1) } q)^k }{N!} \sum_{k_1, \cdots, k_N \ge 0  \atop \sum_{a=1}^N k_a=k}  \sum_{i_1, \cdots, i_N=1}^M \oint_{x_a=z_{i_a}} \prod_{a=1}^N \frac{d x_a}{2\pi {\rm i} x_a}  \left( \prod_{l=1}^n \mathcal{O}_{\text{3d}.l} (x,z)  \right) 
  \nonumber \\
&
\times \left( \prod_{a=1}^N x_{a}^{N k_a - \sum_{b=1}^N k_b } \right) 
 \prod_{1 \le a \neq b \le N} (1-x_a x^{-1}_{b})
\prod_{i=1}^M \prod_{a=1}^N\left( \frac{ 1 }{1-x_a z^{-1}_i  } \right)^{k_a +1}. 
\label{eq:genus0quasi}
\end{align}
of the CS--matter theory on $S^1 \times S^2$
with the choices \eqref{eq:pareid} and \eqref{eq:parameter2}
gives the generating function of indices of quasimap spaces.

For example,
when $N=1$,
the degree $k$ quasimap space is given by $\mathbb{P}^{(k+1)M-1}$,
and \eqref{eq:genus0quasi} can be written as the equivariant 
Euler characteristics;
the expectation value of
$
 \widehat{\mathcal{O}}_{\text{3d}}=\hat{x}^l
$
is given by
\begin{align}
\langle  \widehat{\mathcal{O}}_{\text{3d}} \rangle_{S^1 \times S^2} 
&= \sum_{k=0}^{\infty}  q^k    \oint  \frac{d x}{2\pi {\rm i} x}  x^{l} 
\prod_{i=1}^M \left( \frac{ 1 }{1-x z^{-1}_i  } \right)^{k +1} \nonumber \\
&=\sum_{k=0}^{\infty} q^k \chi_{T} (\mathbb{P}^{(k+1)M-1} ; \mathcal{O} (-l) ),
\end{align}
where $\chi_{T}(\mathbb{P}^{(k+1)M-1} ; \mathcal{O} (-l) ) $ is the $T$-equivariant Euler characteristic
and $x$ is identified with $\mathcal{O}(-1)$.

\subsubsection*{\underline{Seiberg-like duality for CS--matter theories}}
We explain Seiberg-like duality between  $U(N)$ and $U(M-N)$ topologically twisted CS--matter theories with $M$ chiral multiplets.
Let $\{z_i \}_{i=1}^{M}$, (resp. $\{z^{\prime}_{i} \}_{i=1}^{M}$ ) be the flavor Wilson loops in $U(N)$ CS--matter theory, (resp. $U(M-N)$ CS--matter theory). 
When $\beta=-1$, 
\begin{align}
\ominus t^{\prime}=\left(\frac{-t_M}{1- t_M}, \frac{-t_{M-1}}{1-t_{M-1}}, \cdots, \frac{-t_1}{1- t_1} \right).
\label{eq:tprime}
\end{align}
 Then \eqref{eq:levelrankg} and \eqref{eq:npointg} lead to the following equality of 
correlation functions between two  CS--matter theories:
\begin{align}
\langle  \prod_{l=1}^n \mathcal{W}_{\lambda_l} \rangle_{U(N), S^1 \times \Sigma_g} 
=\langle  \prod_{l=1}^n \mathcal{W}_{\lambda^{\prime}_l} \rangle_{U(M-N), S^1 \times \Sigma_g} \, ,
\end{align}
under the identification of flavor Wilson loops $z^{\prime}_i=z_{M-i+1}^{-1}$ for $i=1,\cdots, M$ following from \eqref{eq:tprime}.
$\langle \cdots \rangle_{U(N), S^1 \times \Sigma_g}$ denotes correlation functions of $U(N)$ CS--matter theory on $S^1 \times \Sigma_g$ with the CS levels and parameters 
taken as \eqref{eq:pareid} and \eqref{eq:parameter2}.
$\langle \cdots \rangle_{U(M-N), S^1 \times \Sigma_g}$ denotes correlation functions of $U(M-N)$ CS--matter theory on $S^1 \times \Sigma_g$ with the CS levels and parameters 
taken as
\begin{align}
&\kappa^{(1)} =-N+\frac{M}{2}, \, \, \kappa^{(2)}=-1, \, \, \kappa^{(3)}_{i}=\frac{1}{2} , \nonumber \\
&r=0, \quad n_i=0, \quad  \kappa^{(4)}=\frac{M}{2}, \quad \kappa^{(5)}_i=-\frac{M-N}{2}.
\end{align}

\subsection{CS--matter theory with $\Omega$-background and  $K$-theoretic $I$-function}
\label{sec:CSMsec2}
In the previous section 
we  studied  the correspondence between the algebra of Wilson loops in topologically twisted CS--matter theories without $\Omega$-background and the quantum $K$-ring of Grassmannians.
When $g=0$,  i.e.,  $S^1 \times S^2$,  a fugacity $\mathfrak{q}$ ($\log \mathfrak{q}$ is proportional to  the  $\Omega$-background parameter) for the rotation of $S^2$ 
can be included in the partition function (supersymmetric index) or the correlation functions of supersymmetric Wilson loops. 
The supersymmetric localization computation was performed in \cite{Benini:2015noa}.  In this section we study the geometric aspects of 
topologically twisted CS--matter theory with the $\Omega$-background and relate it to the $K$-theoretic $I$-function of Grassmannians.

For 2d A-twisted GLSM on $S^2$ with the $\Omega$-background parameter ($\Omega$-deformed A-twisted GLSM, also called equivariant A-twist), the supersymmetric localization computation 
 was performed in \cite{Closset:2015rna}.
 The supersymmetric localization formula of 2d $\Omega$-deformed A-twisted GLSMs has a nice mathematical interpretations; it was shown in \cite{Ueda:2016wfa, 2016arXiv160708317K} that the partition function of $\Omega$-deformed A-twisted GLSM on $S^2$ factorizes to a pair of   Givental  $I$-function \cite{MR1354600} for the Grassmannians%
\footnote{For the  Grassmannians,  the $I$-function is same as the $J$-function.}
 \cite{MR2110629}, and also shown in \cite{2016arXiv160708317K} that supersymmetric localization fromula is equivalent to 
the generating function of  integral over the graph spaces of Higgs branch vacuum manifold%
\footnote{This factorization in mathematical literature is known as Givental's double construction lemma.
See, e.g., \cite[Lemma 11.2.12]{MR1677117}.}. 

We will show  a similar result in three dimensions; the partition function of topologically twisted CS--matter theory with $\Omega$-background 
factorizes to a pair of functions,  which are related to the small $K$-theoretic $I$-function of Grassmannian%
\footnote{
The factorization formulas  for 3d A-type linear quiver CS--matter theories were shown in  \cite{Hwang:2017kmk}. 
The physical reason why the factorization occurs is that, by choosing  a $\mathcal{Q}$-exact term appropriately,  point-like vortices and anti-vortices 
exist on two antipodal points of $S^2$ that lead to a pair of  3d vortex partition functions \cite{Fujitsuka:2013fga, Benini:2013yva}. 
The 3d vortex partition function coincides with the specialization of the $K$-theoretic $I$-function for $x_a =z_{i_a}$.}.
We choose  an R-charge and flavor charges  same as before; $r=n_i=0$ . 
On the other hand we take  generic  CS levels $\kappa^{(1)}$, $\kappa^{(2)}$, $\kappa^{(3)}_{i}$, $\kappa^{(4)}$  and $\kappa^{(5)}_{i}$
  in order to compare  physical results 
with  $K$-theoretic $I$-functions with level structures \cite{Ruan:2018v1, Ruan:2018bdg, Wen2019}.  It is convienient to introduce $\tilde{\kappa}^{(1)}, \tilde{\kappa}^{(2)}$, $\tilde{\kappa}^{(3)}_{i}$, $\tilde{\kappa}^{(4)}$ and 
$\tilde{\kappa}^{(5)}_{i}$ to express the factorization
defined by
\begin{align}
&\tilde{\kappa}^{(1)}:= \kappa^{(1)}-\left(N-\frac{M}{2}\right), \quad \tilde{\kappa}^{(2)}:=\kappa^{(2)}+1,  \nonumber \\
&\tilde{\kappa}^{(3)}_i:=\kappa^{(3)}_i-\frac{1}{2}, \quad \tilde{\kappa}^{(4)}:=\kappa^{(4)}-\frac{M}{2}, \quad  \tilde{\kappa}^{(5)}_{i}:=\kappa^{(5)}_{i}+\frac{N}{2}.
\end{align}
$\tilde{\kappa}^{(1)}=\tilde{\kappa}^{(2)}=\tilde{\kappa}^{(3)}_i=\tilde{\kappa}^{(4)}=\tilde{\kappa}^{(5)}_i=0$ corresponds to the $U(N)$ CS--matter theory in which  the Wilson loop algebra is isomorphic to 
the quantum $K$-ring.

In the presence of $\Omega$-background, the one-loop determinants of the vector and chiral multiplets are modified to
\begin{align}
Z^\Omega_{\mathrm{3d.vec}} 
&= \prod_{a > b}^N (-1)^{k_a-k_b} ( (x_a x_b^{-1} \mathfrak{q}^{\frac{k_a-k_b}{2}})^{-\frac{1}{2}} -(x_a x_b^{-1} \mathfrak{q}^{\frac{k_a-k_b}{2}})^{\frac{1}{2}} ) 
\nonumber \\
& \qquad \qquad \qquad \times 
( (x_a^{-1} x_b \mathfrak{q}^{\frac{k_a-k_b}{2}})^{-\frac{1}{2}} -(x_a^{-1} x_b \mathfrak{q}^{\frac{k_a-k_b}{2}})^{\frac{1}{2}} ) \nonumber \\
&=\left(\prod_{a=1}^N x_a^{-N k_a+\sum_{b=1}^N k_b} \right) \prod_{a  \neq b}^N  \frac{\prod_{l=-\infty}^{k_b-k_a}
\left( 1-x^{-1}_a x_b \mathfrak{q}^{-l-\frac{1}{2}(k_a-k_b)}  \right)}
{\prod_{l=-\infty}^{-1} \left( 1-x^{-1}_a x_b \mathfrak{q}^{-l-\frac{1}{2}(k_a-k_b)}  \right) }, 
\label{eq:vecomega}\\
Z^\Omega_{\text{3d.chi}}&=\prod_{i=1}^{M} \prod_{a=1}^N \prod_{j=-\frac{k_a}{2}}^{\frac{k_a}{2} } 
 \frac{ 1 }{ (x_a z^{-1}_i \mathfrak{q}^{j} )^{-\frac{1}{2}}-(x_a z^{-1}_i \mathfrak{q}^{j}  )^{\frac{1}{2} } } \nonumber \\
&=\left(\prod_{a=1}^N x_a^{\frac{M}{2}( k_a+1)} \right) \left( \prod_{i=1}^M z_i^{-\frac{1}{2}( N+\sum_a k_a)} \right) \prod_{i=1}^{M} \prod_{a=1}^N \prod_{j=-\frac{k_a}{2}}^{\frac{k_a}{2} } 
\left( \frac{ 1}{1-x_a z^{-1}_i \mathfrak{q}^{j}} \right) .
\label{eq:chiralomega}
\end{align}
When $\Omega$-background is turned off, i.e.,  $\mathfrak{q}=1$, the one-loop determinants is reduced to the one-loop determinant  without
 $\Omega$-background \eqref{eq:z1loopvec} and \eqref{eq:z1loopvchi3} for $g=0$, $n_i=0$ and $r=0$. 
The CS-terms are independent of $\mathfrak{q}$ and  given by  \eqref{eq:CSterms}  for $g=0$. Then the partition function with the $\Omega$-background $\langle 1 \rangle_{S^1 \times S^2_{\Omega}}
$ is rewritten as 
\begin{align}
\langle 1 \rangle_{S^1 \times S^2_{\Omega}}
&= \sum_{1 \le i_1 < i_2< \cdots < i_N \le M } \sum_{b=1}^N \sum_{k_b=0}^{\infty} \sum_{j^{\prime}_b=-\frac{k_b}{2}}^{\frac{k_b}{2}} q^{\sum_{a=1}^{N} k_a} 
\oint_{x_a=z_{i_a} \mathfrak{q}^{-j^{\prime}_a } } \prod_{a=1}^N \frac{d x_a}{2 \pi {\rm i} x_a} Z^{(0)}_{\text{CS}} Z^\Omega_{\mathrm{3d.vec}} Z^\Omega_{\mathrm{3d.chi}}.
\label{eq:omegaindex}
\end{align}
To show the factorization, we introduce $k^{(1)}_a, k^{(2)}_a$ and $x_a^{\prime}$ for $a=1, \cdots, N$ by
\begin{align}
k_a=k^{(1)}_a+k^{(2)}_a, \quad 2 j^{\prime}_a=k^{(2)}_a-k^{(1)}_a, \quad x_a^{\prime}=x_a \mathfrak{q}^{ j^{\prime}_a} .
\end{align}
Sums in \eqref{eq:omegaindex} are written in terms of $k^{(1)}_a$ and $k^{(2)}_a$ as
\begin{align}
&\sum_{k_b=0}^{\infty} \sum_{j^{\prime}_b=-\frac{k_b}{2}}^{\frac{k_b}{2}}=\sum_{k^{(1)}_b=0}^{\infty} \sum_{k^{(2)}_b=0}^{\infty},  
\end{align}
Then we obtain  the factorization of the partition function \eqref{eq:omegaindex}   given by
\begin{align}
\langle 1 \rangle_{S^1 \times S^2_{\Omega}}
&=\sum_{1 \le i_1 < i_2< \cdots < i_N \le M }    \prod_{i=1}^M z_{i}^{\tilde{\kappa}^{(5)}_i}
\oint_{x_a^{\prime}=z_{i_a}   }   \prod_{a=1}^N \frac{d x_a^{\prime}}{2 \pi {\rm i} x^{\prime 1+\tilde{\kappa}^{(4)}}_a}    
\nonumber \\ & \qquad \times 
\frac{\prod_{a  \neq b}^N (1-x_a^{\prime} x^{\prime -1}_b)}{\prod_{i=1}^{M} \prod_{a=1}^N (1-x_a^{\prime} z^{-1}_i)}
  I(x^{\prime} ,z, q, \mathfrak{q},\tilde{\kappa}) I (x^{\prime} ,z, q, \mathfrak{q}^{-1},\tilde{\kappa})  ,
\label{eq:Factri}
\end{align}
where $I$ is given by
\begin{align}
&I(x ,z, q, \mathfrak{q}, \tilde{\kappa}) \nonumber \\
&=\sum_{c=1}^N \sum_{{k}_c=0}^{\infty} ( (-1)^{N-1}q)^{\sum_{a=1}^N k_a} 
\mathfrak{q}^{\frac{1}{2} \left( \sum_{a=1}^N ({\kappa}^{(1)} k^2_a-  {\kappa}^{(4)} k_a)  + {\kappa}^{(2)}  \sum_{a, b=1}^N k_a k_b \right) } \nonumber \\
& \qquad \times    \prod_{i=1}^M z_i^{\kappa^{(3)}_i \sum_{a=1}^M k_a}  \prod_{a=1}^N x_a^{ {\kappa}^{(1)} k_a+{\kappa}^{(2)} \sum_{b=1}^N k_b}  \nonumber \\
 &  \qquad \times  
\frac{\prod_{a > b}^N  ( (x_a x_b^{-1} \mathfrak{q}^{k_a-k_b})^{-\frac{1}{2}} -(x_a x_b^{-1} \mathfrak{q}^{k_a-k_b})^{\frac{1}{2}} )}
{\prod_{a > b}^N  ( (x_a x_b^{-1})^{-\frac{1}{2}} -(x_a x_b^{-1})^{\frac{1}{2}} )
\prod_{i=1}^{M} \prod_{a=1}^N \prod_{l=1}^{k_a} ( (x_a z^{-1}_i \mathfrak{q}^{l} )^{-\frac{1}{2}}-( x_a z^{-1}_i \mathfrak{q}^{l})^{\frac{1}{2}} )} 
\label{eq:Ifunc1} \\
&=\sum_{c=1}^N \sum_{{k}_c=0}^{\infty}  q^{\sum_{a=1}^N k_a} 
\mathfrak{q}^{\frac{1}{2} \left( \sum_{a=1}^N (\tilde{\kappa}^{(1)} k^2_a-  \tilde{\kappa}^{(4)} k_a)  + \tilde{\kappa}^{(2)}  \sum_{a, b=1}^N k_a k_b \right) } 
\prod_{i=1}^M z_i^{\tilde{\kappa}^{(3)}_i \sum_{a=1}^M k_a }
\nonumber \\ &   \quad \times
 \prod_{a=1}^N x_a^{ \tilde{\kappa}^{(1)} k_a+\tilde{\kappa}^{(2)} \sum_{b=1}^N k_b} 
\prod_{a  \neq b}^N  \frac{  \prod_{l=-\infty}^{k_a-k_b} (1-x_a x^{-1}_b \mathfrak{q}^{l} ) }
{\prod_{l=-\infty}^{0} (1-x_a x^{-1}_b \mathfrak{q}^{l} )   } 
\frac{1}{\prod_{i=1}^{M} \prod_{a=1}^N \prod_{l=1}^{k_a} ( 1-x_a z^{-1}_i \mathfrak{q}^{l} )} .
\label{eq:KIfunction}
\end{align}
We obtained two equivalent expressions  \eqref{eq:Ifunc1} and \eqref{eq:KIfunction} of $I$ associated to two expressions  of  the one-loop determinants \eqref{eq:vecomega} and \eqref{eq:chiralomega}.
If $\tilde{\kappa}^{(1)}$ and $\tilde{\kappa}^{(2)}$  are integers,  all the square roots of $x_a$'s in $I$ are canceled out.

If  CS levels are chosen to  reproduce quantum $K$-ring,  i.e. $\tilde{\kappa}^{(1)}=\tilde{\kappa}^{(2)}= \tilde{\kappa}^{(3)}_i= \tilde{\kappa}^{(4)}=0$,   the function $I$ becomes 
\begin{align}
I_{\mathrm{Gr}(N,M)} & (x ,z, q, \mathfrak{q}):=I(x ,z, q, \mathfrak{q},\tilde{\kappa}) |_{\tilde{\kappa}^{(1)}=\tilde{\kappa}^{(2)}= \tilde{\kappa}^{(3)}_i= \tilde{\kappa}^{(4)}=0} 
\nonumber \\
&  =\sum_{c=1}^N \sum_{{k}_c=0}^{\infty}  q^{\sum_{a=1}^N k_a} 
\prod_{a  \neq b}^N  \frac{  \prod_{l=-\infty}^{k_a-k_b} (1-x_a x^{-1}_b \mathfrak{q}^{l} ) }
{\prod_{l=-\infty}^{0} (1-x_a x^{-1}_b \mathfrak{q}^{l} )   } 
\frac{1}{\prod_{i=1}^{M} \prod_{a=1}^N \prod_{l=1}^{k_a} ( 1-x_a z^{-1}_i \mathfrak{q}^{l} )} .
\label{eq:KIfunction2}
\end{align}
When   $x^{-1}_a, (\text{resp. } x_a)$ for $a=1, \cdots N$ is identified with  the  $K$-theoretic Chern roots of  universal subbundle $S$ (resp. it dual $S^{\vee}$)  of the Grassmannian 
$\mathrm{Gr}(N,M)$, we find that
 $I_{\mathrm{Gr}(N,M)}$ reproduces the  $K$-theoretic $I$-function of  $\mathrm{Gr}(N,M)$  up to a overall factor $(1-\mathfrak{q})$
\footnote{We thank T.~Milanov for telling us the typos of $K$-theoretic $J$-functions of \cite{Taipale:2011pm} and the correct expression.}.
When  $\tilde{\kappa}^{(1)}=-\tilde{\kappa}^{(4)}$ and $\tilde{\kappa}^{(2)}, \tilde{\kappa}^{(3)}_i=0$, \eqref{eq:KIfunction}  reproduces $K$-theoretic $I$-function with level structures.
For example, when $N=1$,  \eqref{eq:KIfunction} is written as
\begin{align}
I(x ,z_i, q, \mathfrak{q})
&= \sum_{{k}=0}^{\infty}         \frac{ q^{k} \mathfrak{q}^{\tilde{\kappa}^{(1)}\frac{k(k+1)}{2}} }
{\prod_{i=1}^{M} \prod_{l=1}^{k} ( 1-x z^{-1}_i \mathfrak{q}^{l} )  } .
\end{align}
This  agrees with the equivariant $K$-theoretic $I$-function of $\mathbb{P}^{M-1}$ with level structures \cite{Ruan:2018v1}. 
%

We comment on  generalizations of  the factorization formula \eqref{eq:Factri} to  the correlation functions with the $\Omega$-background.
 In the presence of the $\Omega$-background, the operators can be inserted only on the 
fixed points of $\Omega$-rotation of $S^2$ which we call the north and south poles of $S^2$, respectively. If we insert a Wilson loop $\widehat{\mathcal{O}}_{3d }$, $(\widehat{\mathcal{O}}^{\prime}_{3d })$ on the north (south) pole, we obtain 
the following expressions 
\begin{align}
\langle \widehat{\mathcal{O}}_{3d } |_{\mathrm{N}} \widehat{\mathcal{O}}^{\prime}_{3d } |_{\mathrm{S}} \rangle_{S^1 \times S^2_{\Omega}}
&= \!\!\!
\sum_{1 \le i_1 < i_2< \cdots < i_N \le M }  \prod_{i=1}^M z_{i}^{\tilde{\kappa}^{(5)}_i}  
\oint_{x_a=z_{i_a}   }  \prod_{a=1}^N \frac{d x_a}{2 \pi {\rm i} x^{1+\tilde{\kappa}^{(4)}}_a}   
\nonumber \\ & \quad \times 
\frac{\prod_{a  \neq b}^N (1-x_a x^{ -1}_b)}{\prod_{i=1}^{M} \prod_{a=1}^N (1-x_a z^{-1}_i)}
  I(x ,z, q, \mathfrak{q},\tilde{\kappa} ; \mathcal{O}_{3d}) I (x ,z, q, \mathfrak{q}^{-1},\tilde{\kappa} ; \mathcal{O}^{\prime}_{3d }) 
,
\end{align}
with
\begin{align}
&I(x ,z, q, \mathfrak{q}, \tilde{\kappa}; \mathcal{O}_{3d}) \nonumber \\
 &:=\sum_{c=1}^N \sum_{{k}_c=0}^{\infty}  q^{\sum_{a=1}^N k_a} \mathcal{O}_{3d } (x \mathfrak{q}^{-k},z)
\mathfrak{q}^{\frac{1}{2} \left( \sum_{a=1}^N (\tilde{\kappa}^{(1)} k^2_a-  \tilde{\kappa}^{(4)} k_a)  + \tilde{\kappa}^{(2)}  \sum_{a, b=1}^N k_a k_b \right) } 
\prod_{i=1}^M z_i^{\tilde{\kappa}^{(3)}_i \sum_{a=1}^M k_a } 
\nonumber \\ &  \quad \times  \prod_{a=1}^N x_a^{ \tilde{\kappa}^{(1)} k_a+\tilde{\kappa}^{(2)} \sum_{b=1}^N k_b} 
\prod_{a  \neq b}^N  \frac{  \prod_{l=-\infty}^{k_a-k_b} (1-x_a x^{-1}_b \mathfrak{q}^{l} ) }
{\prod_{l=-\infty}^{0} (1-x_a x^{-1}_b \mathfrak{q}^{l} )   } 
\frac{1}{\prod_{i=1}^{M} \prod_{a=1}^N \prod_{l=1}^{k_a} ( 1-x_a z^{-1}_i \mathfrak{q}^{l} )} .
\end{align}
Here $|_{\mathrm{N}/\mathrm{S}}$ denotes the operator insertion on the north/south pole, respectively. 
$\mathcal{O}_{3d } (x,z)$, $ (\text{resp. } \mathcal{O}^{\prime}_{3d }  (x,z))$ is the saddle point value of $\widehat{\mathcal{O}}_{3d }$, $ (\text{resp. } \widehat{\mathcal{O}}^{\prime}_{3d })$ without $\Omega$-background. $x \mathfrak{q}^{k}$ expresses $(x_1 \mathfrak{q}^{k_1}, \cdots, x_N \mathfrak{q}^{k_N})$.

\subsection{Comparison with   $K$-theoretic $I$-function from supersymmetric index on $S^1 \times D^2$}
\label{sec:S1xD2}
In \cite{Yoshida:2014ssa}%
\footnote{  
The supersymmetric index in \cite{Yoshida:2014ssa} is different from  the  supersymmetric index (holomorphic block) studied in \cite{Beem:2012mb}. For examples, the vector multiplet contribution
 in these two indices are different. 
And also the definitions of the two indices themselves are different; the index in \cite{Yoshida:2014ssa}   is defined 
by the trace with the insertion of $(-1)^{F}$, where  $F$ is  the fermion number. On the other hand,
 the holomorphic block in \cite{Beem:2012mb} is defined by the trace with the insertion of  $(-1)^R$, where $R$ is  the R-charge. 
Therefore the chiral multiplet contributions are  also different in general. 
The supersymmetric index in \cite{Yoshida:2014ssa} for 3d $\mathcal{N}=4$ gauge theory with a $\mathcal{N}=(2,2)$ boundary condition reproduces vertex functions ($z$-solutions) in \cite{Aganagic:2016jmx} with appropriate identification of global   symmetry parameters .
}%
, a supersymmetric index   $Z_{S^1 \times D^2}$ for 3d $\mathcal{N} \ge 2$ CS--matter theories on $S^1 \times D^2$ was defined and  also 
evaluated in terms of supersymmetric localization techniques. When the Higss branch is $\mathbb{P}^{M-1}$, it was found in \cite{Yoshida:2014ssa} that $Z_{S^1 \times D^2}$ factorizes to products of   
$K$-theoretic $I$-function of $\mathbb{P}^{M-1}$ and  $q$-pochhammer symbol which is regarded as $q$-deformation of Gamma class  
\footnote{Properties of $I$-functions of $\mathbb{P}^{M-1}$ evaluated by $Z_{S^1 \times D^2}$ were studied in \cite{Jockers:2018sfl}. }.
We expect a similar story also holds for $\mathrm{Gr}(N,M)$ 
and study a relation between $Z_{S^1 \times D^2}$ and $I_{\mathrm{Gr}(N,M)}$. 

The supersymmetric localization formula of $Z_{S^1 \times D^2}$  for the CS--matter theory treated  in section \ref{sec:CSMsub1} is written as
\begin{align}
Z_{S^1 \times D^2}(z,q,\mathfrak{q} )
= \frac{1}{N!} \oint \prod_{a=1}^N \frac{d x_a}{2 \pi {\rm i} x_a} Z_{\text{CS}}(x, q,\mathfrak{q}) 
Z_{\text{FI}}(x, q,\mathfrak{q}) Z_{\text{vec}}(x, \mathfrak{q}) 
Z_{\text{chi}. N}(x, \mathfrak{q}) .
\end{align}
Here the contributions from the CS-terms, the FI-term, and the one-loop determinants of vector and chiral multiplets  on $S^1 \times D^2$ are given by
\begin{align}
Z_{\text{CS}}(x, q,\mathfrak{q})  
&=\exp \left( \frac{-1}{2 \log (\mathfrak{q} )} \left[  \sum_{a=1}^N  \right. \right.
 \kappa^{(1)}(\log x_a)^2   + \kappa^{(2)}  \left(\sum_{a=1}^N \log x_a \right)^2 \nonumber \\
& \qquad \qquad +  2 \sum_{i=1}^M \sum_{a=1}^N   \kappa^{(3)}_i \log (x_a) \log (z_i) 
+ \kappa^{(4)}  \log (\mathfrak{q})  \sum_{a=1}^N \log (x_a)  \nonumber \\
& \left. \left.  \qquad  \qquad  + \log (\mathfrak{q}) \sum_{i=1}^M \kappa^{(5)}_i \log (z_i)  +\sum_{i,j=1}^M \kappa^{(6)}_{i j} \log (z_i) \log (z_j) \right] \right), \\
Z_{\text{FI}}(x, q, \mathfrak{q})&=\exp \left(- \sum_{a=1}^N \frac{\log (q) \log (x_a)} {\log (\mathfrak{q})}    \right), \\
Z_{\text{vec}}(x, \mathfrak{q})&=     e^{\frac{ \sum_{a \neq b}^N  (\log (x_a x^{-1}_{b}))^2}{4 \log (\mathfrak{q}) } }  \prod_{a \neq b}^N (x_a x^{-1}_b ;\mathfrak{q})_{\infty} ,
\\
Z_{\text{chi}.N}(x, z, \mathfrak{q})&= \prod_{i=1}^M \prod_{a=1}^N   e^{{E }( \log (x_a z_i^{-1}) ) } (x_a z^{-1}_i ;\mathfrak{q})_{\infty}^{-1} .
\label{eq:oneloopN}
\end{align}
Here  $(a;\mathfrak{q})_{\infty}:= \prod_{i=0}^{\infty }(1-a \mathfrak{q}^i)$, 
\footnote{$\mathfrak{q}$ and $\log(q)/\log(\mathfrak{q})$ in this section correspond to $q^2 $ and $2 \pi r \zeta$ in \cite{Yoshida:2014ssa}, respectively. }
${ E}(a)$ is defined by
\begin{align}
{ E}(a):=-\frac{\log (\mathfrak{q})}{24} -\frac{1}{4} a - \frac{1}{4 \log (\mathfrak{q})} a^2.
\end{align}
There are two possible BPS boundary conditions for each chiral multiplet.  \eqref{eq:oneloopN} is 
the one-loop determinant of chiral multiplets with the Neumann boundary condition.

In the presence of the boundary, CS-terms are not invariant under gauge transformations in general. 
It was observed in \cite{Yoshida:2014ssa} that 
the cancellation of    the logarithmic terms arising from the CS-terms and the ones from the one-loop determinants 
are correlated with  the gauge anomaly cancellation. 
If we choose the gauge and gauge-flavor and gauge-R-symmetry mixed CS levels  as \eqref{eq:pareid} and \eqref{eq:parameter2},  
all the  $\log (x_a)$  and $\log (x_a) \log (x_b)$ terms in $Z_{\text{CS}}$, $Z_{\text{vec}}$ and $Z_{\text{chi}.N}$ are canceled out. 
Moreover, if we choose $\kappa^{(6)}_{i j}=-\frac{N}{2} \delta_{i j}$, $\log(z_i) \log(z_j)$ terms in $Z_{S^1 \times D^2}$ also vanish, which is the signal of the zero effective flavor CS levels. 
Up to overall $\mathfrak{q}$-factors,  
$Z_{S^1 \times D^2}(z,q,\mathfrak{q} )$ can be written as
\begin{align}
Z_{S^1 \times D^2}(z,q,\mathfrak{q} )
=  \frac{1}{N!} 
\oint \prod_{a=1}^N \frac{d x_a}{2 \pi {\rm i} x_a} e^{ -\sum_{a=1}^N \frac{\log(q) \log(x_a)}{\log (\mathfrak{q})}  }
\prod_{a \neq b}^N (x_a x_b^{-1};\mathfrak{q})_{\infty} \prod_{i=1}^M \prod_{a=1}^N \frac{1}{(x_a z_{i}^{-1}; \mathfrak{q})_{\infty}} .
\end{align}
The residues are evaluated at $x_a=z_{i_a} \mathfrak{q}^{-k_a} $ with $i_a=1, \cdots, M$ and $k_a \in\mathbb{N}_{\ge 0}$. 
We obtain the following expression of the supersymmetric index ond $S^1 \times D^2$;
\begin{align}
Z_{S^1 \times D^2}(z,q,\mathfrak{q} )&=
\frac{1}{N!} \sum_{c=1}^N  \sum_{i_c =1}^M \sum_{k_c =0}^{\infty}
\oint_{x_a=z_{i_a}} \prod_{a=1}^N \frac{d x_a}{2 \pi {\rm i} x_a} e^{ -\sum_{a=1}^N \frac{\log(q) \log(x_a)}{\log (\mathfrak{q})}  } q^{\sum_{a=1}^N k_a} \nonumber \\
& \qquad \qquad \times \prod_{a \neq b}^N (x_a x_b^{-1} \mathfrak{q}^{k_b-k_a} ;\mathfrak{q})_{\infty} \cdot \prod_{i=1}^M \prod_{a=1}^N \frac{1}{(x_a \mathfrak{q}^{-k_a} z_{i}^{-1}; \mathfrak{q})_{\infty}} \nonumber \\
&= \frac{1}{N!} \sum_{i_1,\cdots, i_N=1}^M
\oint_{x_a=z_{i_a}} \prod_{a=1}^N \frac{d x_a}{2 \pi {\rm i} x_a} e^{ -\sum_{a=1}^N \frac{\log(q) \log(x_a)}{\log (\mathfrak{q})}  }
 \nonumber \\
& 
\qquad \times \prod_{a \neq b}^N (x_a x_b^{-1};\mathfrak{q})_{\infty} \cdot \prod_{i=1}^M \prod_{a=1}^N \frac{1}{(x_a z_{i}^{-1}; \mathfrak{q})_{\infty}}  \cdot I_{\mathrm{Gr}(N,M)}(x ,z, q, \mathfrak{q}^{-1}).
\end{align}
Therefore we find the $K$-theoretic $I$-function of the Grassmannians can be derived from the supersymmetric index on $S^1 \times D^2$.
In  other choices of CS levels, we have to introduce the elliptic genus of 2d $\mathcal{N}=(0,2)$  multiplets on the boundary of $S^1 \times D^2$ to cancel the gauge anomaly  by anomaly inflow mechanism.

\section{Levels in CS--matter theory and deformations of $U(N)$ Verlinde algebra}
\label{sec:Verlinde}
The correlation functions of Wilson loops in $G$ Chern--Simons theory on $S^1 \times \Sigma_g$,
described by
the Verlinde formula \cite{Verlinde:1988sn, Witten:1988hf},
are indices of line bundles
on moduli stacks of $G_{\mathbb{C}}$-bundles
on $\Sigma_g$.
Similarly,
we expect that
correlation functions in CS--matter theories
are indices of vector bundles on moduli stacks
of $G_{\mathbb{C}}$-bundles with sections.
In this section,
we briefly discuss choices of CS levels in
$U(N)$ CS--matter theories with $M$ fundamental chiral multiplets
in connection with variants of Verlinde algebra. 
Precise relations  between CS--matter theories
and indices on moduli stacks
will be studied elsewhere.

\subsection{CS--matter theory and quantum cohomology of Grassmannian}
First we study  the connection between  $U(N)$ CS--matter theories with $M$ fundamental chiral multiplets and a deformation of Verlinde algebra, which is 
 isomorphic to   quantum cohomology of Grassmannian. 
It was shown in \cite{Witten:1993xi} that Verlinde algebra of $U(N)$ Chern--Simons theory or equivalently $U(N)
/U(N)$ gauged WZW model  with the level 
$\kappa=M-N$ have
 the same structure constants of quantum cohomology of Grassmannian $\mathrm{Gr}(N,M)$ or equivalently A-twisted GLSM  treated in section \ref{sec:GLSM},
 by setting  quantum parameter $q=1$. 
But these two algebras have different Frobenius bilinear forms.  
If we choose the parameters of CS--matter theory as
\begin{align}
&r=0, \quad n_i=0, \quad \kappa^{(1)}=-\frac{M}{2}, \quad \kappa^{(2)}=0,  \nonumber \\
&\kappa^{(3)}_{i}=-\frac{1}{2}, \quad \kappa^{(4)}=\frac{M}{2}-N, \quad \kappa^{(5)}_i=\frac{N}{2}  .
\label{eq:parameter22}
\end{align}
The correlation function \eqref{eq:CSresum2}  of CS--matter theory on $S^1 \times \Sigma_g$ is written as
\begin{align}
\langle \prod_{l} \widehat{\mathcal{O}}_{\mathrm{3d}, l}  \rangle_{S^1 \times  \Sigma_g}
&=
\sum_{x } \left( \prod_{l=1}^n \mathcal{O}_{\mathrm{3d}, l}(x,z_i) \right)  
\left( \frac{ \prod_{a >b}^N ( x_a-x_b)^2 }{\prod_{i=1}^M\prod_{a=1}^N ( x_a-z_i )}
 \left[ \prod_{a=1}^N  \left(\sum_{i=1}^M \frac{1}{x_a-z_i} \right) \right]^{-1}\right)^{1-g}
\end{align}
where the sum is taken over the disctinct roots of   the 3d saddle point equations 
\begin{align}
1=\exp \left(   \frac{\partial W_{\text{3d.eff}}}{\partial \log x_a} \right) \Leftrightarrow \prod_{i=1}^M (z_i-x_a)=(-1)^{N-1} q.
\end{align}
We find that correlation functions  in the CS--matter theory on $S^1 \times \Sigma_g$ agree with correlation functions in A-twisted GLSM (up to overall sign) on $\Sigma_g$ with $r=0$;
\begin{align}
\langle \prod_{l} \widehat{\mathcal{O}}_{\mathrm{3d}, l}  \rangle_{S^1 \times  \Sigma_g}=
\langle \prod_{l} \widehat{\mathcal{O}}_{\mathrm{2d}, l}  \rangle_{ \Sigma_g} \Big|_{r=0} ,
\end{align}
if the saddle point values of 2d and 3d operators are same ${\mathcal{O}}_{\mathrm{3d}, l}={\mathcal{O}}_{\mathrm{2d}, l} $ under the following identification of variables 
\begin{align}
\sigma_a \equiv 1-x_a, \quad m_i \equiv 1-z_i.
\end{align}
Therefore we obtain the CS--matter theory in which the algebra of Wilson loop  has the same structure constant of $U(N)$ Verlinde algebra with level $M-N$,
but a different Frobenius bilinear form.

\subsection{CS--matter theory and  Telemann--Woodward index}

For a line bundle $L$ on a Riemann surface $\Sigma$,
an admissible line bundle $\mathcal{L}$
on the moduli stack $\mathcal{M}$ of $G_{\mathbb{C}}$-bundle on $\Sigma$,
and representations $V_1, \ldots, V_M, U$ of $G$,
the index formula of Telemen--Woodward  \cite{MR2552100} gives
\begin{align}
&\mathrm{Ind}\left(\mathcal{M}; \mathcal{L}  \otimes \mathcal{E}  \otimes E^*_x U \right) 
\nonumber  \\
&\qquad =\sum_{f \in F^{\mathrm{reg}}/W} \theta_{\vec{t}} (f_{\vec{t}})^{1-g} \left(\prod_{i=1}^M \prod_{\rho} \left(1- t_i e^{\rho}(f_{\vec{t}}) \right)^{g-1-\mathrm{deg}(L)} \right) \mathrm{Tr}_U f_{\vec{t}}
\label{eq:TW}
\end{align}
for
\begin{align}
 \mathcal{E} = \otimes_{i=1}^M \exp \left[ -\sum_{p >0}\frac{t^p_i}{p^2} E_{\Sigma}(\psi^p  V_i) \right]  
\otimes_{i=1}^M (E^*_x \lambda_{t_i} V_i)^{\otimes (1-g+\mathrm{deg}(L))},
\end{align}
where $\psi^p$ is the $p$-th Adams operation
and $\rho$ runs over the set of weights of $V_i$.
It was shown in \cite{Halpern-Leistner:2016uay, Andersen:2016hoj}
that 
the index \eqref{eq:TW} 
for $M=1$ and $V=\mathfrak{g}_{\mathbb{C}}$
agrees with the correlation functions of Wilson loops in $G$ CS--matter theory with 
a chiral multiplet in the adjoint representation and R-charge $r=\deg(L)$.

When $G=U(N)$ and
$V_i$ is the fundamental representation of $G$ for all $i$,
we have
\begin{align}
 \theta_{\vec{t}} (f_{\vec{t}})&= \frac{\prod_{a \neq b}^N ( 1-e^{\xi_a -\xi_b } )}{N! \det H}
\label{eq:insertion}
\end{align}
where $\det H$ is the Hessian of
\begin{align}
 \frac{l_1+N}{2} \sum_{a=1}^N \xi^2_a
+\frac{l_2}{2} \left( \sum_{a=1}^N \xi_a \right)^2+ \sum_{i=1}^M \sum_{a=1}^N \mathrm{Li}_2 (e^{\xi_a} t_i )+ \pi {\rm i} (N-1) \sum_{a=1}^N \xi_a
\label{eq:Hss}
\end{align}
with respect to $\xi_a$
and $(l_1, l_2)$ is the level of $\mathcal{L}$.
The sum in \eqref{eq:TW} runs over the saddle points of \eqref{eq:Hss}.
If one has $q=1$,  $\kappa^{(3)}_1=\kappa^{(3)}_2 =\cdots =\kappa^{(3)}_M$, and $\prod_{i=1}^M z_i=1$,
then \eqref{eq:Hss} matches the 3d twisted superpotential
by setting $e^{\xi_a} = x_a$,  $t_i = z^{-1}_i$,
$l_1+N = \kappa^{(1)}+\frac{M}{2}$, and $l_2 = \kappa^{(2)}$,
and the index \eqref{eq:TW} agrees
with the genus $g$ correlation function of the CS--matter theory.

\section*{Acknowledgements}
We are grateful to Todor~Milanov and Yaoxiong~Wen for sending us their notes on $K$-thereotic $I$-function of Grassmannians and $I$-function with level structures.
We also thank the anonymous referee for suggestions for improvements.
KU is supported by JSPS KAKENHI Grant Numbers
15KT0105,
16H03930, and
16K13743.
YY  is supported  by JSPS KAKENHI Grant Number JP16H06335 and also by World Premier International Research Center Initiative (WPI), MEXT Japan.

\appendix
\section{Definitions of symbols}
\label{appendix1}
We give the definitions of symbols used in  section \ref{sec:section2}.
\begin{align}
x \oplus y&:= x+y + \beta x y,  \\
x \ominus y&:= \frac{x - y}{1+ \beta  y}, \label{eq:ominus}
\\
(x_a | t)^r &:= \prod_{i=1}^r x_a \oplus t_i, \\
\Pi (x)& := \prod_{a=1}^N (1+\beta x_a),  \\
\Pi (t_{\lambda})& :=\prod_{a=1}^N (1+\beta t_{\lambda_a+N+1-a}),  \\
\ominus t&:=(0\ominus t_{1}, 0\ominus t_{2}, \cdots, 0\ominus t_{M-1}, 0 \ominus t_M ) \nonumber \\
  &= \left(\frac{-t_1}{1+ \beta t_1}, \frac{-t_2}{1+ \beta t_2}, \cdots, \frac{-t_M}{1+ \beta t_M} \right),
\label{eq:ominust} \\
\ominus t^{\prime}:&=(0\ominus t_{M}, 0\ominus t_{M-1}, \cdots, 0\ominus t_2, 0 \ominus t_1 ). 
\label{eq:ominustprime} 
\end{align}

\section{The components of RTT relation}
\label{sec:RTT}
From the RTT relation, the  ${\sf A, \, B, \, C, \, D}$ in monodromy matrix ${\sf M}$ satisfy  the following  sixteen relations;
\begin{align}
&  X(x|t)X(y|t)=X(y|t)X(x|t), \quad X={\sf A, \, B, \, C, \, D}, \\
&  {\sf C}(x|t) {\sf A}(y|t)={\sf C}(y|t) {\sf A}(x|t), \\
&  {\sf D}(x|t) {\sf B}(y|t)={\sf D}(y|t) {\sf B}(x|t), \\
&  {\sf A}(x|t) {\sf B}(y|t)=(1+ \beta y \ominus x) {\sf A}(y|t) {\sf B}(x|t), \\
&  {\sf C} (x|t) {\sf B}(y|t)=(1+ \beta y \ominus x) {\sf C}(y|t) {\sf B}(x|t), \\
&  {\sf C}(x|t) {\sf D}(y|t)=(1+ \beta y \ominus x) {\sf C}(y|t) {\sf D}(x|t), \\
&  ( x \ominus y) {\sf A}(x|t) {\sf B}(y|t)={\sf B}(y|t) {\sf A}(x|t)-{\sf B}(x|t) {\sf A}(y|t), 
\label{eq:BA} \\
&  (x \ominus y) {\sf C}(x|t) {\sf B}(y|t) 
={\sf A}(x|t) {\sf D}(y|t)-{\sf A}(y|t) {\sf D}(x|t),
\label{eq:CB} \\
&  ( y \ominus x) {\sf C}(y|t) {\sf B}(x|t)={\sf D}(x|t) {\sf A}(y|t)- {\sf D}(y|t){\sf A}(x|t), \\
&(y \ominus x) {\sf C}(y|t) {\sf D}(x|t)={\sf D}(x|t) {\sf C}(y|t)-{\sf D}(y|t) {\sf C}(x|t),
 \\
& (y \ominus x)  {\sf C}(x|t) {\sf A}(y|t)={\sf A}(y|t) {\sf C}(x|t)-{\sf A}(x|t) {\sf C}(y|t),  \\
&  ( y \ominus x) {\sf D}(x|t) {\sf A}(y|t)+(1+ \beta y \ominus x) {\sf B}(x|t) {\sf C}(y|t) \nonumber \\
& \qquad \qquad \qquad ={\sf B}(y|t) {\sf C}(x|t)+(y \ominus x){\sf A}(y|t) {\sf D}(x|t), \\
&  (y \ominus x) {\sf D}(x|t) {\sf B}(y|t)+(1+ \beta y \ominus x) {\sf B}(x|t) {\sf D}(y|t)
={\sf B}(y|t) {\sf D}(x|t). \label{eq:DB} 
\end{align}

\section{Examples of quantum $K$-ring of Grassmannian}
\label{sec:Appstr}
We list  the quantum products of  quantum $K$-ring $QK_{T}(\mathrm{Gr}(N,M))$ for small $N,M$ evaluated   in terms of 
supersymmetric localization formula of three point correlation functions of the Wilson loops $\{ \mathcal{W}_{\lambda} \}_{\lambda\in \mathcal{P}_{N, M}}$ on $S^1 \times S^2$.
A Wilson loop $\mathcal{W}_{\lambda}$ is identified with $K$-theory class of a structure sheaf $\mathcal{O}_{\lambda}$. To shorten  expressions, we write the $K$-theory class $[\mathcal{O}_{\lambda}]$ as $\mathcal{O}_{\lambda}$ and $\mathcal{O}_{\emptyset}=\mathcal{O}=1$.
The isomorphism $QK_{T}(\mathrm{Gr}(N,M)) \simeq QK_{T}(\mathrm{Gr}(M-N,M))$ is given by $\mathcal{O}_{\lambda} \mapsto \mathcal{O}_{\lambda^{\prime}}$ and  $z_i \mapsto z^{-1}_{M-i+1}$ for $i=1,\cdots, M$, where $\lambda^{\prime} \in \mathcal{P}_{M-N,M}$ is the transpose of $\lambda \in \mathcal{P}_{N,M}$ and $z_i$'s are the flavor Wilson loops identified with the equivariant parameters $z_{i} \equiv e^{\varepsilon_{M-i+1}}$.
We list quantum the $K$-ring of $\mathrm{Gr}(N,M) $ with $N \le M-N$.

\subsubsection*{\underline{$QK_T ( \mathbb{P}^1)$}}
The structure sheaves of the Schurbert varieties of $\mathbb{P}^1$:$\{1, \mathcal{O}_{(1)} \} $.
\begin{align}
\mathcal{O}_{(1)} \star \mathcal{O}_{(1)}&=q z_2 z_1^{-1} +\left(1-z_2 z_1^{-1} \right)\mathcal{O}_{(1)}.
\end{align}

\subsubsection*{\underline{$QK_T ( \mathbb{P}^2) \simeq QK_T ( \mathrm{Gr}(2,3))$}}
The structure sheaves of the Schurbert varieties of $\mathbb{P}^2$:$\{1, \mathcal{O}_{(1)}, \mathcal{O}_{(2)} \}$.
\begin{equation}
\begin{aligned}
\mathcal{O}_{(1)} \star \mathcal{O}_{(1)}&=\left(1-z_2 z_1^{-1} \right)\mathcal{O}_{(1)}+ z_2 z_1^{-1}\mathcal{O}_{(2)}, \\
\mathcal{O}_{(1)} \star \mathcal{O}_{(2)}&=q z_3  z_1^{-1}+\left(1-z_3 z_1^{-1} \right)\mathcal{O}_{(2)}, \\
\mathcal{O}_{(2)} \star \mathcal{O}_{(2)}&=q (1-z_3 z_2^{-1}) z_3 z_1^{-1}+q z_3 z_2^{-1}\mathcal{O}_{(1)} +(1-z_3 z_1^{-1}) (1-z_3 z_2^{-1})\mathcal{O}_{(2)} .
\end{aligned}
\end{equation}

\subsubsection*{\underline{$QK_T ( \mathbb{P}^3) \simeq QK_T ( \mathrm{Gr}(3,4))$}}
The structure sheaves of the Schurbert varieties of $\mathbb{P}^3$:
$\{1, \mathcal{O}_{(1)}, \mathcal{O}_{(2)}, \mathcal{O}_{(3)} \}$.
\begin{equation}
\begin{aligned}
\mathcal{O}_{(1)} \star \mathcal{O}_{(1)}&=\left(1-z_2 z_1^{-1} \right)\mathcal{O}_{(1)}+ z_2 z_1^{-1}\mathcal{O}_{(2)}, \\
\mathcal{O}_{(1)} \star \mathcal{O}_{(2)}&=\left(1-z_3 z_1^{-1} \right)\mathcal{O}_{(2)}+ z_3  z_1^{-1}\mathcal{O}_{(3)}, \\
\mathcal{O}_{(1)} \star \mathcal{O}_{(3)}&= q z_4  z_1^{-1}+\left(1-z_4 z_1^{-1} \right)\mathcal{O}_{(3)}, \\
\mathcal{O}_{(2)} \star \mathcal{O}_{(2)}&=q  z_3 z_4 (z_1 z_2)^{-1} +(1-z_3 z_1^{-1}) (1-z_3 z_2^{-1})\mathcal{O}_{(2)} + (z_1 z_2)^{-1} z_3 (z_1+ z_2-  z_3-  z_4) \mathcal{O}_{(3)}, \\
\mathcal{O}_{(2)} \star \mathcal{O}_{(3)}&=q  (1-z_4 z_2^{-1}) z_4 z_1^{-1}+ q z_4 z_2^{-1}\mathcal{O}_{(1)} + (1-z_4 z_1^{-1}) (1-z_4 z_2^{-1}) \mathcal{O}_{(3)}, \\
\mathcal{O}_{(3)} \star \mathcal{O}_{(3)}&=q  (1-z_4 z_2^{-1})(1-z_4 z_3^{-1}) z_4 z_1^{-1}+ q (1 -z_4 z_3^{-1}) z_4 z_2^{-1}\mathcal{O}_{(1)} 
 \\
& \qquad 
 +q z_4 z^{-1}_3 \mathcal{O}_{(2)}+ (1-z_4 z_1^{-1}) (1-z_4 z_2^{-1}) (1-z_4 z_3^{-1}) \mathcal{O}_{(3)}.
\end{aligned}
\end{equation}

\subsubsection*{\underline{$QK ( \mathrm{Gr}(2,5)) \simeq QK ( \mathrm{Gr}(3,5))$}}
The structure sheaves of the Schurbert varieties of $\mathrm{Gr}(2,5)$:$\{ \mathcal{O}_{\lambda}\}_{\lambda \in \mathcal{P}_{2,5} } $.
\begin{align*}
\mathcal{O}_{(1)} \star \mathcal{O}_{(1)}&=\mathcal{O}_{(1,1)}+\mathcal{O}_{(2)}-\mathcal{O}_{(2,1)}, &        \mathcal{O}_{(1)} \star \mathcal{O}_{(1,1)}&=\mathcal{O}_{(2,1)}, \\
\mathcal{O}_{(1)} \star \mathcal{O}_{(2)}&=\mathcal{O}_{(2,1)}+\mathcal{O}_{(3)}-\mathcal{O}_{(3,1)}, &    \mathcal{O}_{(1)} \star \mathcal{O}_{(2,1)}&=\mathcal{O}_{(2,2)}+\mathcal{O}_{(3,1)}-\mathcal{O}_{(3,2)}    , \\
\mathcal{O}_{(1)} \star \mathcal{O}_{(2,2)}&=\mathcal{O}_{(3,2)}, &    \mathcal{O}_{(1)} \star \mathcal{O}_{(3)}&=\mathcal{O}_{(3,1)}  , \\
\mathcal{O}_{(1)} \star \mathcal{O}_{(3,1)}&=q -q \mathcal{O}_{(1)}+\mathcal{O}_{(3,2)} &      \mathcal{O}_{(1)} \star \mathcal{O}_{(3,2)}&=q \mathcal{O}_{(1)} -q \mathcal{O}_{(2)}
+\mathcal{O}_{(3,3)}  , \\
\mathcal{O}_{(1)} \star \mathcal{O}_{(3,3)}&=q \mathcal{O}_{(2)}, &  \mathcal{O}_{(1,1)} \star \mathcal{O}_{(1,1)}&=\mathcal{O}_{(2,2)}  , \\
\mathcal{O}_{(1,1)} \star \mathcal{O}_{(2)}&=\mathcal{O}_{(3,1)}, &    \mathcal{O}_{(1,1)} \star \mathcal{O}_{(2,1)}&=\mathcal{O}_{(3,2)}            , \\
\mathcal{O}_{(1,1)} \star \mathcal{O}_{(2,2)}&=\mathcal{O}_{(3,3)}, &        \mathcal{O}_{(1,1)} \star \mathcal{O}_{(3)}&=q        , \\
\mathcal{O}_{(1,1)} \star \mathcal{O}_{(3,1)}&=q \mathcal{O}_{(1)}, &       \mathcal{O}_{(1,1)} \star \mathcal{O}_{(3,2)}&=q \mathcal{O}_{(2)}         , \\
\mathcal{O}_{(1,1)} \star \mathcal{O}_{(3,3)}&=q \mathcal{O}_{(3)}, &      \mathcal{O}_{(2)} \star \mathcal{O}_{(2)}&= \mathcal{O}_{(2,2)}+\mathcal{O}_{(3,1)}-
\mathcal{O}_{(3,2)}          , \\
\mathcal{O}_{(2)} \star \mathcal{O}_{(2,1)}&=q-q \mathcal{O}_{(1)}+
\mathcal{O}_{(3,2)}, & \mathcal{O}_{(2)} \star \mathcal{O}_{(2,2)}&=q \mathcal{O}_{(1)}               , \\
\mathcal{O}_{(2)} \star \mathcal{O}_{(3)}&= \mathcal{O}_{(3,2)}, &   \mathcal{O}_{(2)} \star \mathcal{O}_{(3,1)}&=q \mathcal{O}_{(1)}-
q \mathcal{O}_{(2)}+ \mathcal{O}_{(3,3)}             , \\
\mathcal{O}_{(2)} \star \mathcal{O}_{(3,2)}&=q \mathcal{O}_{(1,1)}+
q \mathcal{O}_{(2)}-q \mathcal{O}_{(2,1)} &  \mathcal{O}_{(2)} \star \mathcal{O}_{(3,3)}&=q \mathcal{O}_{(2,1)}   , \\
\mathcal{O}_{(2,1)} \star \mathcal{O}_{(2,1)}&=q \mathcal{O}_{(1)} -
q \mathcal{O}_{(2)}+ \mathcal{O}_{(3,3)} ,
& \mathcal{O}_{(2,1)} \star \mathcal{O}_{(2,2)}&=q \mathcal{O}_{(2)}  , \\ ,
\mathcal{O}_{(2,1)} \star \mathcal{O}_{(3)}&=q \mathcal{O}_{(1)}&  \mathcal{O}_{(2,1)} \star \mathcal{O}_{(3,1)}&=q \mathcal{O}_{(1,1)}+q \mathcal{O}_{(2)}
-q \mathcal{O}_{(2,1)}   , \\
  \mathcal{O}_{(2,1)} \star \mathcal{O}_{(3,2)}&=q \mathcal{O}_{(2,1)}+q \mathcal{O}_{(3)}
-q \mathcal{O}_{(3,1)} ,&  \mathcal{O}_{(2,1)} \star \mathcal{O}_{(3,3)}&=q \mathcal{O}_{(3,1)} , \\
\mathcal{O}_{(2,2)} \star \mathcal{O}_{(2,2)}&=q \mathcal{O}_{(3)}  ,&  \mathcal{O}_{(2,2)} \star \mathcal{O}_{(3)}&=q \mathcal{O}_{(1,1)}  , \\
\mathcal{O}_{(2,2)} \star \mathcal{O}_{(3,1)}&=q \mathcal{O}_{(2,1)} ,&   \mathcal{O}_{(2,2)} \star \mathcal{O}_{(3,2)}&=q \mathcal{O}_{(3,1)}  , \\
 \mathcal{O}_{(2,2)} \star \mathcal{O}_{(3,3)}&=q^2 ,&  \mathcal{O}_{(3)} \star \mathcal{O}_{(3)}&=\mathcal{O}_{(3,3)}  , \\
 \mathcal{O}_{(3)} \star \mathcal{O}_{(3,1)}&=q \mathcal{O}_{(2)} ,&  \mathcal{O}_{(3)} \star \mathcal{O}_{(3,2)}&=q \mathcal{O}_{(2,1)}  , \\
\mathcal{O}_{(3)} \star \mathcal{O}_{(3,3)}&=q \mathcal{O}_{(2,2)} ,& \mathcal{O}_{(3,1)} \star \mathcal{O}_{(3,1)}&=q \mathcal{O}_{(2,1)}+q \mathcal{O}_{(3)}
-q \mathcal{O}_{(3,1)}    , \\
\mathcal{O}_{(3,1)} \star \mathcal{O}_{(3,2)}&=q \mathcal{O}_{(2,2)}+q \mathcal{O}_{(3,1)}
-q \mathcal{O}_{(3,2)} ,&   \mathcal{O}_{(3,1)} \star \mathcal{O}_{(3,3)}&=q \mathcal{O}_{(3,2)}  , \\
\mathcal{O}_{(3,2)} \star \mathcal{O}_{(3,2)}&=q^2-q^2 \mathcal{O}_{(1)}+q \mathcal{O}_{(3,2)} ,&    \mathcal{O}_{(3,2)} \star \mathcal{O}_{(3,3)}&=q^2 \mathcal{O}_{(1)}
, \\
\mathcal{O}_{(3,3)} \star \mathcal{O}_{(3,3)}&=q^2 \mathcal{O}_{(1,1)}.& &
\end{align*}

\bibliography{refs}

\end{document}